\documentclass[titlepage]{article}
\usepackage[margin=1in]{geometry}
\usepackage[utf8]{inputenc}

\usepackage{hyperref}

\usepackage{amsmath}

\usepackage{multirow}
\usepackage{placeins}

\usepackage[capposition=bottom]{floatrow}

\usepackage{graphicx} % Required for inserting images

\usepackage[
backend=biber,
style=apa,
maxcitenames=2, %use a max of 2 names, then et. al
uniquelist=false, %for ambiguous keys (which we have)
]{biblatex}
\title{An Approximate Monte Carlo Simulation Method for Estimating Uncertainty and Constructing Confidence Intervals for 2020 Census Statistics}

\author{Robert Ashmead$^1$, Michael B. Hawes$^2$, Mary Pritts$^3$, Pavel Zhuravlev$^2$, and Sallie Ann Keller$^4$\\
\\
$^1$NORC, formerly U.S. Census Bureau\\
$^2$U.S. Census Bureau\\
$^3$New Light Technologies\\
$^4$U.S. Census Bureau, University of Virginia}
\date{Draft, May 12, 2024}

\begin{document}

\maketitle

%200 word limit on abstract

%currently 198

\begin{abstract}
To protect the confidentiality of the 2020 Census, the U.S. Census Bureau adopted a statistical disclosure limitation framework based on the principles of differential privacy. A key component was the TopDown Algorithm, which applied differentially-private noise to an extensive series of counts from the confidential 2020 Census data and transformed the resulting noisy measurements into privacy-protected microdata which were then tabulated to produce official data products. Though the codebase is publicly available, currently there is no way to estimate the uncertainty for statistics included in the published data products that accounts for both the noisy measurements and the post-processing performed within the TopDown Algorithm. We propose an approximate Monte Carlo Simulation method that allows for the estimation of statistical quantities like mean squared error, bias, and variance, as well as the construction of confidence intervals. The method uses the output of the \emph{production} iteration of the TopDown Algorithm as the input to additional iterations, allowing for statistical quantities to be estimated without impacting the formal privacy protections of the 2020 Census. The results show that, in general, the quantities estimated by this approximate method closely match their intended targets and that the resulting confidence intervals are statistically valid.

\vspace{0.5cm}

%up to 5

\vspace{0.5cm}

\noindent Statistics reported in this paper have been cleared for public release by the Census Bureau's Disclosure Review Board (DRB clearance number: CBDRB-FY24-DSEP-0002).

\end{abstract}

% Statistics reported were released under (DMS Project ID: P-7502798, Clearance numbers: CBDRB‑FY23‑0240 and CBDRB-FY22-DSEP-004)

%Note, Graphics for this paper are currently located in:
%\begin{verbatim}
%dms-p0-992:\ashme001\fpguide\paper_analysis\graphics_appendix 
%\end{verbatim}

\section*{Introduction}

The application of differentially-private methods in the 2020 Census data products represented one of the largest deployments of a disclosure avoidance system (DAS) based on the principles of formal privacy to date and the first by a national statistical office for one of its  primary statistical products. To accomplish the goal of utilizing a differential privacy framework, the U.S. Census Bureau developed the TopDown Algorithm (TDA) \parencite{Abowd20222020}, which was used to protect the statistics included in the 2020 Census Redistricting Data (P.L. 94-171) Summary File as well as the 2020 Census Demographic and Housing Characteristics File (DHC). The TDA produces privacy-protected microdata: person- and housing unit-level untabulated records. These records are referred to as Privacy-Protected Microdata Files (PPMFs) and are scheduled for public release in 2024.

One of the advantages of utilizing a differentially-private framework is that the methods and parameters of the disclosure limitation methodology can be made fully transparent to the public, meaning that the exact specifications of the algorithms can be released without compromising the formal privacy protections. By contrast, the integrity of the swapping mechanism used to protect the 2010 Census data products was predicated on keeping the swap rate and other key parameters (and their impacts on data utility) confidential. 

One major benefit of the transparency afforded by differentially-private methods is that the information on the random noise distributions utilized as part of the disclosure limitation algorithm can be released to help data users understand the expected level of uncertainty in the published data products. For many differentially-private methods, measures of uncertainty like variance, mean-squared error, bias, and confidence intervals can be directly calculated from the random noise distributions and resulting algorithm outputs. However, this is not the case for statistics protected using the TDA because, while the algorithm uses known random noise distributions to achieve its privacy protections, it subsequently post-processes the noisy measurements along with invariant totals (statistics that are purposely left unchanged, such as state total population counts) and other constraints, to produce statistics that are non-negative integers and that are internally and hierarchically consistent within the person statistics and within the housing statistics. This consistency ensures that the same statistic calculated from different person tables will always yield the same results, such as counts of males and females summing to the total population count, and the same statistic calculated from different housing tables will always yield the same results, such as county totals summing to state totals. To achieve these properties, the TDA utilizes numerical optimization methods to solve optimization problems with equality and inequality constraints, including non-negativity constraints, in its post-processing steps. Consequently, the resulting TDA-protected statistics do not have closed-form formulas for their corresponding measures of uncertainty. 

A further consequence of using equality and inequality constraints in the TDA optimization process is that the disclosure avoidance-related variance and bias of the estimators will depend on the underlying data itself rather than solely depending on the randomness of the noisy measurements, which means that applying the algorithm with exactly the same parameters to two different datasets could result in different expected levels of disclosure avoidance-related error. One important category of inequality constraints used in the TDA is the non-negativity constraints. Because the TDA estimate cannot be negative, if a count is zero or close to zero in the underlying confidential data, the TDA estimate  will not have the same range of randomness as it would have had if the underlying confidential count was larger. This likely will result in a smaller variance and slight bias in the positive direction. Similarly, because the TDA kept state total population counts invariant, the small positive bias for small counts can result in a corresponding small negative bias for larger counts.

There is currently no generalized way to estimate the statistical properties of specific queries %\footnote{The term ``query" here is used in its general sense to mean any statistical calculation, and should not be confused with more specific uses of the term in other discussions of the 2020 Census Disclosure Avoidance System where ``query" often refers to the specific tabulations taken of the confidential Census data to which differentially-private noise is added (e.g., ``the TDA query strategy" or ``privacy-loss budget allocation by query"). To avoid confusion, we will refer to this more specific latter TDA-specific usage as "noisy measurements" and use "query" exclusively in the more general sense.} 
tabulated from the results of the TopDown 
Algorithm. The Noisy Measurement Files (NMFs), which contain the noisy measurements observed in the initial stages of TDA iterations, are publicly available for both the Redistricting Data File and the Demographic and Housing Characteristics file 2020 production iterations \parencite{abowd2023PLNMF,abowd2023DHCNMF}. These NMFs can be used to generate point estimates and confidence intervals for arbitrary query answers and arbitrary geographic entities composed of block geographic units, but these estimates do not leverage the accuracy improvements provided by the subsequent post-processing steps used by the TDA. The goal of this work is to propose a method that can approximate the statistical properties and uncertainty of any statistic tabulated from the TDA's post-processed privacy-protected microdata output.

The remainder of the paper is organized as follows. First, we motivate and describe the approximate Monte Carlo simulation method and propose methods to create estimates of statistical quantities and confidence intervals from the simulation iterations. Then, we test the methodology on the 2010 Census Redistricting data and Demographic and Housing Characteristics data, respectively. In the final section we summarize and discuss possible future work.

\section*{The Approximate Monte Carlo (AMC) Simulation Method} \label{sec:methodology}

The primary confidential data input for the Census application of the TDA is the Census Edited File (CEF), the final version of the confidential census data before disclosure avoidance is applied.  The CEF consists of records for all housing units and occupied group quarters facilities, and for all persons residing in those housing units and group quarters. For the purposes of estimating the statistical properties of the TDA in this article, we use the CEF as the standard of truth\footnote{Note that other USCB publications have calculated error relative to the 2010 Census Hundred-percent Detail File (HDF), which is the 2010 Census internal microdata file with the 2010 disclosure avoidance methodology (swapping) applied and from which all published 2010 Census data products were tabulated.}, ignoring other sources of error that may be present in the CEF, such as coverage or measurement error relating to the Census enumeration itself.

Theoretically, we could use a standard Monte Carlo simulation to estimate the statistical properties of the TDA. This would involve running the TDA many times using the same CEF input file, each time taking new draws of the noisy query measurements from the confidential CEF and independently post-processing those noisy measurements into multiple, independent privacy-protected microdata files. In practice, however, taking additional noisy measurements of the confidential data would contribute to the total privacy loss (multiplying the privacy loss by the number of simulations used), significantly weakening the confidentiality protections provided to census respondents. 

The concept behind the proposed \textit{approximate} Monte Carlo (AMC) simulation method is relatively simple. Instead of repeatedly executing the TDA using the confidential CEF data as input, we repeatedly execute the TDA using the already-privacy-protected  PPMF \textit{output} of the algorithm as the \textit{input} to the TopDown Algorithm used in the AMC simulations. Because using a realization of the output of the algorithm as the input to further Monte Carlo simulations does not involve accessing the confidential data, the AMC method does not contribute to additional privacy loss.

The motivation here is that if we consider the output of the TDA to be a sufficiently accurate reflection of the CEF itself, then we should get similar statistical properties when using it as the input to the TDA compared with using the CEF as the input. In the limiting case, if it was exactly equivalent to the CEF, then we would get exactly the same distribution. We hypothesize that the less accurate the input to the TopDown Algorithm is relative to the CEF (\textit{e.g.}, the lower the privacy-loss parameter(s) used to generate the noisy measurements from the CEF), the less accurate the AMC simulation method will be. 

\subsection*{Mathematical Description}

Let $\theta^{p}_{CEF}$ and $\theta^{h}_{CEF}$ be the multidimensional fully saturated contingency tables that are counts of the number of persons and households respectively with specific combinations of characteristics by block in the CEF.  $\theta^{p}_{CEF}$ and $\theta^{h}_{CEF}$ represent the data unadjusted for disclosure avoidance. We can therefore think of $\theta^{p}_{CEF}$ and $\theta^{h}_{CEF}$ as the parameters we are estimating in the TDA. Let $TDA(\mathbf{X^p},\mathbf{X^h})$ represent the TDA with a fixed set of parameters relating to the privacy-loss budget and invariants.  Let $\mathbf{X^p}$ and $\mathbf{X^h}$ represent the input person and household datasets. A realization of the TDA applied to the CEF tables is represented as $\{\hat{\theta}^p, \hat{\theta}^h \} = TDA(\theta^p_{CEF}, \theta^h_{CEF})$ where the dimensionality of $\hat{\theta}^p$ and $\hat{\theta}^h$ will match that of the input data.  
Let $q()$ represent a generic univariate query (e.g. the number of voting-age persons in Alaska; the number of Hispanic persons in Franklin County, Ohio; the persons per household in a specific census tract) so that $q(\theta^p_{CEF}, \theta^h_{CEF})$ would represent the answer to the query using the CEF persons and/or housing data.  

\subsubsection*{Standard Monte Carlo Estimates} 

Consider a number ($m$) of independent realizations of the TopDown Algorithm from the CEF data

$$\{\hat{\theta}^p_i, \hat{\theta}^h_i \} \sim TDA(\theta^p_{CEF}, \theta^h_{CEF}), \; i=1, \cdots, m.$$

\noindent Monte Carlo empirical estimates of bias, variance, and MSE for any query $q(\cdot)$ could be calculated as

\begin{equation}
\widehat{BIAS}_{MC} = \frac{1}{m}\sum_{i=1}^{m} q(\hat{\theta}^p_i, \hat{\theta}^h_i) - q(\theta^p_{CEF}, \theta^h_{CEF}) \end{equation}

\begin{equation}
\widehat{VAR}_{MC} = \frac{1}{m-1}\sum_{i=1}^{m} \left( q(\hat{\theta}^p_i, \hat{\theta}^h_i) - \overline{q(\hat{\theta}^p_i, \hat{\theta}^h_i)} \right)^2,
\end{equation}
where $\overline{q(\hat{\theta}^p_i, \hat{\theta}^h_i)} = \frac{1}{m}\sum_{i=1}^{m}q(\hat{\theta}^p_i, \hat{\theta}^h_i),$ and

\begin{equation}
\widehat{MSE}_{MC} = \frac{1}{m}\sum_{i=1}^{m} \left( q(\hat{\theta}^p_i, \hat{\theta}^h_i) - q(\theta^p_{CEF}, \theta^h_{CEF}) \right)^2,
\end{equation}

\noindent
respectively. As $m$ goes to infinity, assuming finite variance and appropriate finite moments, these calculations will be asymptotically equal to the bias, variance, and MSE of the query for the algorithm applied to the CEF. 

\subsubsection*{Approximate Monte Carlo (AMC) Simulation Estimates}

Let $\{\hat{\theta}^p_0, \hat{\theta}^h_0 \} = TDA(\theta^p_{CEF}, \theta^h_{CEF})$ be a single realization from the TopDown Algorithm that represents the production iteration.\footnote{This is comparable to the single realization that is used to create the PPMF and associated published tabulations.} Next, assume we simulate $s$ independent realizations of the algorithm using $\{\hat{\theta}^p_0, \hat{\theta}^h_0 \}$ in place of the CEF as input into the TDA, meaning we have

$$\{\tilde{\theta}^p_i, \tilde{\theta}^h_i \} \sim TDA(\{\hat{\theta}^p_0, \hat{\theta}^h_0 \}), \; i=1, \cdots, s.$$

\noindent Then, Monte Carlo empirical estimates of bias, variance, and MSE for any query $q(\cdot)$ \textit{relative to} $\hat{\theta}^p_0, \hat{\theta}^h_0$ could be calculated as, respectively

\begin{equation}
\widehat{BIAS}_{AMC} = \frac{1}{s}\sum_{i=1}^{s} q(\tilde{\theta}^p_i, \tilde{\theta}^h_i ) - q(\hat{\theta}^p_0, \hat{\theta}^h_0 ), \end{equation}

\begin{equation}
\widehat{VAR}_{AMC} = \frac{1}{s-1}\sum_{i=1}^{s} \left( q(\tilde{\theta}^p_i, \tilde{\theta}^h_i ) - \overline{q(\tilde{\theta}^p_i, \tilde{\theta}^h_i)} \right)^2,
\end{equation}
where $\overline{q(\tilde{\theta}^p_i, \tilde{\theta}^h_i)} = \frac{1}{s}\sum_{i=1}^{s}q(\tilde{\theta}^p_i, \tilde{\theta}^h_i),$ and 

\begin{equation}
\widehat{MSE}_{AMC} = \frac{1}{s}\sum_{i=1}^{s} \left( q(\tilde{\theta}^p_i, \tilde{\theta}^h_i) - q(\hat{\theta}^p_0, \hat{\theta}^h_0) \right)^2.
\end{equation}

\noindent The proposed AMC simulation method relies on the assumption that for a given query of interest, $BIAS_{AMC} \approx BIAS_{MC}$,  $VAR_{AMC} \approx VAR_{MC}$, and $MSE_{AMC} \approx MSE_{MC}$ when \{$\theta^p_{CEF}, \theta^h_{CEF}\} \approx \{\hat{\theta}^p_0, \hat{\theta}^h_0\}$. If those quantities are similar, we can utilize the approximate quantities for describing the statistical properties of the algorithm and constructing confidence intervals, all without adding to the privacy loss.

\subsection*{Constructing Confidence Intervals From the Simulations}

For a given query $q(\cdot),$ we want to construct a confidence interval that reflects the uncertainty of $q(\hat{\theta}^p_0, \hat{\theta}^h_0)$, the query answer from the production iteration of the algorithm. We examine two basic strategies for constructing confidence intervals from the approximate Monte Carlo simulations. The first is a Wald-type interval, assuming a Gaussian or Student's $t$ distribution for the error. Given that the noisy measurements from the TDA were generated using a discrete Gaussian distribution, this would seem to be a reasonable assumption, though one we will examine further.  The second is a quantile-based confidence interval based on empirical quantiles of the approximate Monte Carlo simulations.  Additionally, we propose and examine bias-adjusted confidence intervals to better account for bias in the TDA estimates as measured by the AMC simulations.  

\subsubsection*{Wald-type Confidence Intervals} \label{sec:wald_cis}

If we were to assume that the query estimates produced by the TDA are unbiased and approximately normally distributed, then a Wald-type interval given by 

\begin{equation}
q(\hat{\theta}^p_0, \hat{\theta}^h_0) \pm \mathcal{Z}_{1-\frac{\alpha}{2}} \sqrt{\widehat{VAR}_{AMC}}    
\end{equation}
would be the best choice. As acknowledged above, however, we know that in some cases the TDA yields biased query estimates which will decrease the accuracy and coverage of our confidence intervals. Replacing the estimated variance with the estimated mean-squared error results in a more statistically conservative interval.  When the estimated bias is small, the width of the interval will be the same or only slightly larger as if we had used the variance.  For larger estimated bias, the interval width will increase proportionally.  That confidence interval is given by      

\begin{equation}
\label{RMSEci}
q(\hat{\theta}^p_0, \hat{\theta}^h_0) \pm \mathcal{Z}_{1-\frac{\alpha}{2}} \sqrt{\widehat{MSE}_{AMC}}.    
\end{equation}

In addition to the Gaussian distributions for the margins-of-error, we also consider whether a heavier-tailed distribution like a Student's $t$ distribution is more appropriate. We chose to use five degrees of freedom based on examining the results of our experiments;\footnote{The critical value for a 90\% confidence would be 2.015 vs. 1.645 for the Gaussian distribution, an approximately 25\% increase.} however, we do not have a theoretical rationale for this value. Lastly, if the query of interest is a count, we propose to apply a floor and ceiling function to the endpoints of the confidence intervals respectively and truncate at zero to account for the outcome space being integer and non-negative.

\subsubsection*{Quantile-based Confidence Intervals}

If we have a large enough number of simulations, another strategy for constructing confidence intervals would be to use a non-parametric method based on the quantiles of the simulated query answers. Let $p_{\alpha}( q(\tilde{\theta}^p_i, \tilde{\theta}^h_i )) $ represent the $\alpha$th percentile of the approximate Monte Carlo Values from the simulated estimates of the quantile.  Then, for example, a $90\%$ approximate confidence interval could be estimated as   

\begin{equation}
\left( p_{05}( q(\tilde{\theta}^p_i, \tilde{\theta}^h_i)),
p_{95}( q(\tilde{\theta}^p_i, \tilde{\theta}^h_i))
\right)
\end{equation}

If the query of interest is a count, we propose to apply a floor and ceiling function to the endpoints of the confidence intervals, respectively, to account for the outcome space being integer. A major downside of the quantile-based method is that in order to calculate relatively precise quantiles, it is necessary to have a large number of simulations.  Given that the TDA is computationally expensive to run\footnote{A single run of the TDA will generally cost around \$1,500 for the redistricting data schema, and around \$10,000 for the DHC data schema}, this is a major limitation.

\subsubsection*{Bias-corrected Confidence Intervals}

We know that in some cases, the TDA estimator is biased.  Therefore, if the estimate of bias from the AMC method is accurate, then an improved point estimate and confidence interval could be constructed for the Wald-type intervals by subtracting the estimated bias from the point estimate and using that as the pivotal value for the confidence interval. For quantile-based intervals, we propose subtracting the estimated bias as constructed by using the median of the iterations relative to $\{\hat{\theta}^p_0, \hat{\theta}^h_0 \}$ from each of the endpoints of the confidence interval.

\subsection*{Comparisons With Existing Literature} 

Several authors have proposed bootstrap-related methods which are similar to the proposed AMC method in that they attempt to estimate the statistical properties of a differentially-private estimator empirically rather than using closed-form formulas based on the noise distribution. Brawner and Honacker \parencite*{brawner2018bootstrap} proposed a bootstrap method in which the original dataset is bootstrapped before noisy queries are generated and then the average and standard deviation of the bootstrapped answers is used to construct a confidence interval. 
 Others have extended these ideas \parencite{wang2022differentially, chadha2024resampling}.  Relatedly, parametric bootstrap methods   \parencite{ferrando2022parametric, WangDebiasedPB} have also been proposed, which apply when a distributional form has been assumed for the underlying data.

Preliminary work on the AMC simulation method proposed in this article applied an early version of the TDA to the 1940 Decennial Census and found that the method did well in estimating variance \parencite{Ashmead2019JSM}. While our methods share some of the same ideas and motivations as the bootstrap methods, our specific experiment setup is unique in several important ways. First, the TDA is designed to produce microdata. As a consequence, the input and output parameter space of the algorithm are equivalent, allowing us to use the algorithm output as the input to our simulations. Second, many of the bootstrap methods rely on incorporating the methodology as part of the design of the differentially private algorithm. In the census use case, the methodology for creating confidence intervals was applied afterwards. Lastly, the TDA is computationally expensive, so it is necessary to keep the number of simulations or bootstraps to a modest number.
 
\section*{Experimental Results Using 2010 Census Redistricting Data}\label{sec:redistrict}

To test the validity of the AMC method, we conducted Monte Carlo simulations utilizing the same parameters from the 2020 TDA redistricting production settings applied to the 2010 Decennial Census data. The choice of employing the 2010 dataset was deliberate, as it enabled a comparison between the TDA output and the confidential input data. This approach was adopted to safeguard the privacy assurances associated with the 2020 data. We argue that the validity of the results obtained from the 2010 dataset extends to the 2020 data, given we are using the same TDA parameters and the two datasets are generally similar in structure.

We first completed 100 independent simulations of the TDA using the confidential 2010 CEF as input.  These simulations represent the sampling distribution of the TopDown Algorithm applied to the CEF and serve as a gold standard to the AMC method. Second, using a single privatized PPMF output\footnote{Technically the output of the TopDown Algorithm is called the microdata detail file (MDF), the format and metadata of which differ slightly from published PPMFs.  We used the MDF for our experiments, but all of the necessary detail is available on the PPMF.} (which we call $\text{PPMF}_0$) from the first set of iterations in place of the CEF, we completed 100 additional simulations of the algorithm to be used for the AMC estimates.

In order to test the methodology, we selected all queries (301 in total) from the redistricting data summary file \parencite{pl94tables} and tabulated them using  the simulations at multiple geographic levels. Queries were taken from the following tables:

\begin{itemize}
\item P1. Race
\item P2. Hispanic or Latino, and not Hispanic or Latino by Race
\item P3. Race for the Population 18 Years and Over
\item P4. Hispanic or Latino, and not Hispanic or Latino by Race for the Population
18 Years and Over
\item P5. Group Quarters Population by Major Group Quarters Type
\item H1. Occupancy Status (Housing)
\end{itemize}

Given the large number of geographies at the block level, we selected a stratified sample of blocks to keep the experiment analysis more manageable. The strata were created by grouping blocks by CEF total population. We randomly sampled 5 blocks for each state within each strata.

This experiment has three major goals. The first is to determine how well the approximate quantities match their empirical counterparts.  We do this by comparing the empirical RMSE, bias, and standard deviation estimates graphically. The second is to determine if confidence intervals based on the approximate quantities have the appropriate coverage. In order to properly test the coverage, it would be necessary to repeat the entire simulation many times starting with creating a new $\text{PPMF}_0$ followed by the AMC simulations. This is too computationally expensive for this use case, so instead we consider the proportion of confidence intervals that contain the CEF query value across different queries. Lastly, we examine if it is feasible to utilize the method with a fewer number of AMC simulations.

\subsection*{Results Comparing RMSE, Bias, and Standard Deviation Estimates} \label{sec:rmse_bias_sd}

We created graphics to compare RMSE, bias, and standard deviation estimates for each of the three statistics and for queries within each of the following geographic levels: U.S., state, county, tract, block, American Indian/Alaska Native (AIAN) area, and elementary school districts. Figures shown here are meant to be representative of larger patterns or to communicate a specific point. Additional results are available in the online appendix (Figures \ref{fig:nat_rmse} - \ref{fig:sd_sdev}).  

\begin{figure}[!ht]
\includegraphics[width = \textwidth]{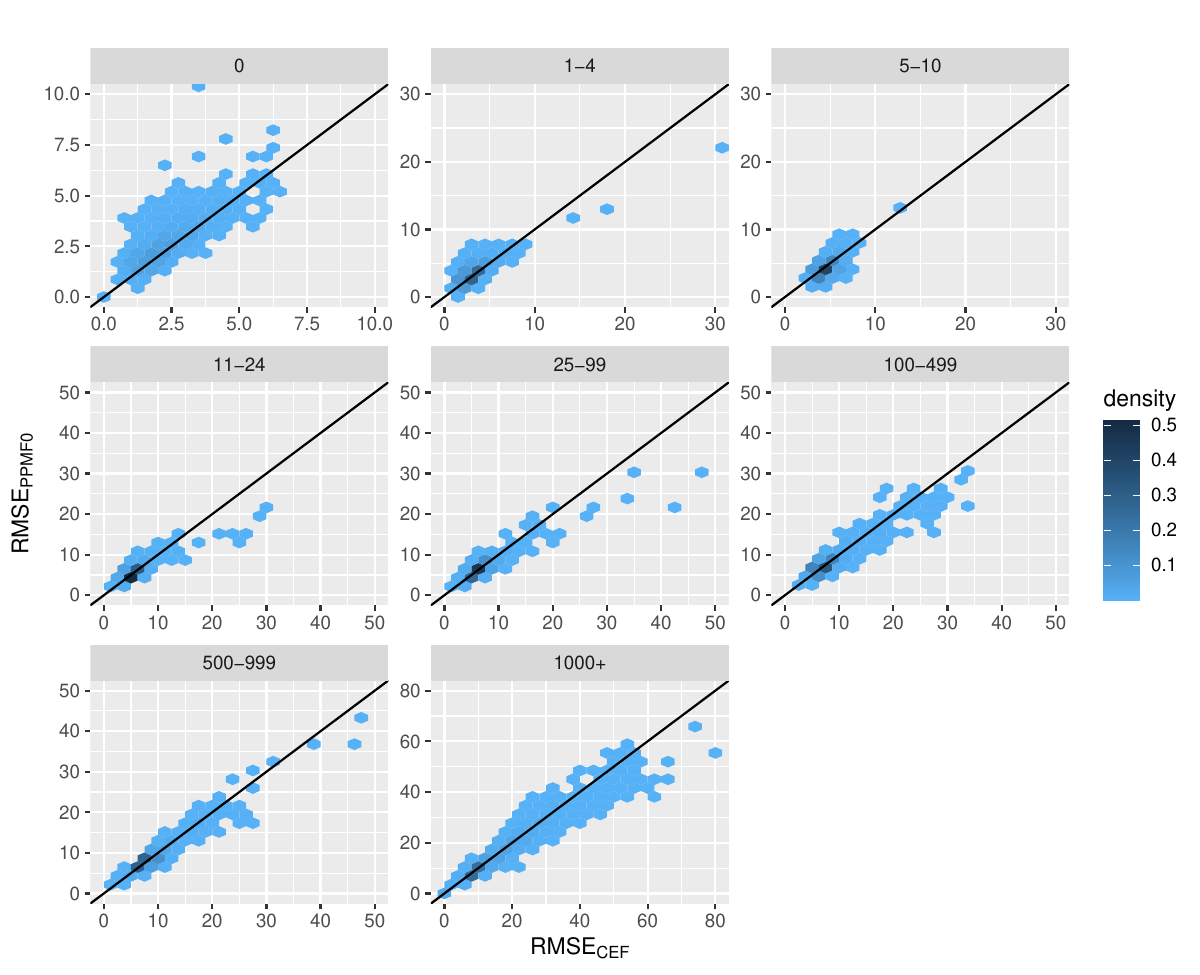}
\caption{State level comparison of estimated RMSE ($RSME_{PPMF0}$ vs. $RMSE_{CEF}$). Each panel corresponds to certain sizes (i.e. counts) of the Query (based on CEF). In the case when $RSME_{PPMF0}$ values are very similar to $RMSE_{CEF}$ values they line up along the diagonal, showing the AMC method to be a good approximation. This is mostly the case with the data shown. Hexagons show location and density of the $15,652$ (301 queries x 52 geographies) points used for the plot.}\label{fig:state_rmse}
\end{figure}

Figure \ref{fig:state_rmse} shows a set of hexplots of the RMSE of the $15,652$ query estimates\footnote{301 queries x 52 geographies} at the state level with the estimate coming from the CEF-generated calculations on the x-axis and the $\text{PPMF}_0$-generated calculations along the y-axis. Hexplots divide the plotting area into hexagons each reflecting locations and density of the $15,652$ points. The darker the shade, the more individual points were in the hexagon. The plot is faceted by the CEF value of the query and note that the range of the axes changes between subplots. Ideally, we would like to see all the points fall along the diagonal which would indicate that the AMC-based values are equivalent to their CEF-based counterparts and could therefore be substituted for one another freely.  For the most part, we see that the estimates closely match one another. Most of the data points are close to the diagonal line; however, in the subplots for CEF query size 0 and 1-4, we see some points were above the diagonal line, indicating the RMSE from the AMC was slightly larger than from the CEF in some cases.

Looking at the bias from the standard Monte Carlo method (which used the CEF as input to the TDA) compared to the bias from the AMC method, we saw the estimates generally match one another across geographies and statistics; however, as we get to smaller geographies (tract, block, and elementary school district) we start to see some underestimation of the bias from the AMC method when it is further from zero.  See for example Figure \ref{fig:sd_bias}. This is most apparent for queries in the larger groupings.  As a result, the RMSE estimates of those queries for the AMC are also slightly underestimated.

\subsection*{Are TDA Estimates Approximately Normally Distributed?}

\begin{figure}[!ht]
\includegraphics[width = \textwidth]{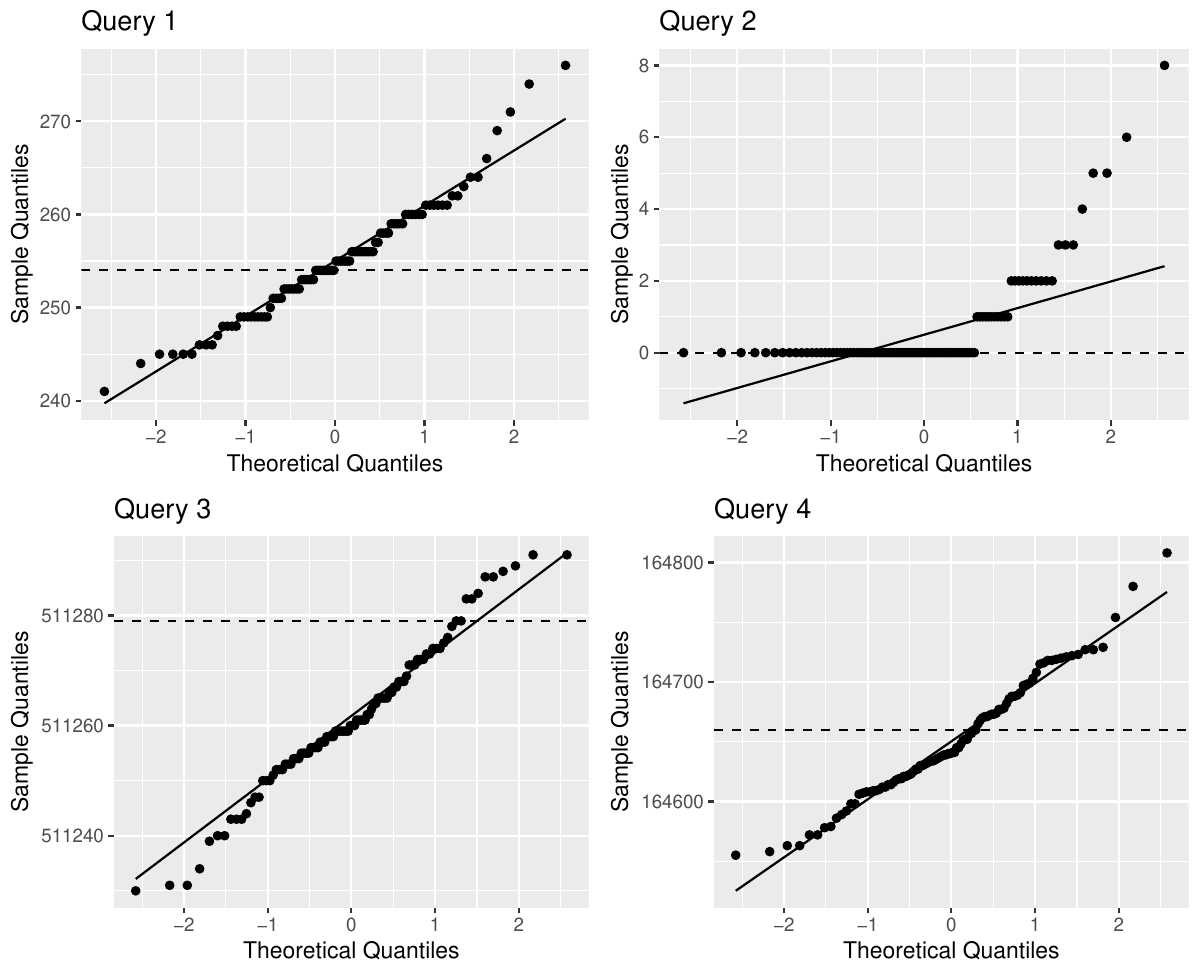}
\caption{Quantile-quantile example plot of four example 2010 redistricting queries at the state level. The plot shows the distribution over 100 AMC iterations in order to assess if the distribution is approximately normally distributed. Query 1 is the P0040020 query (Population of Two Races: Black or African American; Native Hawaiian and Other Pacific Islander) for Arizona; Query 2 is the P0040064 query (Population of Four Races: American Indian and Alaska Native; Asian; Native Hawaiian and Other Pacific Islander; Some Other Race) for New Jersey; Query 3 is the P0010003 query (Population of One Race: White Alone) for Wyoming; Query 4 is the H0010003 query (Vacant Units) for Maine.   For each query, there are 100 query counts from the AMC iterations plotted against the quantiles of the theoretical normal distribution. Horizontal dashed lines represent $\text{PPMF}_0$ value.  Theoretical quantiles are those of a Gaussian distribution using the estimated mean and variance from the 100 iterations. Sample quantiles for Queries 1, 3 and 4 correspond very closely to theoretical normal distribution quantiles (less closely at the tails). Query 2 had PPMF$_0$ value (count) of zero, and non-negativity constraints rendered the distribution obviously non-Gaussian.}\label{fig:qqplot}

\end{figure}

In order for the Wald-type confidence intervals to be valid, the distribution of the TDA estimates for a query needs to be approximately normally distributed. We examine this assumption by looking at the empirical distributions of 100 TDA runs on four selected queries in Figure \ref{fig:qqplot} through a quantile-quantile plot, which compares the two distributions to each other.  Queries 1, 3, and 4 have distributions that very closely resemble a normal distribution. In each of those cases, there are some points on the tail that are slightly more extreme than the theoretical quantiles. The TDA estimates from Query 3 are also biased in that the $\text{PPMF}_0$ value is about 20 larger than the center of the distribution.  Query 2 is a case where the TDA distribution is obviously not normally distributed because of restriction to non-negative values.  In this case the $\text{PPMF}_0$ value of the query was zero and most of the resulting TDA estimates were also zero, but a few end up being larger.  While the results from queries like Query 2 are clearly not normally distributed, we we will see later on from our experiments that the confidence intervals still appear to be valid (and actually statistically more conservative than for larger queries) in terms of their coverage.  These queries are representative of what we found in our exploratory results and while these examples are in no way exhaustive of the possible queries, they show that, for queries that are substantially larger than zero, a normal approximation or a similar distribution with slightly thicker tails is a reasonable assumption to use when constructing confidence intervals about the TDA estimates.   

\FloatBarrier

\subsection*{Confidence Intervals for Redistricting Data}

A valid 90\% confidence interval estimator will contain the true population parameter at least 90\% of the time upon repeated sampling. In an ideal experiment we could calculate the coverage for each individual query across multiple iterations of the process. In this scenario, that would include multiple iterations of both the $\text{PPMF}_0$ creation and the subsequent additional AMC iterations. This was judged to be infeasible given the computational cost of running the full TDA at scale.  Instead, we compute the proportion of the confidence intervals that include the population parameter (here the confidential CEF-based count) for different groups of queries and for a single iteration $\text{PPMF}_0$. This should not be considered the actual coverage of the confidence intervals because confidence intervals for different queries are correlated to varying degrees; however, we do think it is still adequate in assessing the validity of the methodology given that many queries (especially across geographies and about different attributes) should have small correlation.   

In this experiment, we considered eight different types of nominal $90\%$ confidence intervals.\footnote{$90\%$ confidence intervals are the standard of the U.S. Census Bureau} Labels for the CI type are in parentheses. First, we considered the non-parametric quantile-based method (np) and a bias-corrected non-parametric quantile-based method (BCnp) which subtracted the estimated bias (constructed from the difference between the median value across the iterations and the $\text{PPMF}_0$ value) from the interval. Next, we constructed a RMSE-based confidence interval assuming a normal distribution like Equation \ref{RMSEci} (z) and a Student's $t$ distribution with 5 degrees of freedom (t). Confidence interval types (BCz) and (BCt) are bias-corrected versions of z and t intervals in which the estimated bias was subtracted from the interval. Lastly, versions of the confidence intervals were constructed using a conditional approach in which the bias-corrected interval was used only if some criteria was met and otherwise the non-bias-corrected versions were utilized. These were created for both the Gaussian (cz) and Student's $t$ (ct) based CIs.

The criteria for using the bias-corrected versions were 

\begin{itemize}
\item The $\text{PPMF}_0$ value was greater than 5; and
\item The absolute value of the estimated bias divided by the estimated standard deviation of the AMC estimates was greater than or equal to 0.5; and
\item The estimated bias was either negative or the $\text{PPMF}_0$ value was at least 25. 
\end{itemize}

The rules for the conditional intervals were determined based on the observations that the estimated bias for very small counts was often not very reliable and that the estimated bias for small counts was almost always positive. The rule concerning the size of the bias relative to the standard deviation was motivated by the desire to only make the adjustment when the estimated bias would be impactful. The thresholds for the three criteria were selected based on our empirical testing.  With the conditional rules, the majority of the time the criteria was not met, but when it was it typically made the confidence intervals more accurate. In summary, the eight confidence interval types used and the labels that we will use to refer to them are:
\begin{enumerate}
\item np: Non-parametric quantile-based CI;
\item BCnp: Bias-corrected non-parametric quantile-based CI;
\item z: RMSE-based Wald CI with normal distribution;
\item t: RMSE-based Wald CI with Student's $t$ distribution;
\item BCz: Bias-corrected RMSE-based Wald CI with normal distribution;
\item BCt: Bias-corrected RMSE-based Wald CI with Student's $t$ distribution;
\item cz: Conditionally bias-corrected RMSE-based Wald CI with normal distribution; and
\item ct: Conditionally bias-corrected RMSE-based Wald CI with Student's $t$ distribution.
\end{enumerate}

\begin{figure}[!ht]
\caption{The proportion of 780,493 90\% confidence intervals that contained the CEF value for 2010 block level PL94 queries aggregated by query size (the bracketed number in the lower right corner indicates the proportion of the queries in that size category, e.g. in this figure most, 0.91, have value of 0). Non-parametric CIs perform the most poorly with respect to containing the CEF value. For the Wald-based CIs, the ones using Student's $t$ distribution performed slightly better than the ones using the normal distribution. Bias correction slightly improves performance only in the last two panels (largest query size (count) groups). Conditional bias correction (ct) combines the best properties of bias-uncorrected and bias-corrected CIs.}
\includegraphics[width = \textwidth]{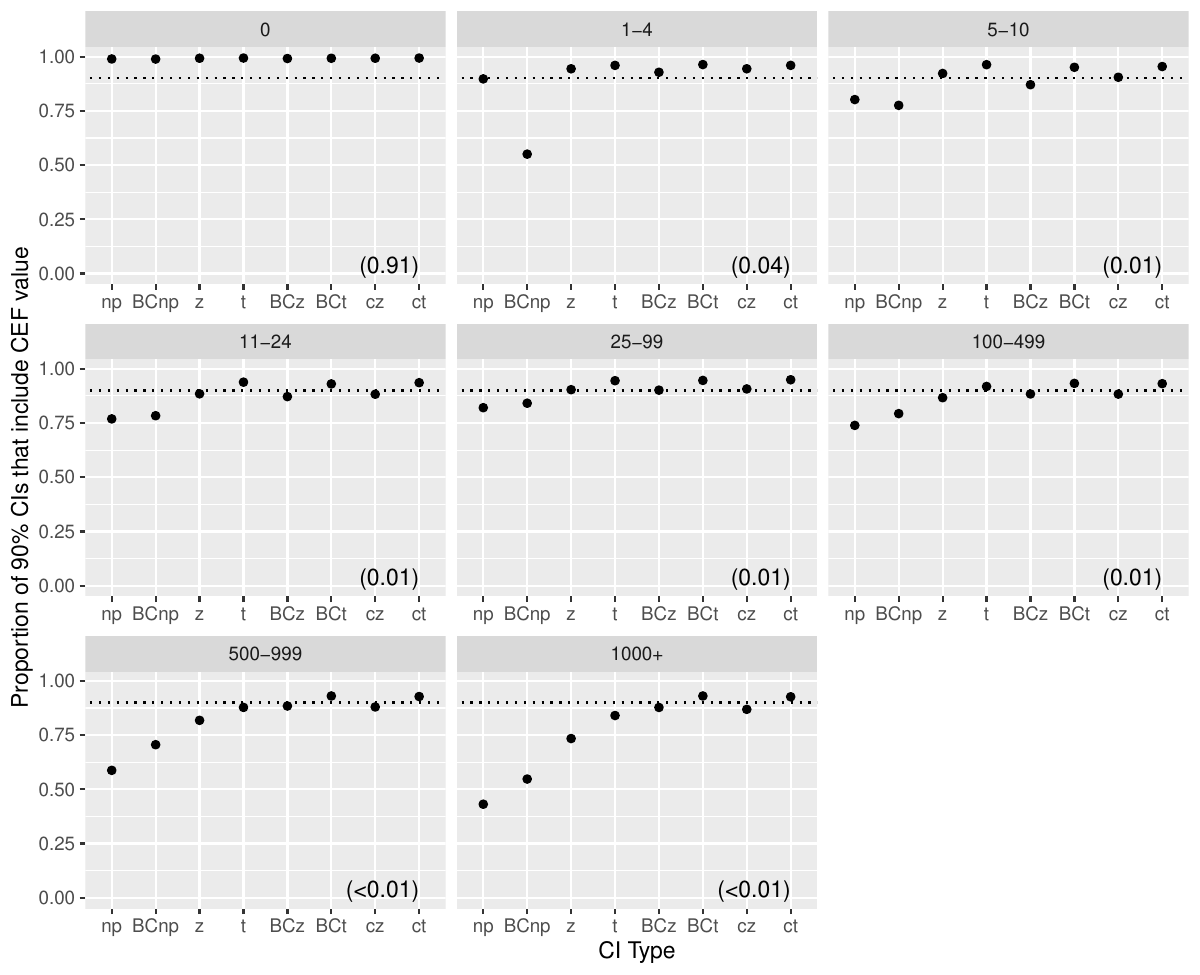}
\label{fig:block_pl94_cis}
\end{figure}

Figure \ref{fig:block_pl94_cis} shows the proportions of the confidence intervals that include the CEF values by query size (count) groupings for the block-level. Similar graphics are reported in Figures \ref{fig:nat_pl94_cis} - \ref{fig:sd_pl94_cis} at other geographic levels. Across all the geographies, the two non-parametric methods (np and BCnp) were the least likely to meet the 0.9 threshold, especially for larger queries.  Conversely, the Wald-type intervals were consistently able to meet or surpass the 0.9 threshold. The intervals using the Student's $t$ quantile (t, BCt, ct) are wider and thus have slightly higher coverage than the (z, BCz, cz) intervals using a normal quantile. For the Wald-type intervals, the bias-correction mostly does not make a difference and sometimes slightly decreases the proportion of the CIs that contain the CEF value.  In a few cases, like the two largest query size (count) groups in Figure \ref{fig:block_pl94_cis}, the bias-correction makes a positive and important difference. The conditional bias-corrected confidence intervals (cz, ct) do a fairly good job of balancing the best properties of the unadjusted and bias-adjusted intervals.    

%Note - removed this text when removed the additional graphic
% The AIAN area results are similar to those of many of the larger geographies, while the block-level results show the benefit of the bias-correction for specific cases. 

\subsection*{Confidence Interval Width}

\begin{figure}[!ht]
\caption{The distribution of $969,521$ (301 queries x 3221 geographies) 90\% confidence interval widths for 2010 County Level PL94 queries aggregated by query size. As is to be expected, $t$ and BC$t$ intervals are slightly wider than the $z$ and BC$z$ intervals; also the confidence interval width generally increases as the size of the query increases.}
\includegraphics[width = \textwidth]{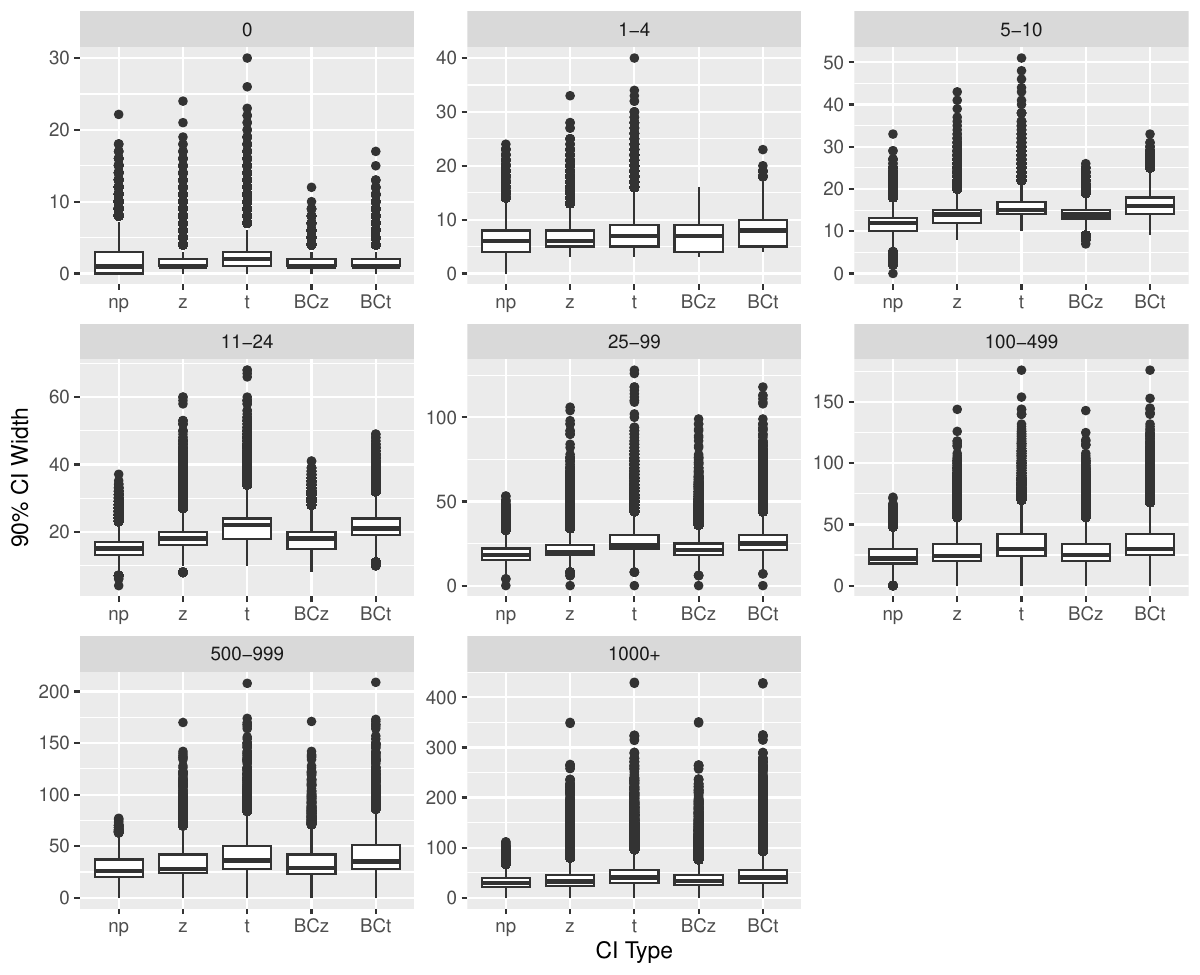}
\label{fig:county_pl94_cis_width}
\end{figure}

We compared summaries of confidence interval widths between the different methods across multiple geographic levels. The summary for the county level queries is shown in Figure \ref{fig:county_pl94_cis_width} and is generally representative of the patterns in the other geographic levels (Figures \ref{fig:nat_pl94_cis_width} - \ref{fig:sd_pl94_cis_width}). We omit the bias-adjusted non-parametric and conditional confidence interval methods in these plots as the widths are similar to the results shown (BCnp is similar to np; ct and cz were similar to z and t, respectively). The widths were generally slightly smaller for the non-parametric CIs which relates to the smaller probability of containing the CEF value that we observed in Figure \ref{fig:block_pl94_cis}. As we would expect, t and BCt intervals are slightly wider than the z and BCz intervals.  In general, the confidence interval width increases as the size of the query increases.  The distributions of widths are relatively similar across geographic levels within the query size groupings. In general, the CI widths appear to be small enough to provide meaningful information to users, with the exception of some outliers.

\subsection*{The Effect of the Number of AMC Simulation Iterations on the Estimates}

\begin{figure}[!ht]
\caption{The proportion of 90\% ct confidence intervals that contained the CEF value by geography, query size, and number of iterations. Color reflects the number of simulations in the analysis, that is, the different color points clustered together mean no significant dependence on the number of runs. The ordinate axis (Y) is the CI coverage and the abscissa axis (X) is the value of the query. The panels correspond to different geolevels. Reducing to 25 iterations decreased the proportion slightly in most of the groupings compared with using a higher number of iterations, but still generated intervals that met the 90\% threshold.}
\includegraphics[width = \textwidth]{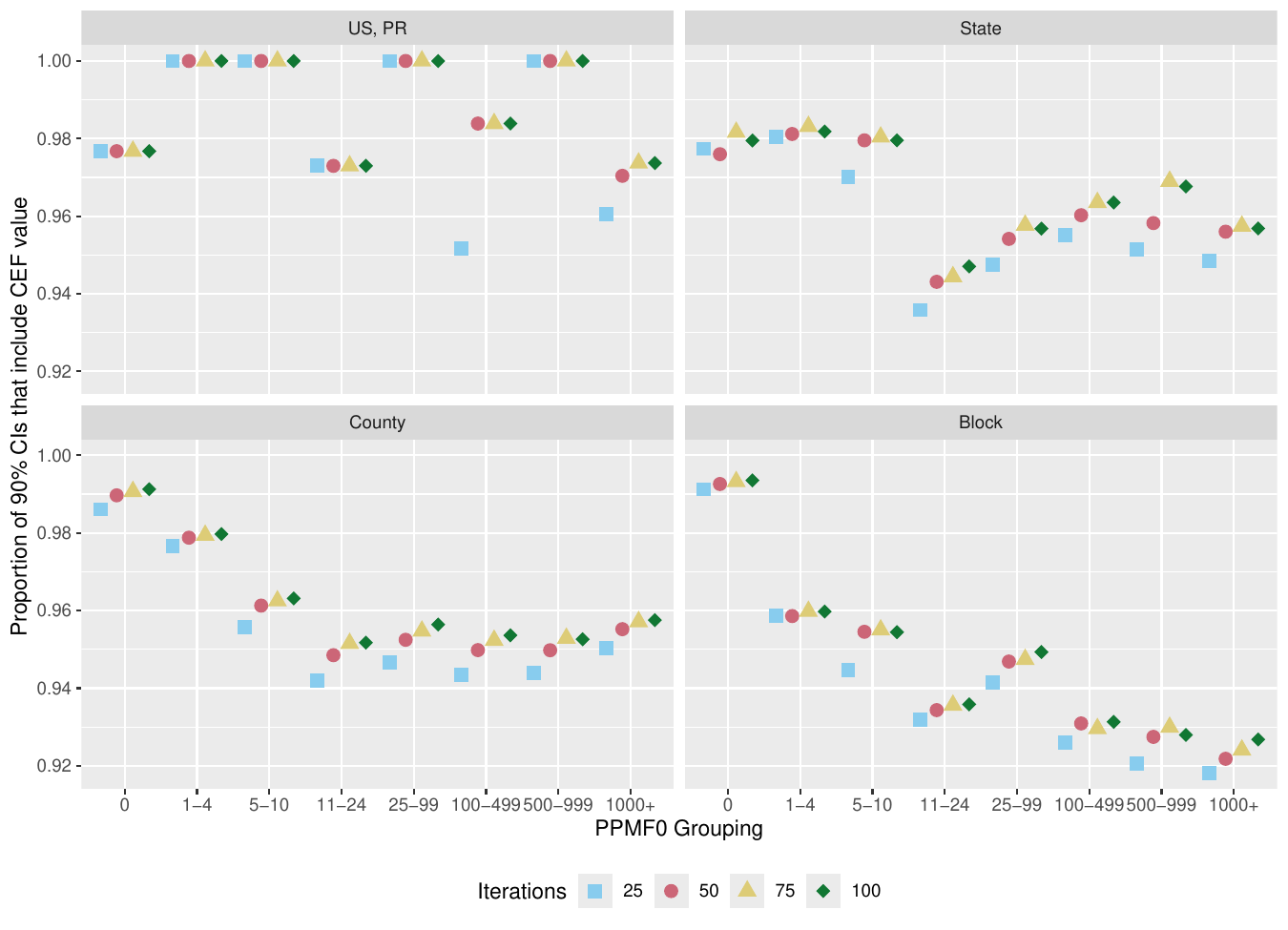}
\label{fig:diff_n_runs}
\end{figure}

Running the TDA incurs a significant computational cost and being able to reduce the number of iterations needed to estimate statistical quantities used for the confidence intervals is beneficial. Non-parametric quantile methods were not considered in the remaining analysis because they did not perform sufficiently well based on 100 iterations. In order to test whether fewer iterations would be feasible using the Wald-type intervals, we used random subsets of sizes 25, 50, and 75 of the 100 iterations and examined the difference in the proportion of the confidence intervals that contained the CEF value and how the estimated statistical quantities (bias, RMSE) compared to one another. We  found that reducing the iterations to  25  yielded RMSE metrics that were only slightly less precise and only slightly lower on average.  Figure \ref{fig:diff_n_runs} shows the proportions of ct 90\% confidence intervals that contained the CEF value by geographic level and query grouping. Reducing to 25 iterations decreased the proportion slightly in most of the groupings compared with using a higher number of iterations, but still generated intervals that met the 90\% threshold. Given these results, we decided to use only 25 Monte Carlo iterations in our additional experiments below that explore applying our proposed approach to the Demographic and Housing Characteristics PPMFs that are output from the TDA.

\section*{Experimental Results Using 2010 Census Demographic and Housing Characteristics Data}\label{sec:dhc}

The Demographic and Housing Characteristics File (DHC) contains more granular tables than the redistricting tables. The DHC release is split into two data universes. The household universe (DHCH) provides counts of categories of householders and for categories of housing unit and group quarters types. The persons universe (DHCP) provides counts of individuals within population groups.  Additional details on the TDA implemented for the DHC setting can be found in Cumings et al. \parencite*{cumingsmenon2023disclosure}. In this set of experiments we considered 25 simulations of the TDA on the DHC tables using the $\text{PPMF}_0$ as the input. 
% Because of the results of the earlier experiments on the redistricting data, we decided to reduce the total number of simulations from 100 to 25, which greatly saved on computational costs to create the replicates. We also decided that we no longer needed to consider the quantile-based confidence interval methods.

The DHC file contains a very large set of tables.  Instead of trying to consider all the tables, which would be too time-consuming, we selected a subset of tables for our experiments and considered all queries within those tables. The DHCP tables used were 

\begin{itemize}
\item PCT8. Relationship by Age Under 18 Years 
\item PCT9, PCT9A-I. Household Type by Relationship for the Population 65 Years and Over
\item PCT12. Sex by Single-Year Age
\end{itemize}

The tables for the race iterations correspond to A: White alone, B: Black or African American alone, C: American Indian and Alaska Native alone, D: Asian alone, E: Native Hawaiian and Other Pacific Islander alone, F: Some Other Race alone, G: Two or More Races, H: Hispanic or Latino, I: White alone, not Hispanic or Latino. Each DHCP table had multiple levels for a total of 405 unique queries per geography. See \parencite{DHCtables} for additional details on the individual queries associated with the tables listed above.

The DHCH tables used were 

\begin{itemize}
\item H4. Tenure 
\item P19. Households by Presence of People 65 Years and Over, Household Size, and Household Type
\item PCT7, PCT7A-I. Household Type by Household Size
\item PCT10, PCT10A-I. Family Type by Presence and Age of Own Children
\item PCT14, PCT14A-I. Presence of Multigenerational Households
\end{itemize}

Each DHCH table had multiple levels for a total of 405 unique queries per geography.  Similar to the redistricting data experiments, we sampled geographic units by grouping by total population size and sampling 5 geographies per grouping per state in order to make the size of the results more manageable at the block and tract geographic levels. See \parencite{DHCtables} for additional details on the individual queries associated with the tables listed above.      

\subsection*{DHCP Results}

\begin{figure}[!ht]
\caption{The proportion of $1,304,505$ 90\% confidence intervals that contained the CEF value for 2010 county level DHCP queries aggregated by query size (the bracketed number in the lower right corner indicates the proportion of the queries in that size category). Mostly, the proportion is above 0.9 threshold, but generally the proportions are lower than the redistricting queries; also CIs using normal distribution (z, BCz, cz) fail to reach the threshold more often. }
\includegraphics[width = \textwidth]{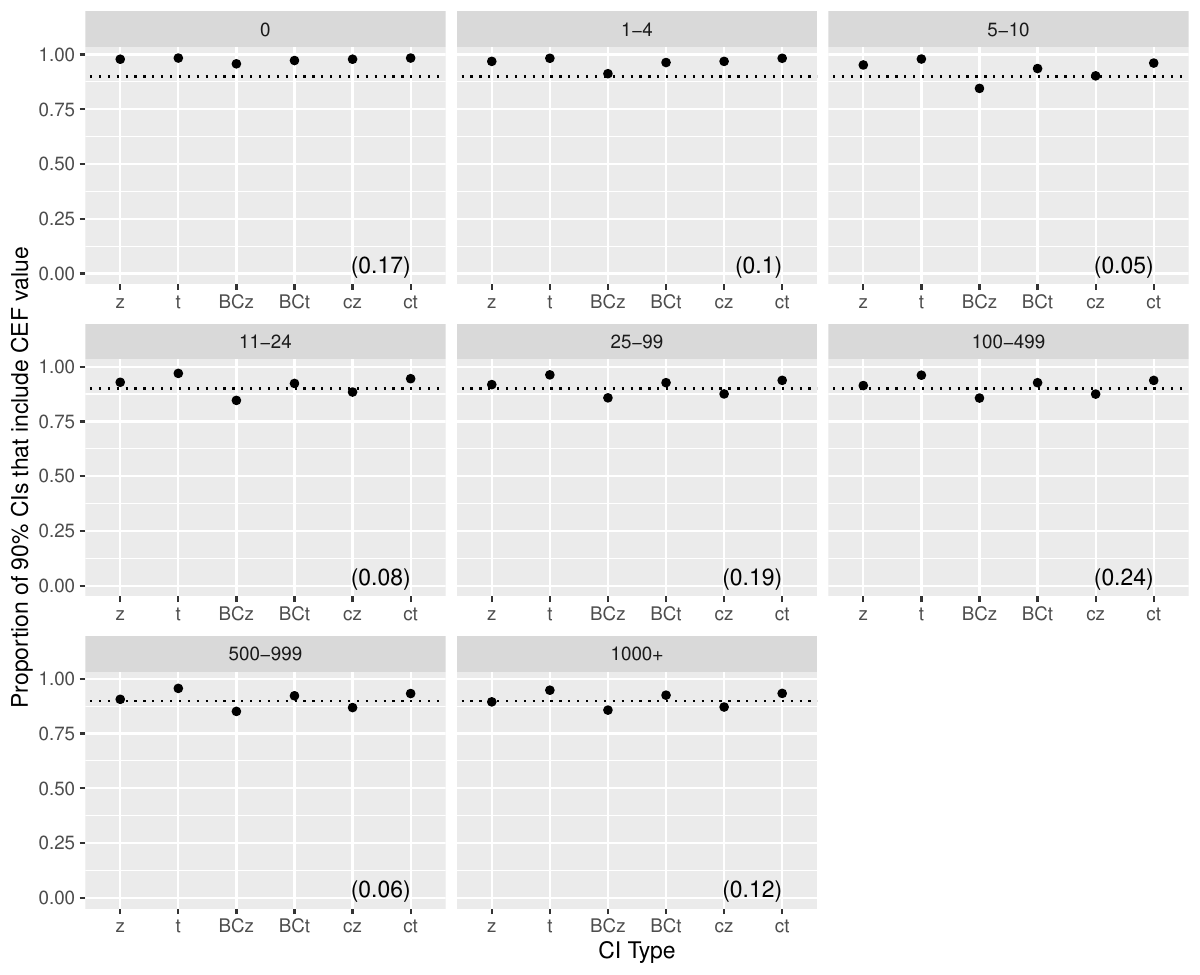}
\label{fig:county_dhcp_coverage}
\end{figure}

Figure \ref{fig:county_dhcp_coverage} shows the proportion of CIs that contained the CEF value by query size at the county level. Additional geographic levels are available in the online appendix (Figures \ref{fig:nat_dhcp_coverage} - \ref{fig:sd_dhcp_coverage}). While overall we saw that the proportion was above the 90\% threshold most of the time for multiple confidence interval types, the number of interval types that reached the threshold was lower relative to the redistricting tables. It was also more often the case that the intervals using the normal distribution quantile (z, BCz, cz) failed to meet the threshold. We observed more substantial estimated bias than the redistricting data experiments, so using bias-adjustments (and using the conditional rule) positively impacted the coverage more often. Among the query groups with higher count answers (500-999 and 1000+) at the U.S. and Puerto Rico level and block level, the bias-corrected confidence intervals were substantially better than the unadjusted intervals. The unadjusted intervals also missed the threshold in these cases. We theorize that this pattern is related to the privacy-loss budget allocation pattern used at these levels. Relatively small percentages of the privacy-loss budget ($1.91\%$ and $0.22\%$) were used at the U.S. and block levels, respectively for the DHCP \parencite{PLBAllocations}.  As a result, those estimated queries would have larger variances and be more susceptible to bias related to enforcing non-negative counts.  This pattern was also seen in the redistricting data, though it was less pronounced there. The redistricting data also used a relatively small portion of the budget at the U.S. and block levels ($2.54\%$ and $4.03\%$).

\subsection*{DHCH Results}

\begin{figure}[!ht]
\caption{The proportion of $1,304,505$  90\% confidence intervals that contained the CEF value for 2010 county level DHCH queries aggregated by query size (the bracketed number in the lower right corner indicates the proportion of the queries in that size (count) category). The proportion of CIs containing the CEF value is above the 0.9 threshold for the t, BCt, and ct intervals}
\includegraphics[width = \textwidth]{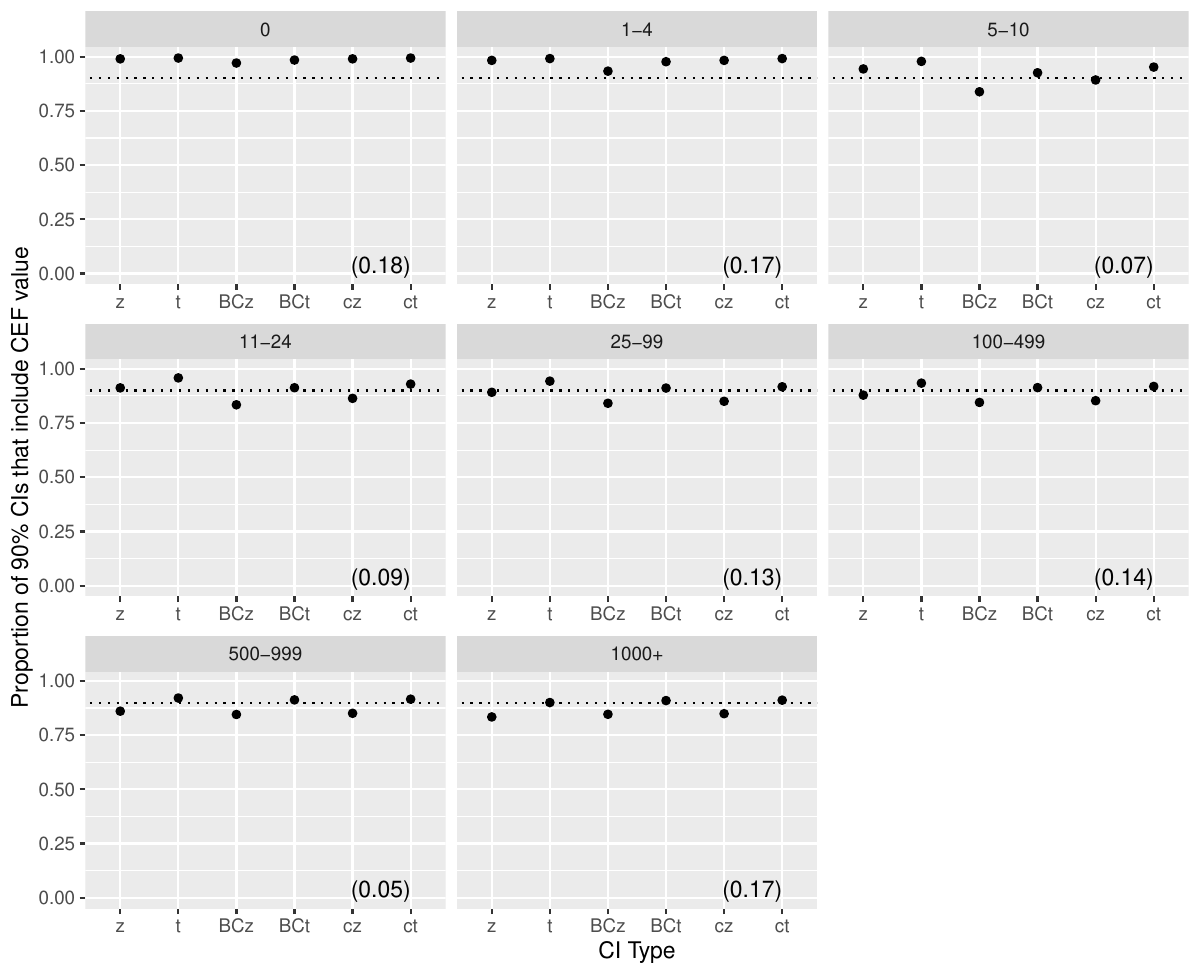}
\label{fig:county_dhch_coverage}
\end{figure}

Among DHCH queries, we observed that the confidence intervals under-performed  within some query size categories at the U.S. and State levels (Figures \ref{fig:nat_dhch_coverage} and  \ref{fig:state_dhch_coverage}) but that pattern did not continue at lower geographies. For example, county level queries (Figure \ref{fig:county_dhch_coverage}) were either very close (z, BCz, cz) or exceeded (t, BCt, ct) the 90\% threshold across all the query size categories.  At the U.S. Level, most of the queries (81\%) were in the largest size category (1000+) where the bias-corrected or conditionally bias-corrected intervals met the 90\% threshold. So overall, queries for which the confidence intervals under-performed represented a small proportion of all queries. We hypothesize that the reason we see worse performance is the relatively small proportion of privacy-loss budget spent at the U.S. level ($6.28\%$) \parencite{PLBAllocations}, which could result in biased TDA query estimates because of the effect of the non-negative count constraint. Relatedly, at the U.S. and Puerto Rico level, the bias correction for the confidence intervals was beneficial for all query size categories 25 and above and made a substantial improvement for the largest queries.  Results for additional geographic levels can be found in Figures \ref{fig:tract_dhch_coverage} - \ref{fig:sd_dhch_coverage}.        

\subsection*{Estimated Bias}

In order to better understand the link between the confidence intervals and the bias, we computed a summary of the estimated bias using the approximate Monte Carlo method (bias relative to the $\text{PPMF}_0$ answers). Table \ref{tab:dhch_bias_tab} shows the 1st, 50th, and 99th percentiles of the estimated bias by geographic level and query size group for DHCH queries (Table \ref{tab:dhcp_bias_tab} in the appendix shows results for DHCP queries).  We see that the 1st and 99th percentiles of estimated bias at the US level for the 1000+ grouping are substantially more extreme than those at the state level. 

\vspace{1cm}

\begin{center}
    [Insert Table 1 about here]
\end{center}

\vspace{1cm}

The estimated bias also effects the width of the confidence intervals because the RMSE is used in the margin of error calculation. As the estimated bias of a query increases, so does the width of the confidence interval. We stated previously that the bias can be underestimated by the AMC methodology, so even when we are using bias-adjustments to the confidence intervals, allowing for a wider margin of error is beneficial in that it will help to make up for underestimation. However, there may be cases in which the magnitude of the estimated bias and its effect on the width of the resulting confidence intervals is so large that the confidence interval is no longer meaningful. For example, the AMC-estimated bias for the U.S. level query concerning the number of households with no one over 65 years of age, as provided in Table P19, is $-41,137$ with a BCt 90\% CI of $(87,517,612 - 87,683,422)$.  In these cases, it may be preferable to utilize the noisy measurement files to get an unbiased estimate of the query and a confidence interval derived from the noisy measurement(s). Estimates and confidence intervals calculated directly from the noisy measurements will not benefit from the error reduction that TDA's post-processing affords. That said, for estimates at higher geographic levels (especially at the U.S. and state level where the benefit of TDA's error reduction is less pronounced) for queries derived from noisy measurements with larger shares of privacy-loss budget, we would anticipate the confidence intervals derived from the noisy measurements to be similar to or even outperform (when the estimated bias is large) those generated via the AMC approach.

\section*{Discussion}\label{sec:discussion}

We proposed a general method for estimating key statistical properties of query estimates from the TDA and how to combine them into meaningful confidence intervals.  While we cannot examine the results for every single possible query at every single level of geography, the results consistently demonstrate that the AMC method performed well using the 2010 data with parameters equivalent to those applied to the 2020 data and a range of queries using as few as 25 simulations replicates. Given the parameters mirrored those used for the 2020 Census and the 2010 data are generally similar in structure and size to 2020, we argue that this methodology will work similarly well when applied to the 2020 Census data.  The Census Bureau intends to release 25 AMC replicates from the 2020 PPMF so that researchers may use them to construct confidence intervals for 2020 queries.  

Among the proposed confidence interval types, we found that Wald-type intervals performed better than the non-parametric quantile intervals.  Within the Wald-type intervals, we found that bias-adjusted intervals were generally less effective overall, and only select cases yielded improved coverage results. We attempted to characterize the select instances in which the bias-adjusted estimator may improve coverage with a conditional rule as an alternative method. 

Considering the results overall, we recommend users utilize the conditional bias-corrected intervals with the Student's $t$ distribution. 
Future work might improve the logic of that rule, but we think the experiments show that it does a good job balancing when to use or not use the bias correction.  Using the more conservative Student's $t$ critical value was often necessary to reach the desired threshold for DHC data table queries, and therefore we recommend its use over the normal distribution critical value.

%TODO find a better place for this
% \footnote{The 2010 $\text{PPMF}_0$ and associated simulations are available at \url{https://registry.opendata.aws/census-2010-amc-mdf-replicates/}. Additional resources including a Jupyter notebook containing a full working code example for estimating the bias-corrected intervals with the Student's $t$ distribution are available via the U.S. Census Bureau at \url{https://www2.census.gov/programs-surveys/decennial/2020/technical-documentation/complete-tech-docs/demographic-and-housing-characteristics-file-and-demographic-profile/data_analysis_resources/1_estimating_conf_intervals_2010_demo/} and \url{https://github.com/uscensusbureau/AMC_Confidence_Intervals}.}
 
A major benefit of this methodology is that it does not incur any additional privacy loss because it does not access confidential data. The method is also relatively simple to apply for a single query or set of queries, assuming the user can compute or retrieve the query value computed for the $\text{PPMF}_0$ and the associated simulations. Since all the inputs of our proposed AMC simulation approach  (\textit{i.e.}, the PPMF,  the TopDown code, and parameters) are public, it is also feasible for an organization outside of the Census Bureau to create additional simulation runs in the event this would be beneficial for a particular use case.      
There are a wide variety of applications or extensions where a version of the proposed methodology could be successfully deployed.  We have have focused here on count queries, but for example ratio queries (e.g. persons per household) may be of interest.  In addition, more complex statistical quantities such as the covariance between two different queries could be estimated using the multiple iterations, or even the uncertainty of a query total could be taken into account in a linear regression through a strategy similar to multiple imputation combination rules that incorporates the variation between and within different replicates \parencite{RubinMI}. Future research is needed to explore a more appropriate way of constructing confidence intervals when the distribution of the query answers across the iterations is not approximately normally distributed (such as when the query count is near 0 or if ratios were considered). While we focused specifically on the feasibility of this methodology for the TDA and the 2020 Census use case, the Approximate Monte Carlo method could apply more broadly to any algorithm whose output produces data that could then be used as the input to the algorithm. 

\section*{Disclosure Statement}
 No rights reserved. This work was authored as part of the authors' official duties as Employees of the United States Government and is therefore a work of the United States Government. In accordance with 17 U.S.C. 105, no copyright protection is available for such works under U.S. law.

\printbibliography
%\bibliography{references}

\newpage

\section*{Tables} %Limit of 8 tables and figures (combined)

%each table should appear as a seperate page
%for each table, include a call-out where they would go "set table 4 about here"

% \begin{table}[h!]
% \begin{tabular}{c|l}
% \hline
% Label & Description \\
% \hline
% np & Non-parametric quantile-based CI \\
% BCnp & Bias-corrected non-parametric quantile-based CI\\
% z & RMSE-based Wald CI with normal distribution\\
% t & RMSE-based Wald CI with Student's $t$ distribution\\
% BCz & Bias-corrected RMSE-based Wald CI with normal distribution\\
% BCt & Bias-corrected RMSE-based Wald CI with Student's $t$ distribution\\
% cz & Conditionally bias-corrected RMSE-based Wald CI with normal distribution\\
% ct & Conditionally bias-corrected RMSE-based Wald CI with Student's $t$ distribution \\
% \hline
% \end{tabular}
%     \caption{Confidence Interval Types Used in Experiments}
%     \label{tab:ci_table}
% \end{table}

\begin{table}[ht!]
\caption{The 1st, 50th and 99th percentile estimated bias of DHCH queries by query size for U.S. \& Puerto Rico, state, county, tract, block, AIAN area, and elementary school district geographies. There are large estimated biases at the extremes for both the U.S. \& Puerto Rico and state levels for the largest query size groups. The estimated bias distributions are more concentrated around 0 for the other geographies.}
\centering
\begin{tabular}{l|rrr|rrr|rrr|rrr}
  \hline
 & \multicolumn{3}{c|}{U.S. \& Puerto Rico} & \multicolumn{3}{c|}{State} & \multicolumn{3}{c|}{County} & \multicolumn{3}{c}{Tract} \\
Query Size & 1st & 50th & 99th &  1st & 50th & 99th &  1st & 50th & 99th &  1st & 50th & 99th \\ 
  \hline
  0 &  &  &  & 0.4 & 1.3 & 3.9 & 0.0 & 0.3 & 3.0 & 0.0 & 0.2 & 2.2 \\ 
  1-4 & -0.4 & 2.6 & 6.4 & -1.3 & 1.2 & 6.1 & -2.2 & 0.1 & 4.6 & -2.3 & -0.2 & 3.5 \\ 
  5-10 & -2.9 & 4.3 & 6.7 & -3.0 & 1.6 & 9.0 & -3.9 & 0.0 & 6.2 & -3.9 & -0.3 & 4.8 \\ 
  11-24 & -4.7 & 6.1 & 20.6 & -4.8 & 1.6 & 15.4 & -5.8 & 0.1 & 8.1 & -5.6 & -0.2 & 6.1 \\ 
  25-99 & -4.9 & 12.0 & 45.6 & -7.5 & 3.4 & 26.1 & -8.8 & 0.0 & 11.5 & -8.2 & -0.3 & 8.1 \\ 
  100-499 & -20.0 & 10.6 & 126.4 & -16.2 & 3.3 & 49.8 & -14.6 & -0.4 & 15.7 & -12.2 & -0.2 & 12.1 \\ 
  500-999 & -30.7 & -0.9 & 225.1 & -28.4 & 1.6 & 52.9 & -21.2 & -0.8 & 21.3 & -14.5 & -0.1 & 12.9 \\ 
  1000+ & -6321.2 & -4.9 & 4460.1 & -636.6 & -1.6 & 510.1 & -62.2 & -0.5 & 55.2 & -16.1 & -0.1 & 11.7 \\ 
   \hline
    & \multicolumn{3}{c|}{Block} & \multicolumn{3}{c|}{AIAN Area} & \multicolumn{3}{c|}{School District} \\
Query Size & 1st & 50th & 99th &  1st & 50th & 99th &  1st & 50th & 99th \\ 
  \hline
  0 & 0.0 & 0.0 & 2.0 & 0.0 & 0.1 & 2.0 & 0.0 & 0.1 & 2.2 \\ 
  1-4 & -3.0 & -0.6 & 5.5 & -2.2 & -0.2 & 4.0 & -2.3 & -0.2 & 4.2 \\ 
  5-10 & -6.0 & -1.3 & 8.1 & -3.9 & -0.2 & 6.0 & -4.1 & -0.2 & 6.2 \\ 
  11-24 & -10.0 & -1.5 & 10.2 & -5.8 & -0.4 & 8.1 & -6.4 & -0.2 & 8.2 \\ 
  25-99 & -14.4 & -0.7 & 15.5 & -9.4 & -0.5 & 12.3 & -10.5 & -0.3 & 11.9 \\ 
  100-499 & -17.8 & 1.6 & 23.2 & -18.7 & -0.7 & 22.3 & -18.1 & -0.4 & 18.7 \\ 
  500-999 & -10.6 & 3.0 & 24.0 & -32.8 & -1.4 & 43.0 & -27.2 & -0.5 & 27.5 \\ 
  1000+ & -9.9 & 0.3 & 4.7 & -123.4 & -2.2 & 107.2 & -53.7 & -0.4 & 46.8 \\ 
   \hline
\end{tabular}
\label{tab:dhch_bias_tab}
\end{table}

% \begin{table}[ht!]
% \caption{The 1st, 50th and 99th percentile estimated bias of DHCH queries by query size for Block, AIAN area, and elementary school district geographies}
% \centering
% \begin{tabular}{l|rrr|rrr|rrr}
%   \hline
%  & \multicolumn{3}{c}{Block} & \multicolumn{3}{c}{AIAN Area} & \multicolumn{3}{c}{School District} \\
% Query Size & 1st & 50th & 99th &  1st & 50th & 99th &  1st & 50th & 99th \\ 
%   \hline
%   0 & 0.0 & 0.0 & 2.0 & 0.0 & 0.1 & 2.0 & 0.0 & 0.1 & 2.2 \\ 
%   1-4 & -3.0 & -0.6 & 5.5 & -2.2 & -0.2 & 4.1 & -2.3 & -0.2 & 4.2 \\ 
%   5-10 & -6.1 & -1.3 & 8.0 & -3.9 & -0.2 & 6.1 & -4.1 & -0.2 & 6.2 \\ 
%   11-24 & -10.0 & -1.4 & 10.2 & -6.1 & -0.4 & 8.7 & -6.4 & -0.2 & 8.3 \\ 
%   25-99 & -14.3 & -0.7 & 15.4 & -10.8 & -0.5 & 13.4 & -10.5 & -0.2 & 12.1 \\ 
%   100-499 & -17.7 & 1.7 & 23.0 & -21.6 & -0.6 & 26.2 & -18.1 & -0.4 & 19.1 \\ 
%   500-999 & -10.7 & 2.9 & 24.1 & -35.1 & -1.4 & 45.5 & -27.3 & -0.5 & 28.2 \\ 
%   1000+ & -9.9 & 0.3 & 4.7 & -97.2 & -1.7 & 88.6 & -102.0 & -0.5 & 76.3 \\ 
%    \hline
% \end{tabular}
% \label{tab:dhch_bias_tab2}
% \end{table}

\FloatBarrier

\pagebreak

\section*{Figures}

%title in initial cap only
%combined title and notes below the figure
%use PDF Figures
%Need alt text for each figure

% \begin{figure}[!ht]
% \includegraphics[width = \textwidth]{graphics/distribution_comparison/sd_bias.pdf}
% \caption{Elementary school district level estimated bias from the PPMF$_0$ vs estimated bias from the CEF, grouped by CEF-based query size. Some underestimation of the bias in the AMC method can be seen (which leads to underestimating RMSE as well).}\label{fig:sd_bias}
% \end{figure}

% \begin{figure}[!ht]
% \caption{The proportion of 90\% CIs that contained the CEF value for 2010 AIAN area level PL94 queries aggregated by query count (size) (the bracketed number in the lower right corner indicates the proportion of the queries in that size category, e.g. in this figure most -- 0.79 -- have value of 0). The Wald-type intervals all do similarly well and mostly surpass 0.9 threshold. Student's $t$ based CIs (t, BCt, ct) are wider and have slightly higher coverage. The most poorly performing are the non-parametric ones (np and BCnp). Not much difference was made by the bias correction.}
% \includegraphics[width = \textwidth]{graphics/pl94cis/aian_pl94_cis_coverage.pdf}
% \label{fig:aian_pl94_cis}
% \end{figure}

% to Sallie's point above, here are the number of queries used for pl94/redistricting

% GeoLevel	tot_n_geo
% U.S.	602 (301 queries x 2 geographies)
% State	15652 (301 queries x 52 geographies)
% County	969521 (301 queries x 3221 geographies)
% Tract	22105741 (301 queries x 73441 geographies)
% Block	780493 (301 queries x 2593 geographies)
% AIAN Area	186921 (301 queries x 621 geographies)
% Elementary SD	693504 (301 queries x 2304 geographies)

%number of queries in both DHCH and DHCP graphics
% GeoLevel	tot_n_geo
% U.S.	810 (405 queries x 2 geographies)
% State	21060 (405 queries x 52 geographies)
% County	1304505 (405 queries x 3221 geographies)
% Tract	1052595 (405 queries x 2599 geographies)
% Block	1050165 (405 queries x 2593 geographies)
% AIAN Area	251505 (405 queries x 621 geographies)
% Elementary SD	933120 (405 queries x 2304 geographies)

% \begin{figure}[!ht]
% \caption{The proportion of 90\% CIs that contained the CEF value for 2010 U.S. and Puerto Rico Level DHCH queries aggregated by query size (the bracketed number in the lower right corner indicates the proportion of the queries in that size category, e.g. in this figure most -- 0.81 -- have values over 1000). US level queries often underperform, likely due to lower privacy budget allocation; bias correction is beneficial for high count query groups (500+).}
% \includegraphics[width = \textwidth]{graphics/DHCcis/nat_dhch_coverage.pdf}
% \label{fig:nat_dhch_coverage}
% \end{figure}

\FloatBarrier

\pagebreak

\section*{Appendix}

% number the figures and tables like A1, A2, etc. 
\setcounter{table}{0}
\renewcommand{\thetable}{A\arabic{table}}

\setcounter{figure}{0}
\renewcommand{\thefigure}{A\arabic{figure}}

\begin{figure}[ht!]
\caption{U.S. and Puerto Rico level comparison of estimated RMSE ($RSME_{PPMF0}$ vs. $RMSE_{CEF}$). Each panel corresponds to certain sizes (i.e. counts) of the query (based on CEF). In the case when $RSME_{PPMF0}$ values are very similar to $RMSE_{CEF}$ values they line up along the diagonal, showing the AMC method to be a good approximation. This is generally the case with the data shown. Circles show location of the $602$ (301 queries x 2 geographies) points used for the plot.}
\includegraphics[width = \textwidth]{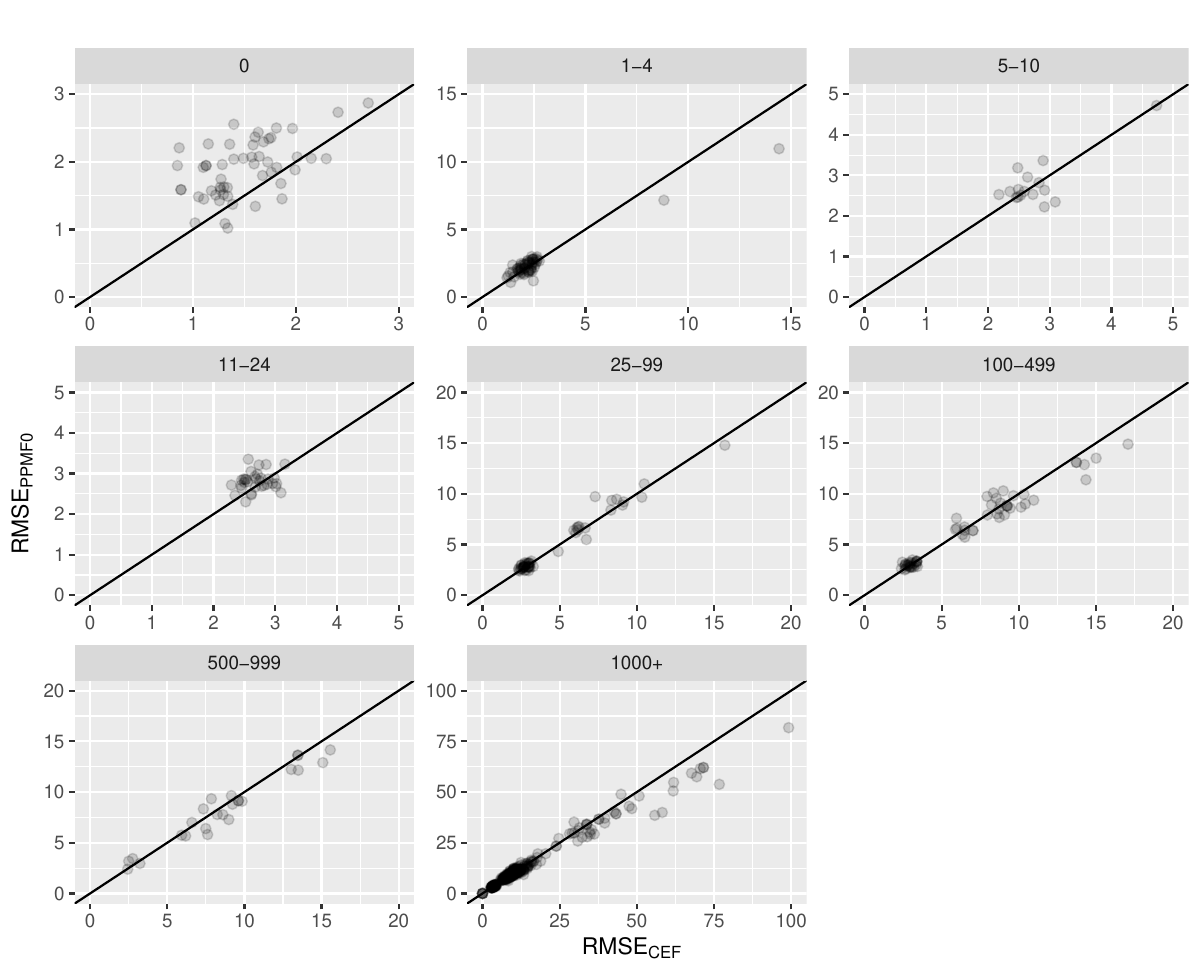}
\label{fig:nat_rmse}
\end{figure}

\begin{figure}[ht!]
\caption{Elementary school district level estimated bias ($Bias_{PPMF0}$ vs. $Bias_{CEF}$), grouped by CEF-based query size. Each panel corresponds to certain sizes (i.e. counts) of the Query (based on CEF). In the case when $Bias_{PPMF0}$ values are very similar to $Bias_{CEF}$ values they line up along the diagonal, showing the AMC method to be a good approximation. For query size of 0, the estimated $Bias_{CEF}$ value can only be positive, due to the non-negativity constraint. For some cases, the estimated $Bias_{CEF}$ value is more strongly negative than the estimated $Bias_{PPMF0}$ value. Hexagons show location and density of the $693,504$ (301 queries x 2,304 geographies) points used for the plot.}
\includegraphics[width = \textwidth]{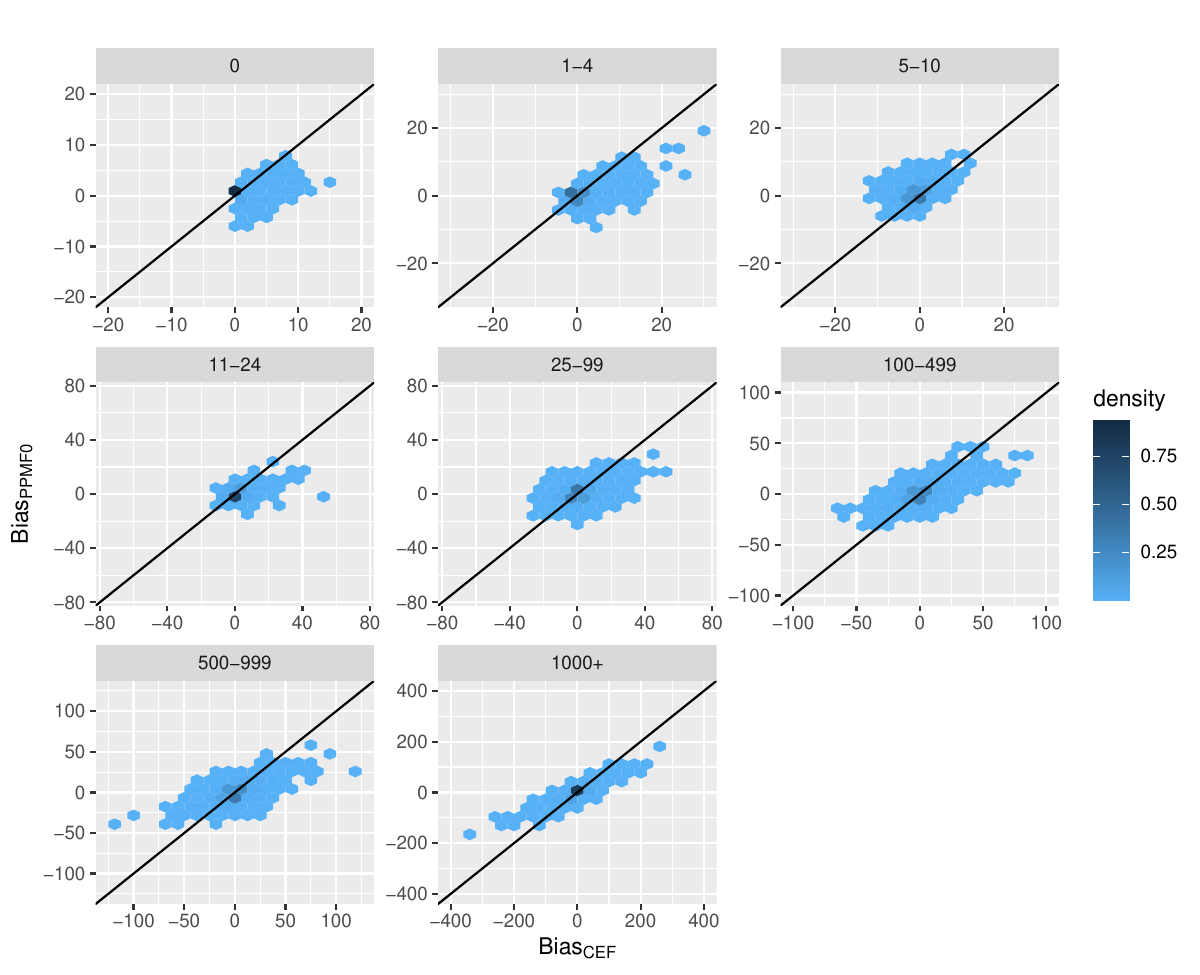}
\label{fig:sd_bias}
\end{figure}

\begin{figure}[ht!]
\caption{Elementary school district level estimated standard deviation ($SD_{PPMF0}$ vs. $SD_{CEF}$) by CEF query size grouping. Each panel corresponds to certain sizes (i.e. counts) of the Query (based on CEF). In the case when $SD_{PPMF0}$ values are very similar to $SD_{CEF}$ values they line up along the diagonal, showing the AMC method to be a good approximation. This is generally the case with the data shown. Hexagons show location and density of the $693,504$ (301 queries x 2,304 geographies) points used for the plot.}
\includegraphics[width =\textwidth]{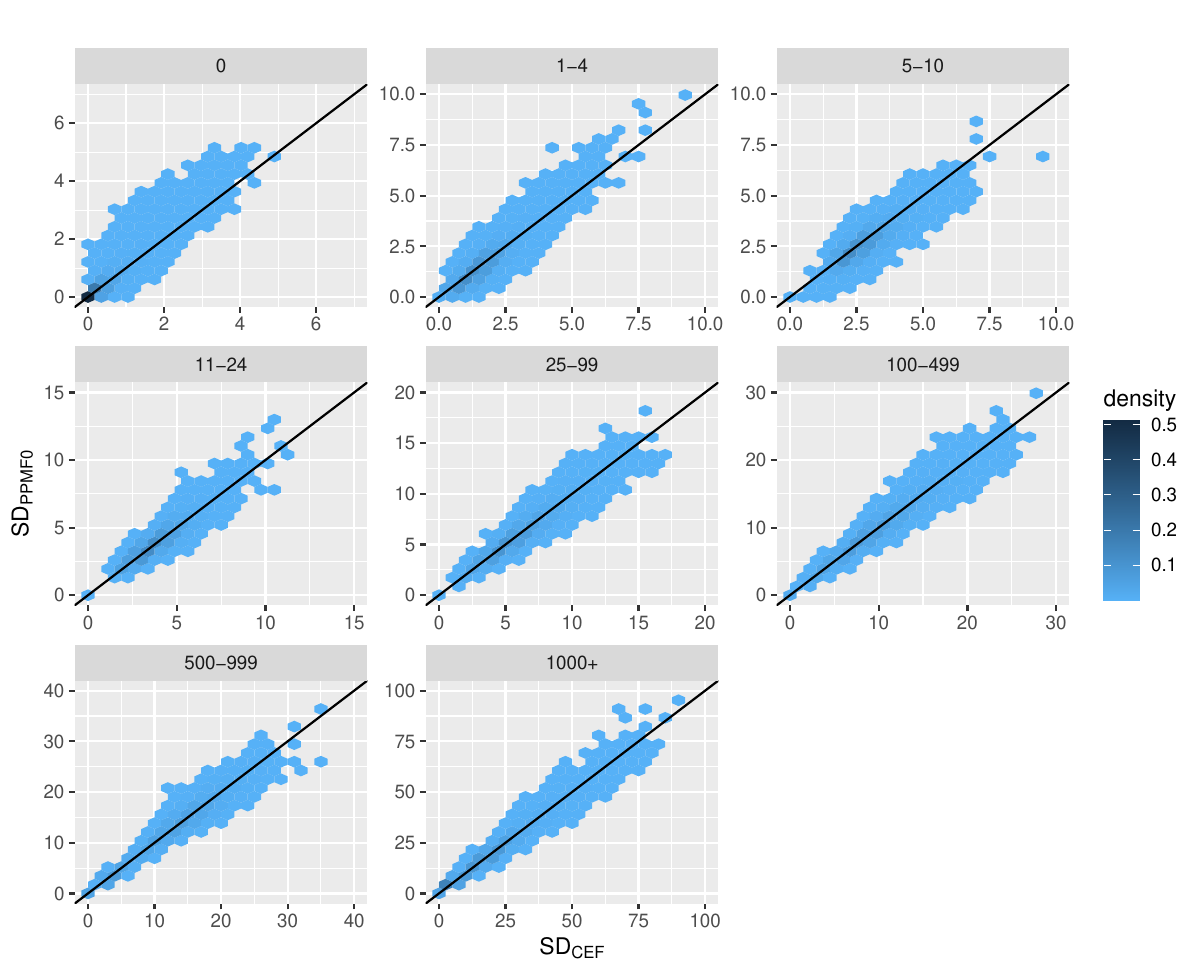}
\label{fig:sd_sdev}
\end{figure}

The following graphics and tables are meant as online supplementary materials.  In the main paper we show results for selected geographies among those in the experiments that we felt had representative results or had results of particular note. In this appendix we have included more extensive results at additional geographies to help demonstrate the validity of the results.   
The first set of graphics (Figures \ref{fig:nat_rmse} - \ref{fig:sd_sdev}) compare the estimated RMSE, bias, and standard deviation for the redistricting data queries.  As a reminder, these are 2020 query definitions applied to 2010 data. The graphics visually compare the AMC-based estimate with the gold standard constructed from the TDA applied to the CEF, by geography and query size.

\begin{figure}[ht!]
\caption{The proportion of $602$ (301 queries x 2 geographies) 90\% confidence intervals that contained the CEF value for 2010 U.S. and Puerto Rico level PL94 queries aggregated by query size (the bracketed number in the lower right corner indicates the proportion of the queries in that size category, e.g. in this figure most, 0.5, have value of 1000 or larger). Non-parametric CIs perform the most poorly with respect to containing the CEF value. For the Wald-based CIs, the ones using Student's $t$ distribution performed slightly better than the ones using the normal distribution. Conditional bias correction (ct) performs similarly to not using bias correction (t) with both meeting the 0.90 target in all query size groupings.}
\includegraphics[width =0.75\textwidth]{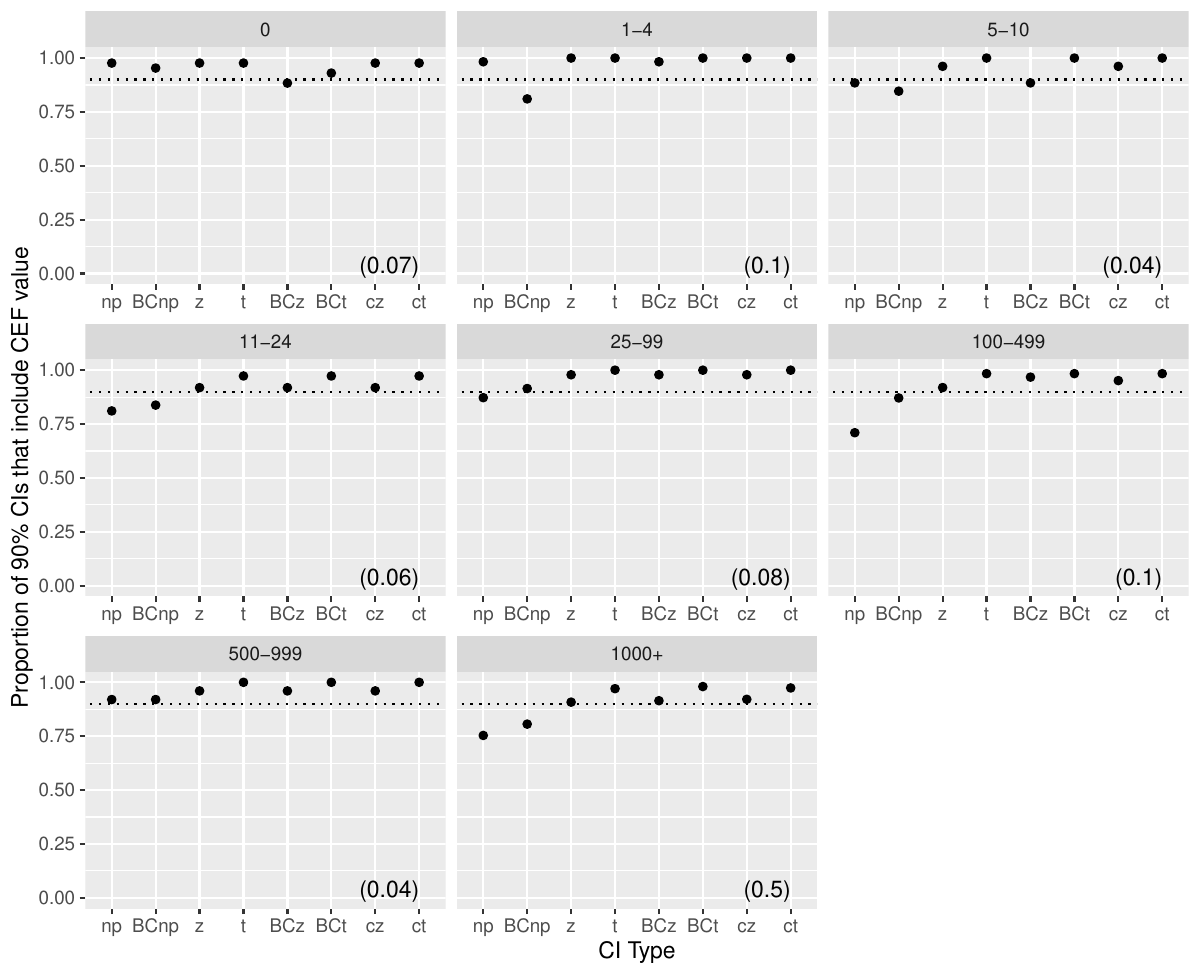}
\label{fig:nat_pl94_cis}
\end{figure}

\begin{figure}[ht!]
\caption{The proportion of $693,504$ (301 queries x 2,304 geographies) 90\% confidence intervals that contained the CEF value for elementary school district area level PL94 queries aggregated by query size (the bracketed number in the lower right corner indicates the proportion of the queries in that size category, e.g. in this figure the majority, 0.71, have value of 0). Non-parametric CIs perform the most poorly with respect to containing the CEF value. For the Wald-based CIs, the ones using Student's $t$ distribution performed slightly better than the ones using the normal distribution. Conditional bias correction (ct) performed similarly to not using bias correction (t) with both meeting the 0.90 target in all query size groupings.}
\includegraphics[width =\textwidth]{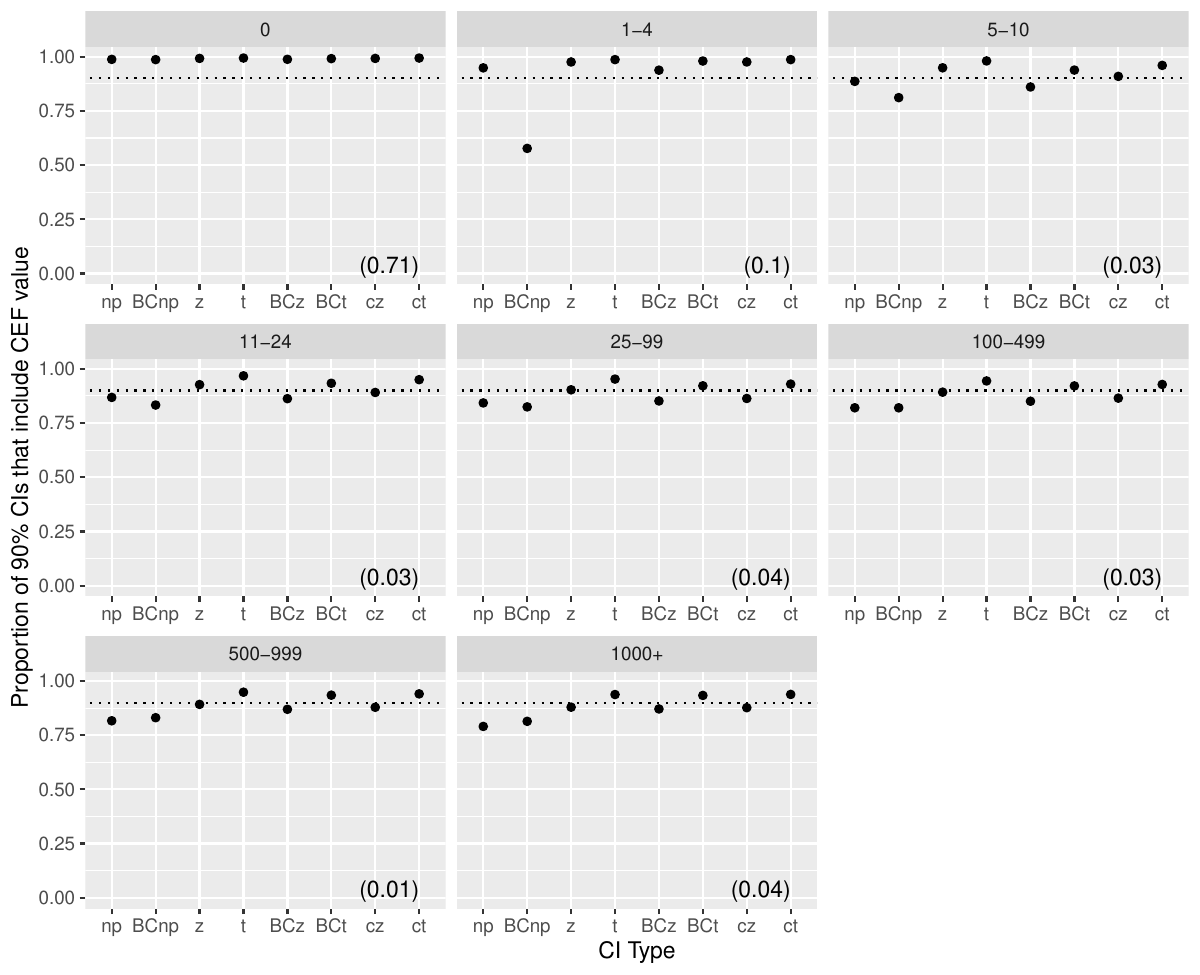}
\label{fig:sd_pl94_cis}
\end{figure}

\begin{figure}[ht!]
\caption{The distribution of $602$ (301 queries x 2 geographies) 90\% confidence interval widths for U.S and Puerto Rico Level PL94 queries aggregated by query size. As is to be expected, $t$ and BC$t$ intervals are slightly wider than the $z$ and BC $z$ intervals; also the confidence interval width generally increases as the size of the query increases.}
\includegraphics[width =\textwidth]{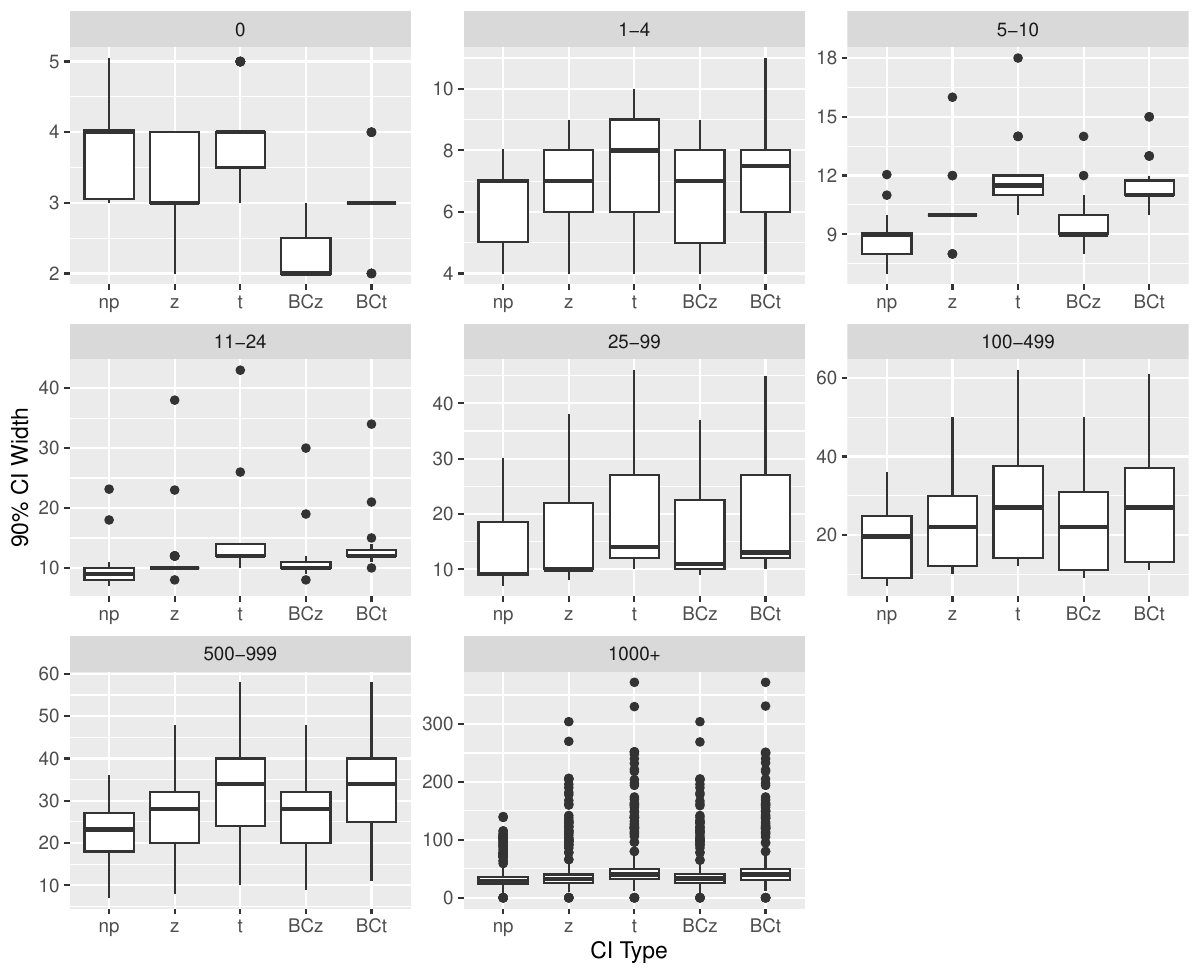}
\label{fig:nat_pl94_cis_width}
\end{figure}

\begin{figure}[ht!]
\caption{The distribution of $693,504$ (301 queries x 2,304 geographies) 90\% confidence interval widths for elementary school district PL94 queries aggregated by query size. As is to be expected, $t$ and BC$t$ intervals are slightly wider than the $z$ and BC $z$ intervals; also the confidence interval width generally increases as the size of the query increases.}
\includegraphics[width =0.75\textwidth]{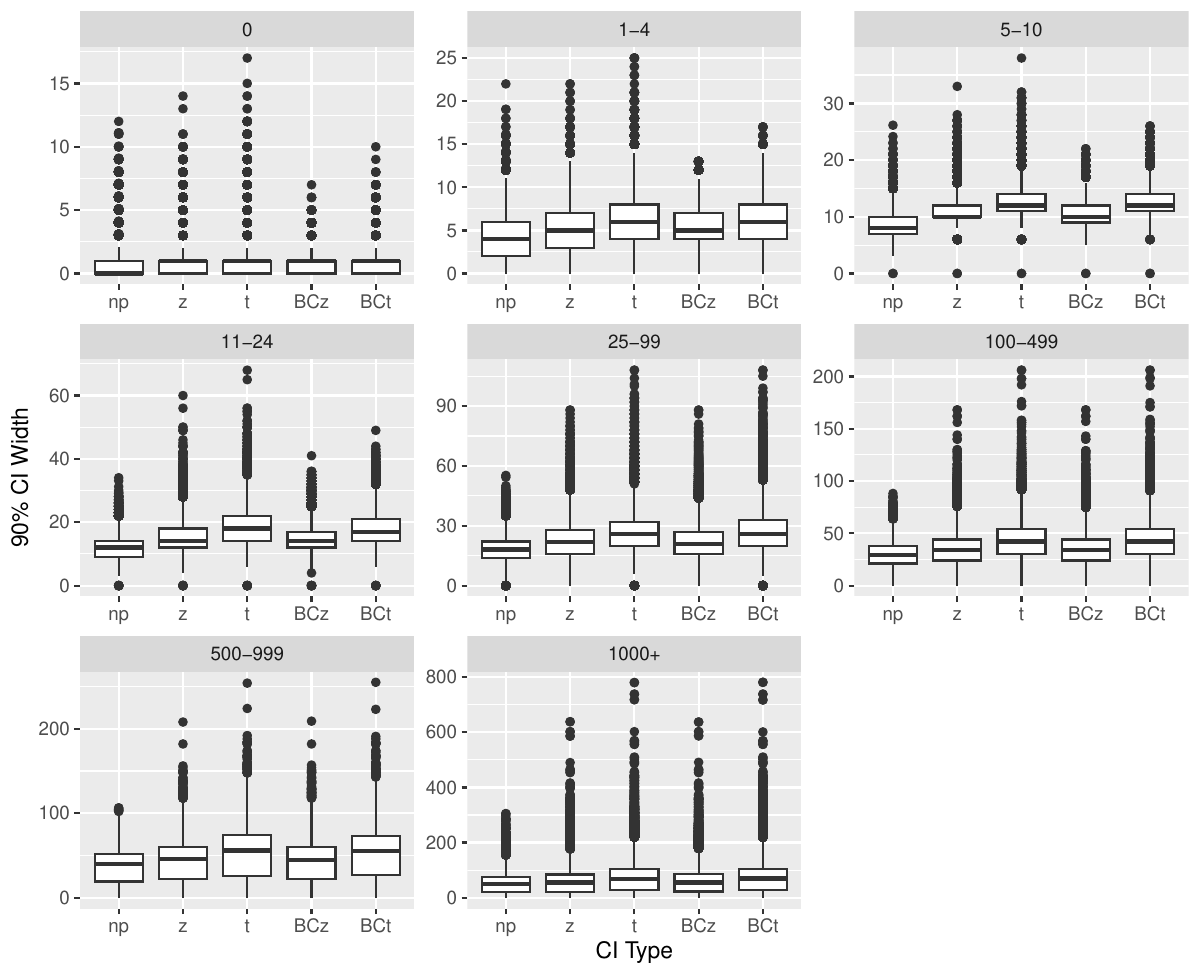}
\label{fig:sd_pl94_cis_width}
\end{figure}

\begin{figure}[ht!]
\caption{The proportion of $810$ (405 queries x 2 geographies) 90\% confidence intervals that contained the CEF value for U.S. and Puerto Rico level DHCP queries aggregated by query size (the bracketed number in the lower right corner indicates the proportion of the queries in that size category, e.g. in this figure the majority, 0.87, have value of 1,000 or larger). The $t$ distribution based confidence intervals performed substantially better than the ones using the normal distribution for the 11-24 and 25-99 panels. Conditional bias correction (ct) performed better than the (t) interval in several panels.}
\includegraphics[width =\textwidth]{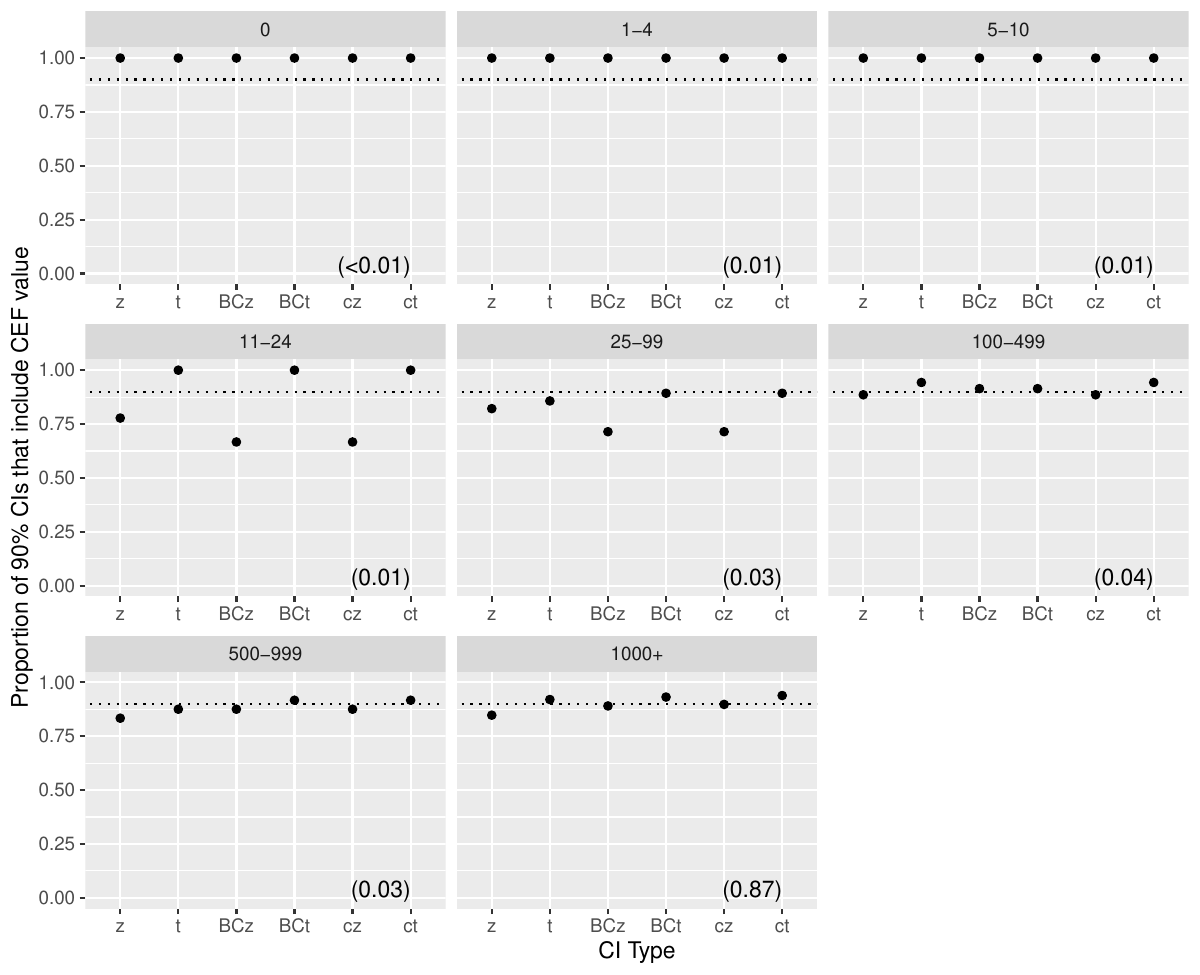}
\label{fig:nat_dhcp_coverage}
\end{figure}

\begin{figure}[ht!]
\caption{The proportion of $933,120$ (405 queries x 2,304 geographies) 90\% confidence intervals that contained the CEF value for elementary school district level DHCP queries aggregated by query size (the bracketed number in the lower right corner indicates the proportion of the queries in that size category, e.g. in this figure the plurality, 0.32, have value of 0). The $t$ distribution based confidence intervals performed better than the ones using the normal distribution. Conditional bias correction (ct) performed similarly to not using bias correction (t) with both meeting the 0.90 target in all query size groupings.}
\includegraphics[width =\textwidth]{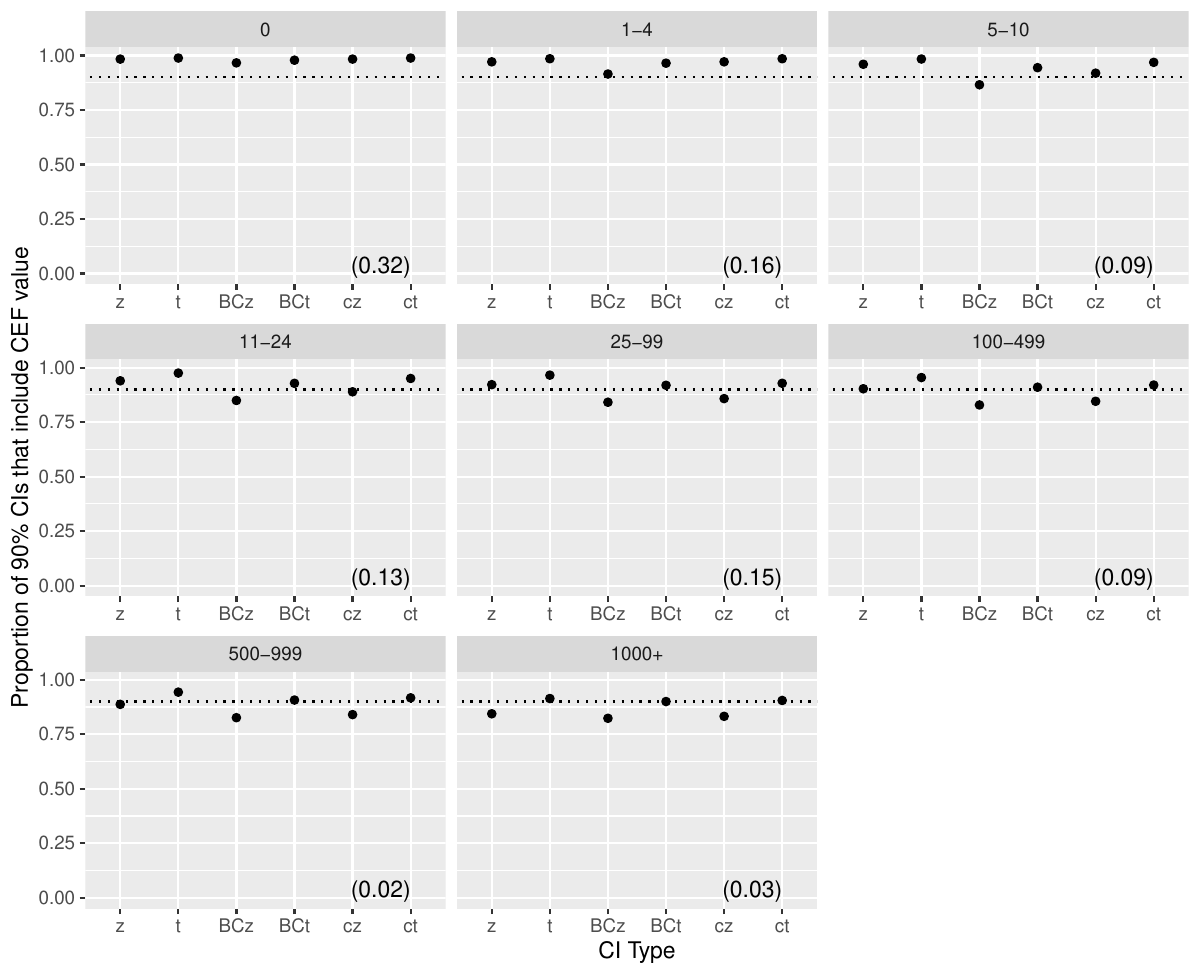}
\label{fig:sd_dhcp_coverage}
\end{figure}

\begin{figure}[ht!]
\caption{The proportion of $810$ (405 queries x 2 geographies) 90\% confidence intervals that contained the CEF value for U.S. and Puerto Rico level DHCH queries aggregated by query size (the bracketed number in the lower right corner indicates the proportion of the queries in that size category, e.g. in this figure the majority, 0.81, have value of 1,000 or larger). Many of the CI types failed to meet the 0.90 target in several of the panels.  None of them met the target in the 11-24, 25-99, 100-499, or 500-999 panels. The $t$ distribution based confidence intervals performed substantially better than the ones using the normal distribution for many of the panels. Conditional bias correction (ct) performed the best overall and met the target for the panel with the majority of queries (1,000+).}
\includegraphics[width =\textwidth]{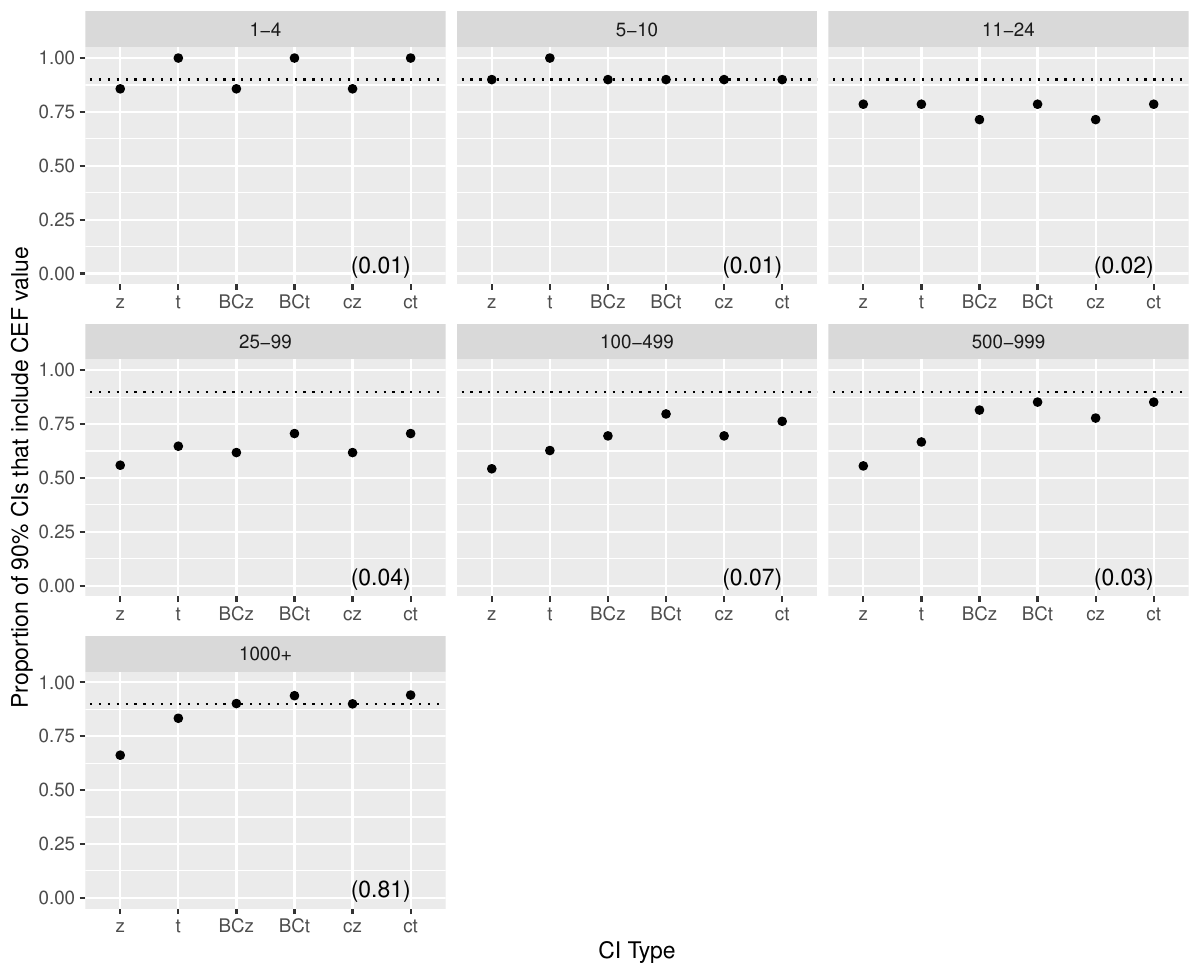}
\label{fig:nat_dhch_coverage}
\end{figure}

\begin{figure}[ht!]
\caption{The proportion of $21,060$ (405 queries x 52 geographies) 90\% confidence intervals that contained the CEF value for state level DHCH queries aggregated by query size (the bracketed number in the lower right corner indicates the proportion of the queries in that (size category, e.g. in this figure the majority, 0.69, have value of 1,000 or larger). The $t$ distribution based confidence intervals performed substantially better than the ones using the normal distribution for many of the panels. The ct interval was at or barely below the $0.90$ target for all panels, performing the best overall.}
\includegraphics[width =\textwidth]{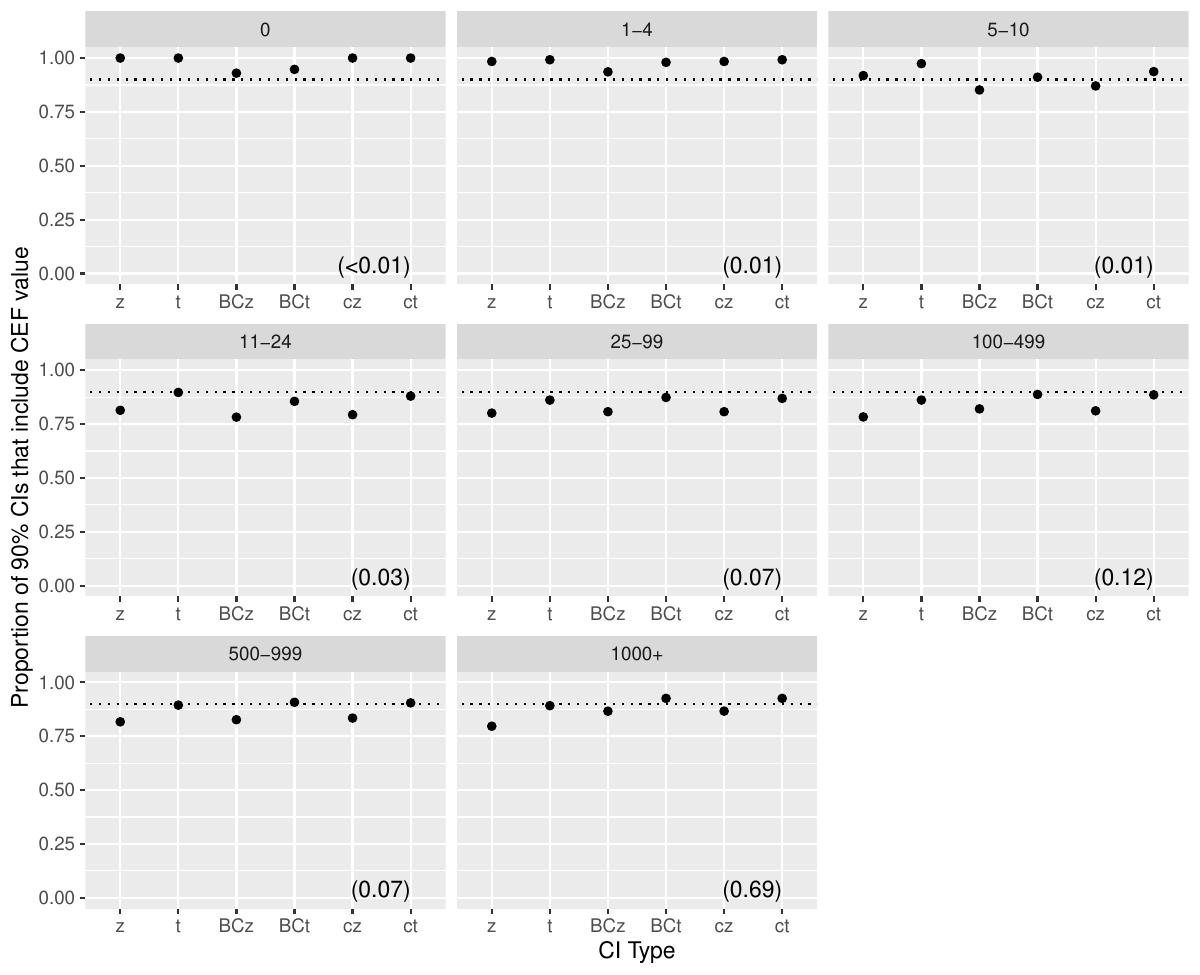}
\label{fig:state_dhch_coverage}
\end{figure}

\begin{figure}[ht!]
\caption{The proportion of $1,052,595$ (405 queries x 2,599 sampled geographies) 90\% confidence intervals that contained the CEF value for tract level DHCH queries aggregated by query size (the bracketed number in the lower right corner indicates the proportion of the queries in that (size category, e.g. in this figure the plurality, 0.33, have value of 0). The $t$ distribution based confidence intervals performed substantially better than the ones using the normal distribution for many of the panels. The t, BCt, and ct were all above the $0.90$ target for all panels.}
\includegraphics[width = \textwidth]{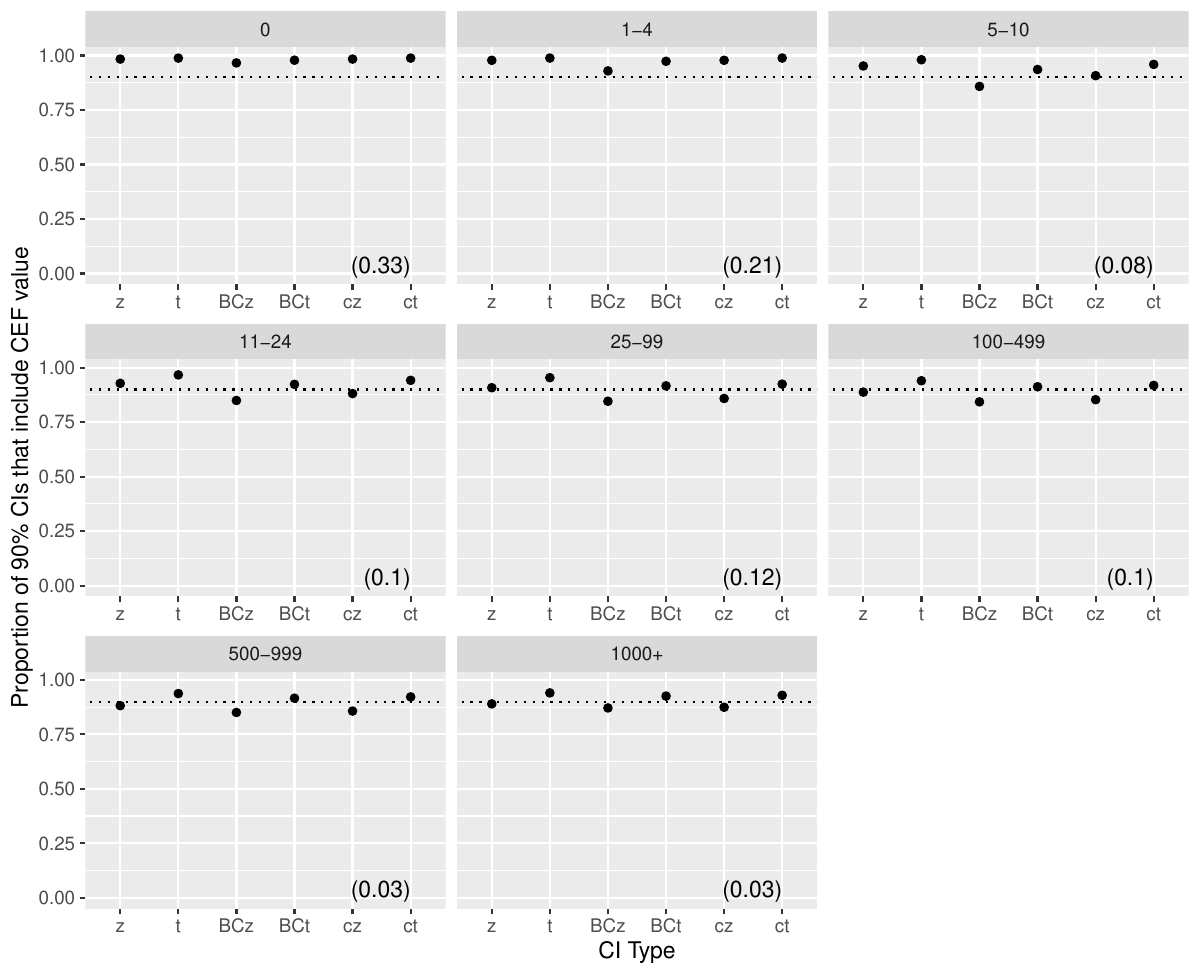}
\label{fig:tract_dhch_coverage}
\end{figure}

\begin{figure}[ht!]
\caption{The proportion of $933,120$ (405 queries x 2,304 geographies) 90\% confidence intervals that contained the CEF value for elementary school district level DHCH queries aggregated by query size (the bracketed number in the lower right corner indicates the proportion of the queries in that (size category, e.g. in this figure the plurality, 0.37, have value of 0). The $t$ distribution based confidence intervals performed better than the ones using the normal distribution for many of the panels. The t, BCt, and ct were all t or near the $0.90$ target for all panels.}
\includegraphics[width =\textwidth]{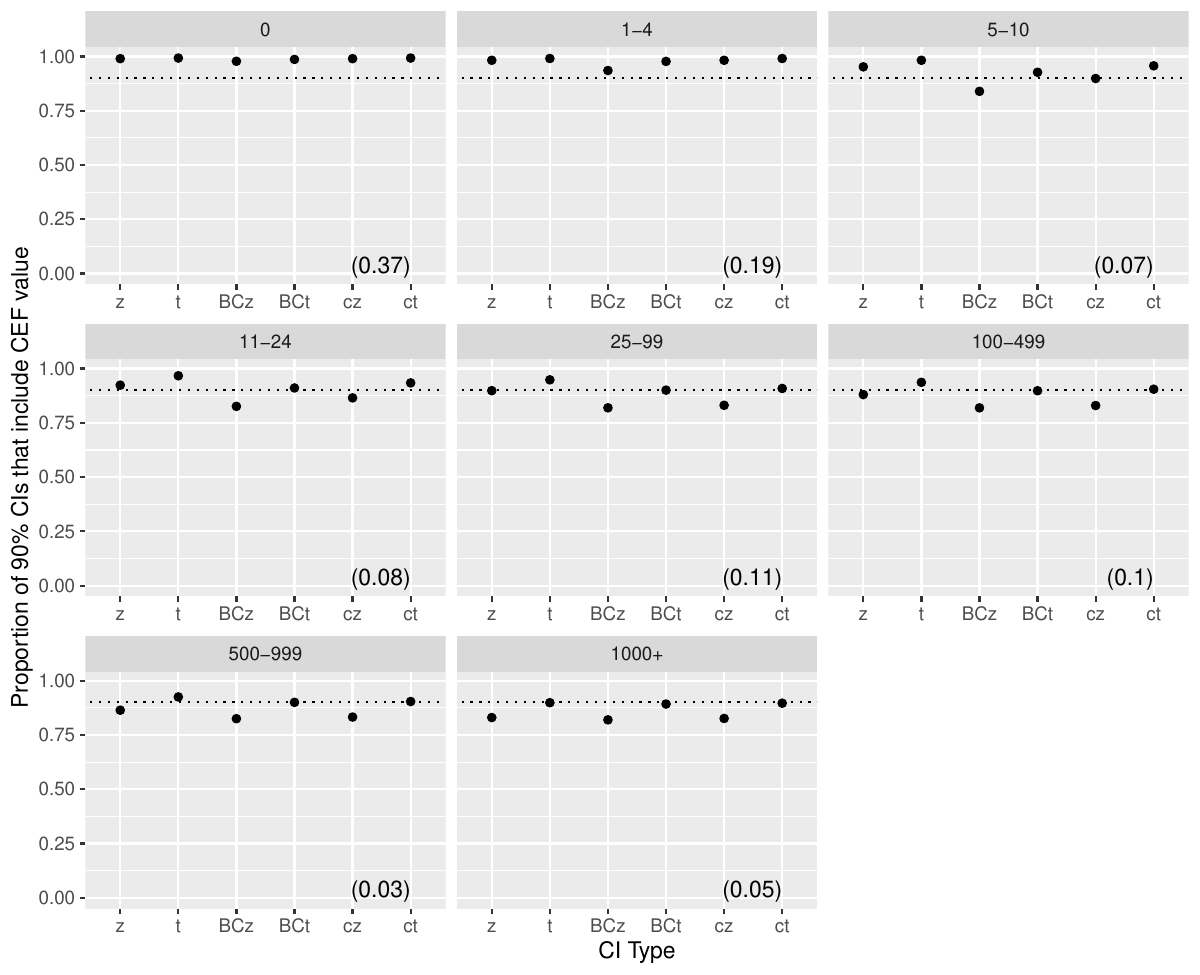}
\label{fig:sd_dhch_coverage}
\end{figure}

The second set of graphics (Figures \ref{fig:nat_pl94_cis} - \ref{fig:sd_pl94_cis}) show the proportion of AMC-based confidence intervals that contained the CEF value among redistricting queries for geographic levels not shown in the main paper.  The third set of graphics (Figure \ref{fig:nat_pl94_cis_width} - \ref{fig:sd_pl94_cis_width}) show the distribution of confidence interval widths for additional geographic levels not shown in the main paper. Figures \ref{fig:nat_dhcp_coverage} - \ref{fig:sd_dhch_coverage} show the proportion of AMC-based confidence intervals that contained the CEF value among DHCH and DHCP queries for geographic levels not shown in the main paper.  Lastly, Table \ref{tab:dhcp_bias_tab} shows the distribution of AMC-estimated bias for DHCP queries by query size and geographic level.

Statistics reported in this paper have been cleared for public release by the Census Bureau's Disclosure Review Board (DRB clearance number: CBDRB-FY24-DSEP-0002).

\begin{figure}[ht!]
\caption{U.S. and Puerto Rico level estimated bias ($Bias_{PPMF0}$ vs. $Bias_{CEF}$), grouped by CEF-based query size. Each panel corresponds to certain sizes (i.e. counts) of the query (based on CEF). In the case when $Bias_{PPMF0}$ values are very similar to $Bias_{CEF}$ values they line up along the diagonal, showing the AMC method to be a good approximation. This is generally the case with the data shown. For query size of 0, the estimated $Bias_{CEF}$ can only be positive, due to the non-negativity constraint. Circles show location of the $602$ (301 queries x 2 geographies) points used for the plot.}
\includegraphics[width = \textwidth]{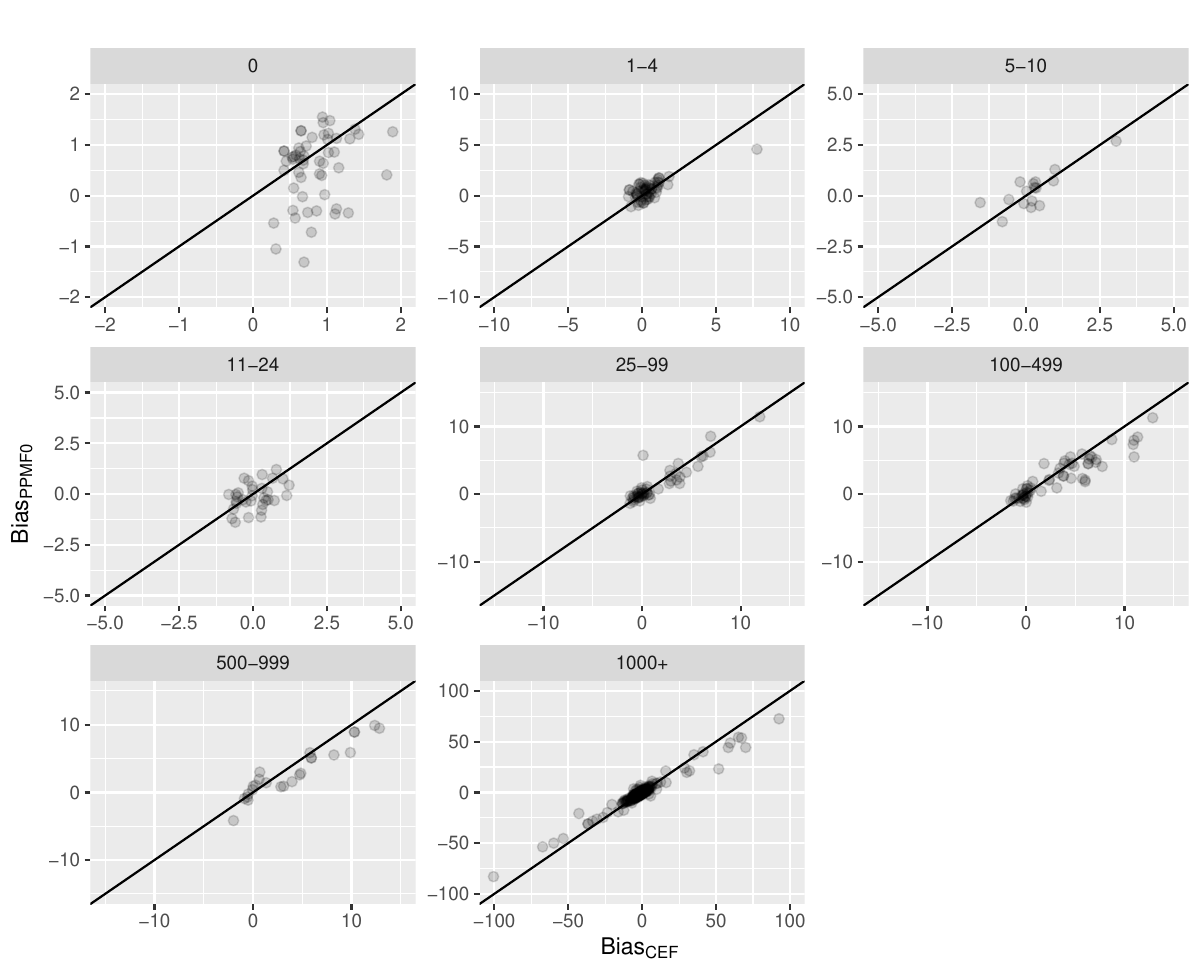}
\end{figure}

\begin{figure}[ht!]
\caption{U.S. and Puerto Rico level estimated standard deviation ($SD_{PPMF0}$ vs. $SD_{CEF}$) by CEF query size grouping. Each panel corresponds to certain sizes (i.e. counts) of the query (based on CEF). In the case when $SD_{PPMF0}$ values are very similar to $SD_{CEF}$ values they line up along the diagonal, showing the AMC method to be a good approximation. This is generally the case with the data shown. For some cases, the $SD_{PPMF0}$ value was very slightly higher than the standard deviation-from-CEF. Circles show location of the $602$ (301 queries x 2 geographies) points used for the plot.}
\includegraphics[width = \textwidth]{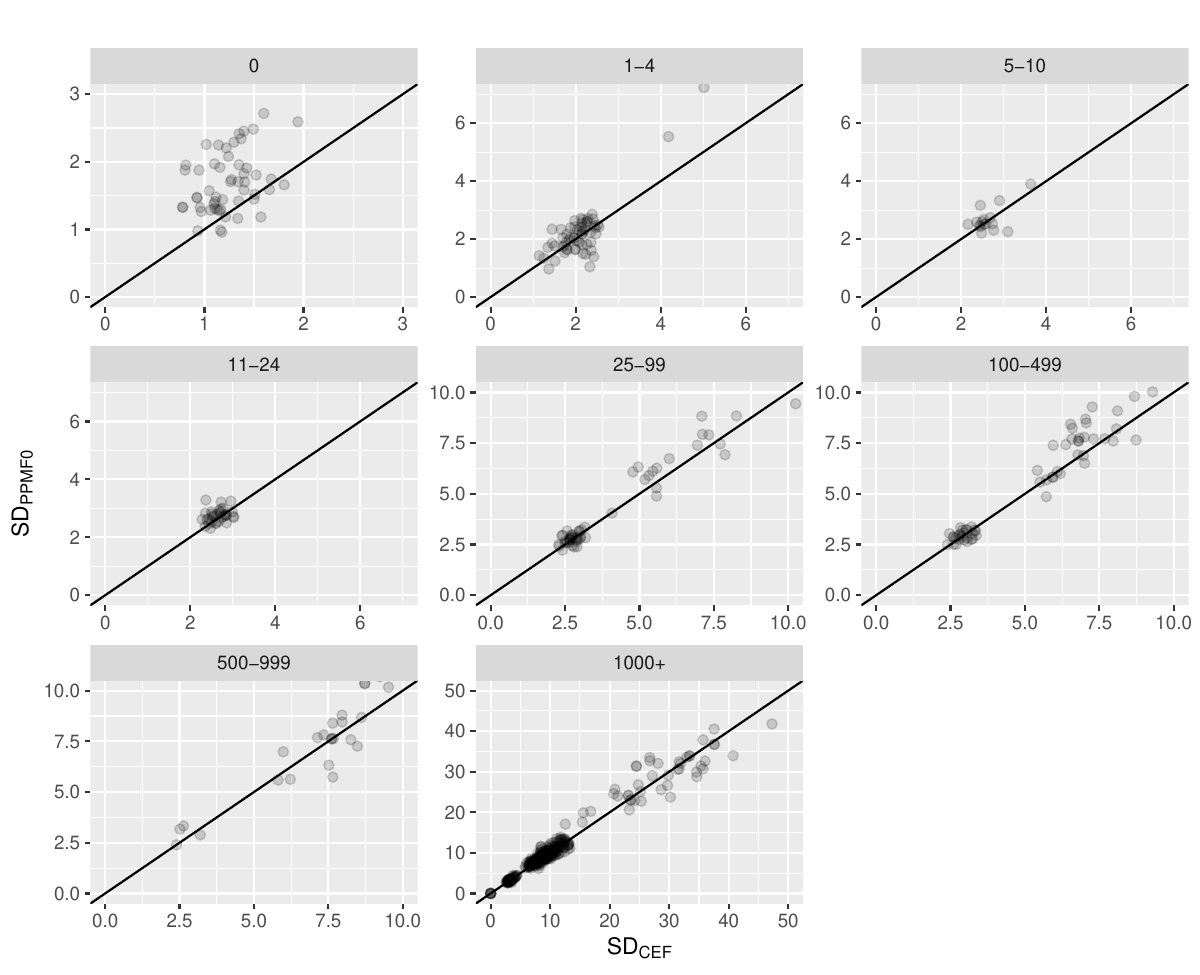}
\end{figure}

%State

%%%%%%%%%
%THIS ONE IS IN THE MAIN TEXT
%%%%%%%%%

% \begin{figure}[ht!]
% \caption{State level estimated RMSE-from-PPMF$_0$ vs estimated RMSE-from-CEF by CEF query size grouping}
% \includegraphics[width = 0.75\textwidth]{appendix_graphics/distribution_comparison/state_rmse.pdf}
% \end{figure}

\begin{figure}[ht!]
\caption{State level comparisons of estimated bias ($Bias_{PPMF0}$ vs. $Bias_{CEF}$), grouped by CEF-based query size. Each panel corresponds to certain sizes (i.e. counts) of the query (based on CEF). In the case when $Bias_{PPMF0}$ values are very similar to $Bias_{CEF}$ values they line up along the diagonal, showing the AMC method to be a good approximation. This is mostly the case with the data shown. For query size of 0, the estimated $Bias_{CEF}$ values can only be positive, due to the non-negativity constraint. Hexagons show location and density of the $15,652$ (301 queries x 52 geographies) points used for the plot.}
\includegraphics[width = \textwidth]{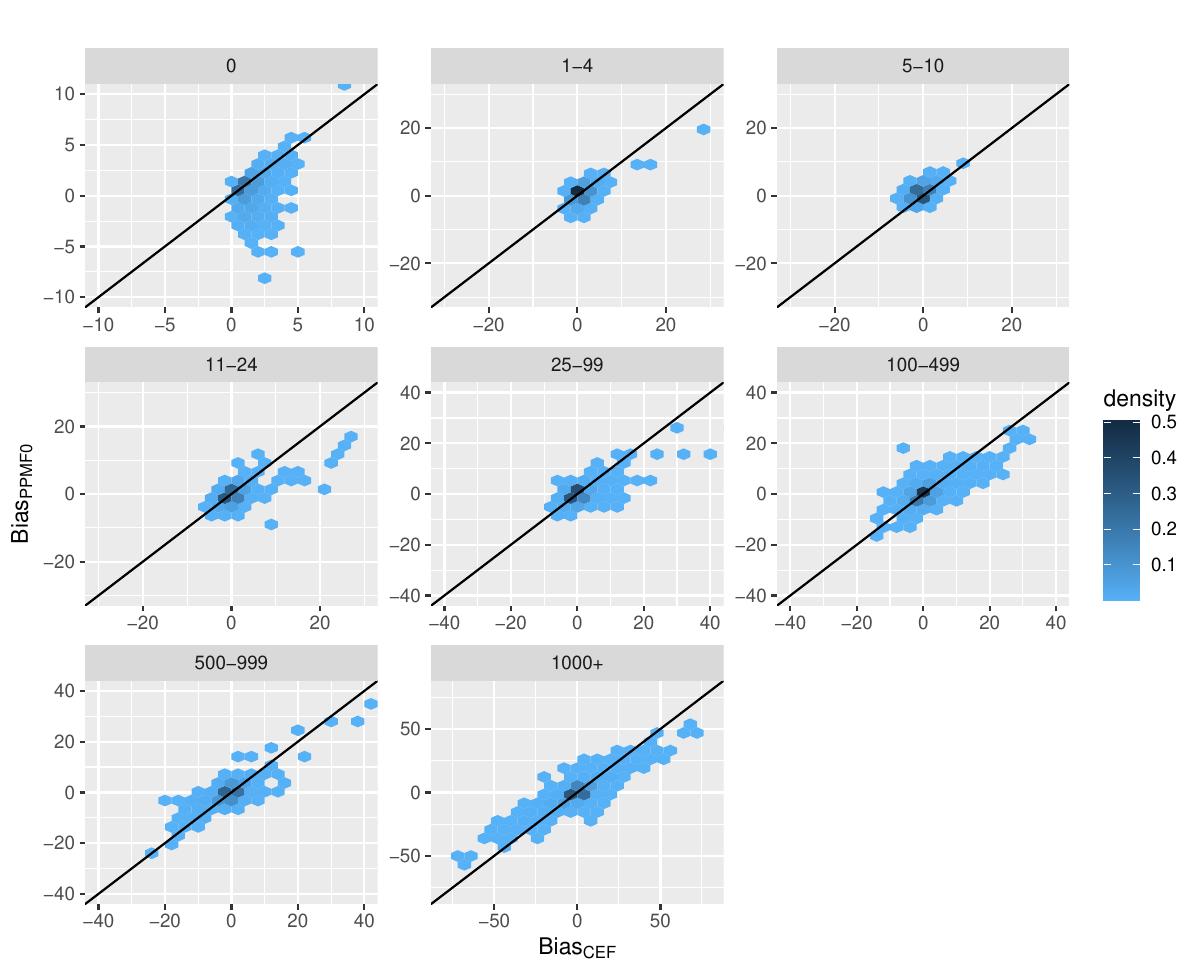}
\end{figure}

\begin{figure}[ht!]
\caption{State level comparisons of estimated standard deviation ($SD_{PPMF0}$ vs. $SD_{CEF}$), grouped by CEF-based query size. Each panel corresponds to certain sizes (i.e. counts) of the Query (based on CEF). In the case when $SD_{PPMF0}$ values are very similar to $SD_{CEF}$ values they line up along the diagonal, showing the AMC method to be a good approximation. This is mostly the case with the data shown. Hexagons show location and density of the $15,652$ (301 queries x 52 geographies) points used for the plot.}
\includegraphics[width = \textwidth]{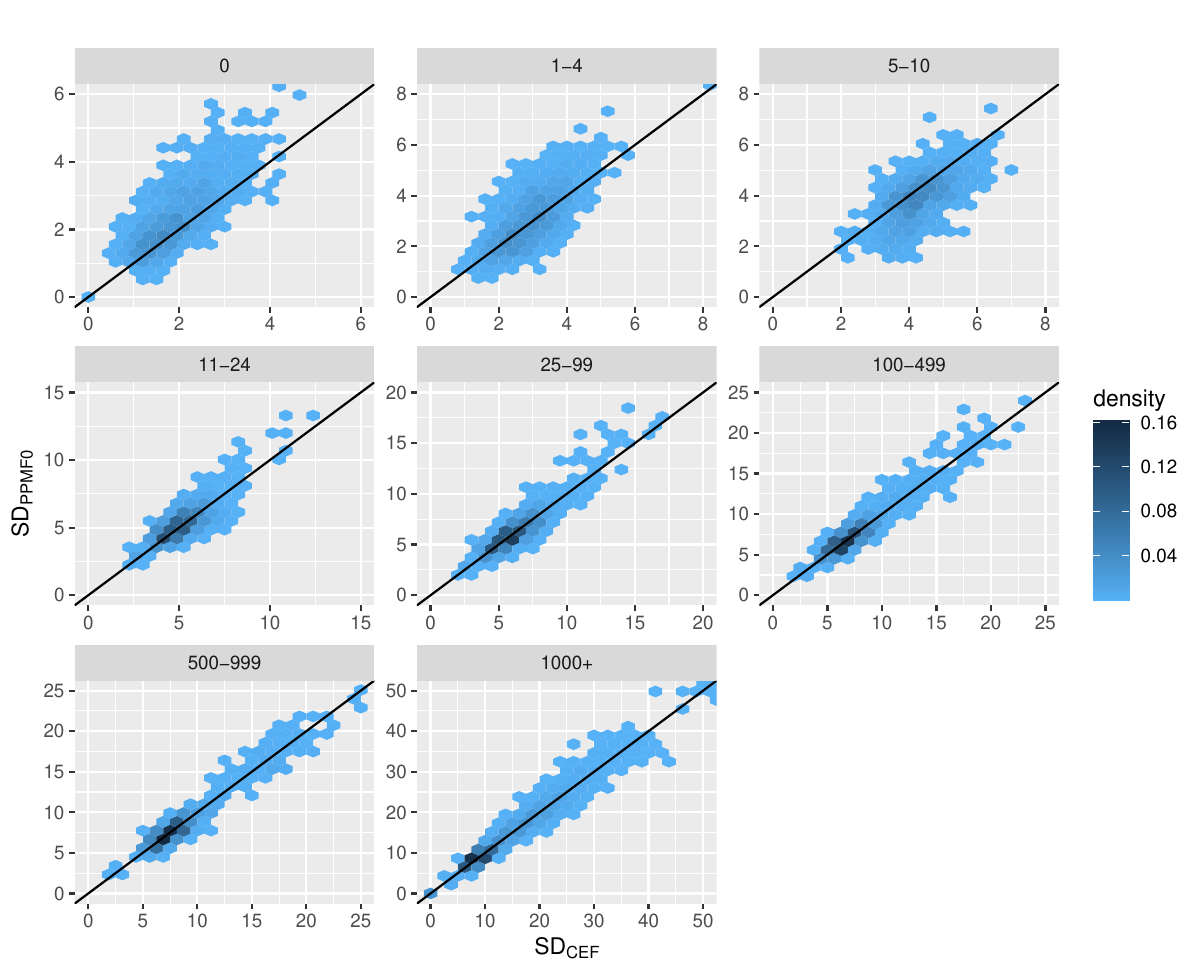}
\end{figure}

%County

\begin{figure}[ht!]
\caption{County level comparisons of estimated RMSE ($RSME_{PPMF0}$ vs. $RMSE_{CEF}$), grouped by CEF-based query size. Each panel corresponds to certain sizes (i.e. counts) of the query (based on CEF). In the case when $RSME_{PPMF0}$ values are very similar to $RMSE_{CEF}$ values they line up along the diagonal, showing the AMC method to be a good approximation. This is mostly the case with the data shown. Hexagons show location and density of the $969,521$ (301 queries x 3,221 geographies) points used for the plot.}
\includegraphics[width = \textwidth]{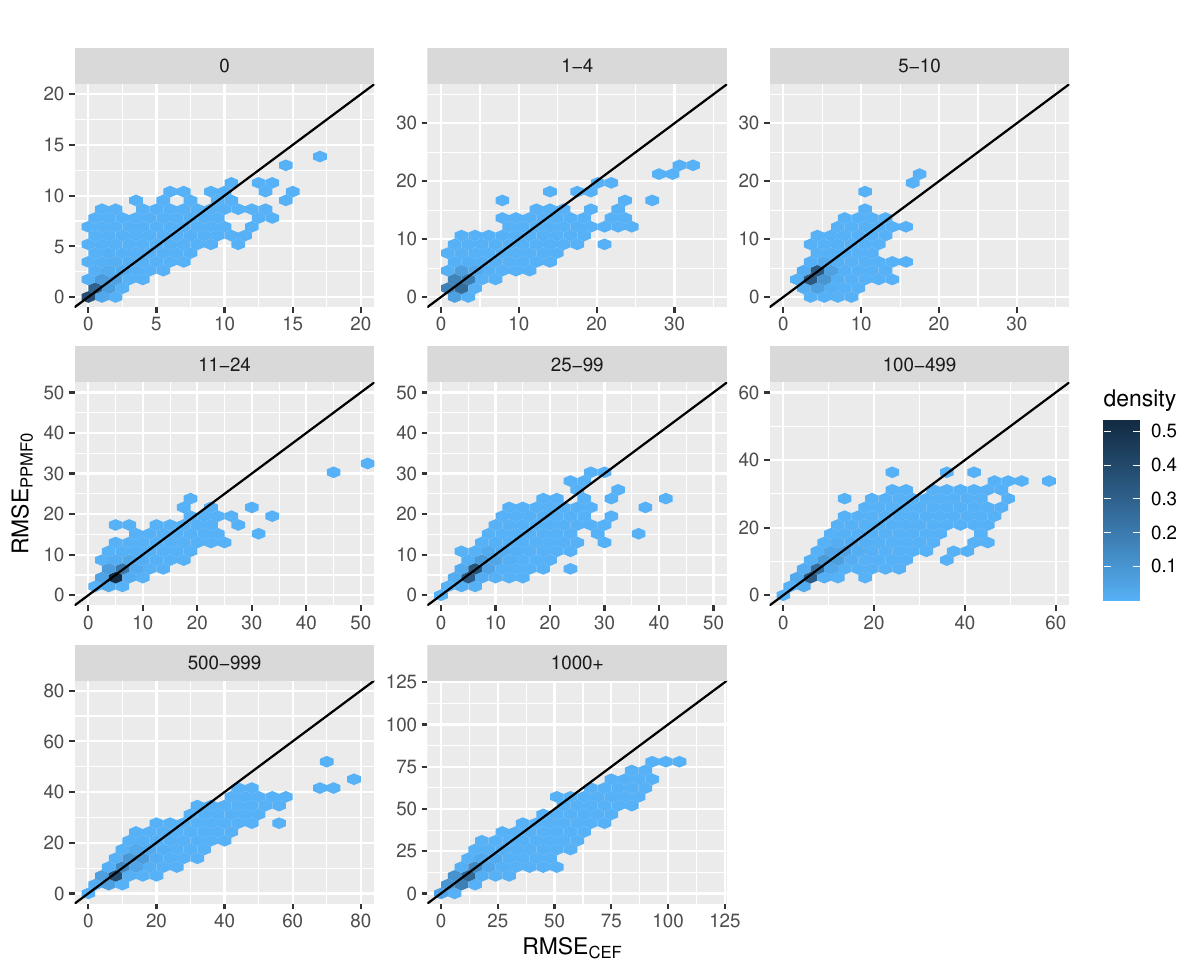}
\end{figure}

\begin{figure}[ht!]
\caption{County level estimated bias ($Bias_{PPMF0}$ vs. $Bias_{CEF}$), grouped by CEF-based query size. Each panel corresponds to certain sizes (i.e. counts) of the query (based on CEF). In the case when $Bias_{PPMF0}$ values are very similar to $Bias_{CEF}$ values they line up along the diagonal, showing the AMC method to be a good approximation. This is mostly the case with the data shown. For query size of 0, the estimated $Bias_{CEF}$ value can only be positive, due to the non-negativity constraint. For some of the larger query sizes (100-1000), sometimes the estimated $Bias_{CEF}$ is more strongly negative than the estimated $Bias_{PPMF0}$. Hexagons show location and density of the $969,521$ (301 queries x 3,221 geographies) points used for the plot.}
\includegraphics[width = \textwidth]{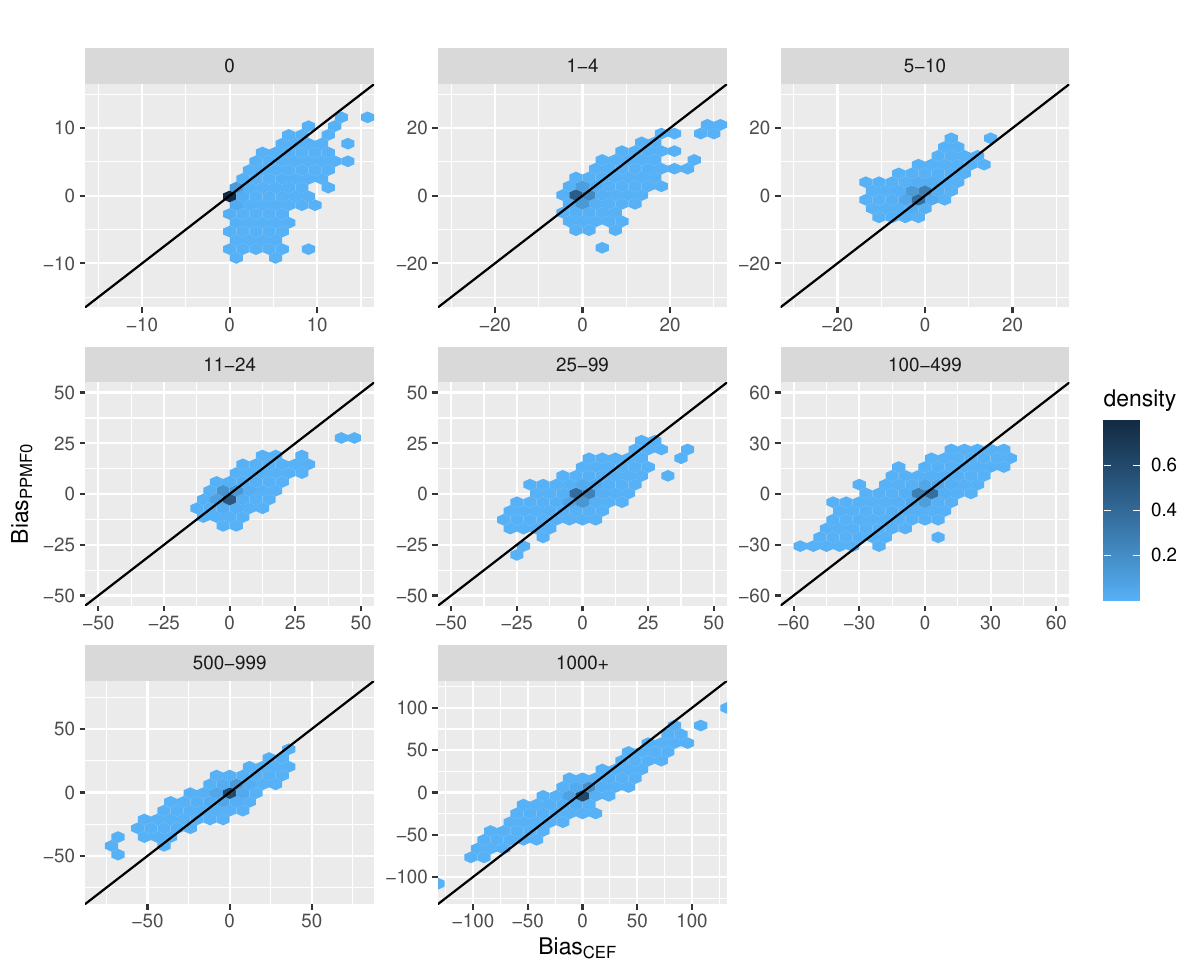}
\end{figure}

\begin{figure}[ht!]
\caption{County level estimated standard deviation ($SD_{PPMF0}$ vs. $SD_{CEF}$) by CEF query size grouping. Each panel corresponds to certain sizes (i.e. counts) of the query (based on CEF). In the case when $SD_{PPMF0}$ values are very similar to $SD_{CEF}$ values they line up along the diagonal, showing the AMC method to be a good approximation. This is mostly the case with the data shown; however, for some cases the $SD_{PPMF0}$ value appears to be very slightly overestimated for query sizes 0-4 and very slightly underestimated for query sizes 5-10. Hexagons show location and density of the $969,521$ (301 queries x 3,221 geographies) points used for the plot.}
\includegraphics[width = \textwidth]{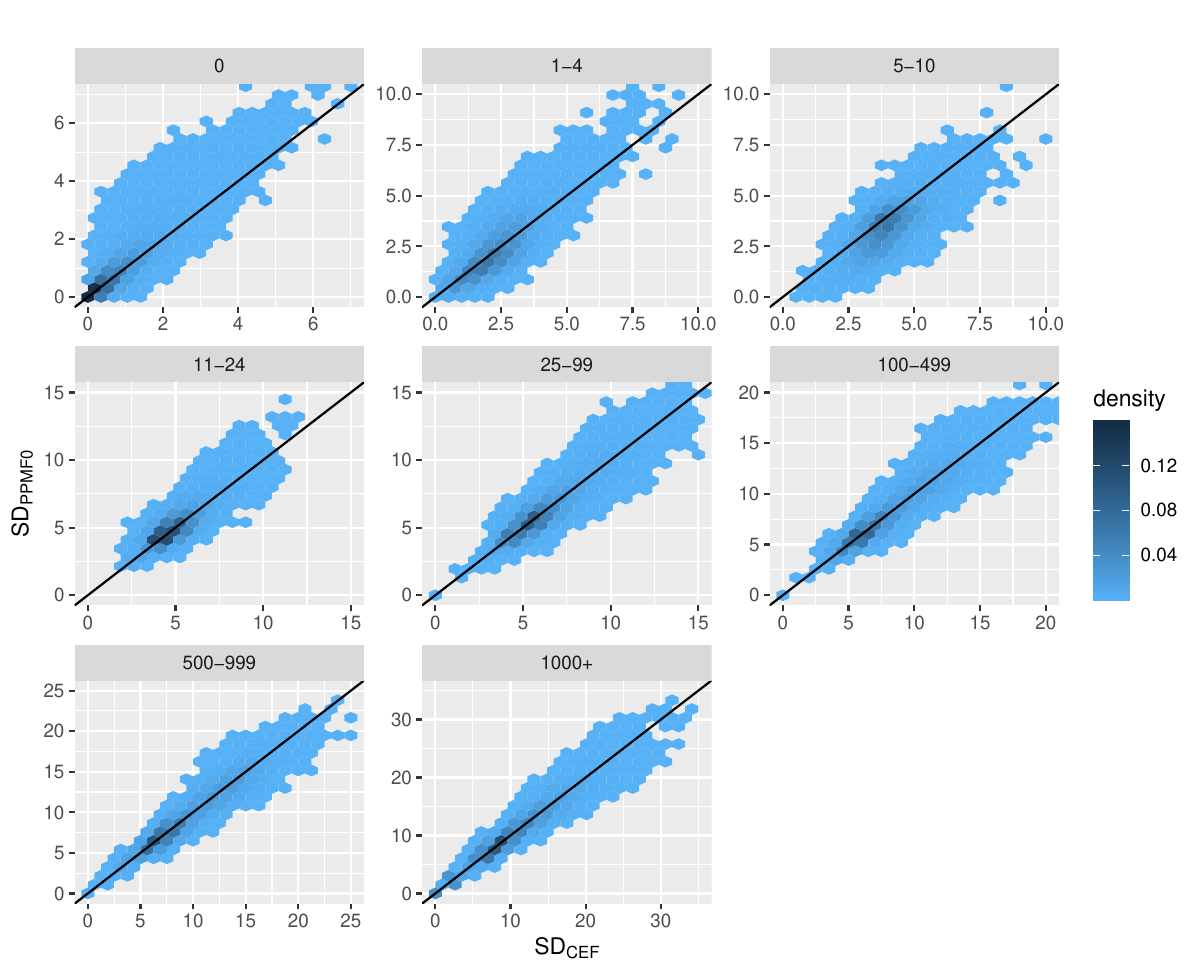}
\end{figure}

%Tract

\begin{figure}[ht!]
\caption{Tract level comparison of estimated RMSE ($RSME_{PPMF0}$ vs. $RMSE_{CEF}$). Each panel corresponds to certain sizes (i.e. counts) of the Query (based on CEF). In the case when $RSME_{PPMF0}$ values are very similar to $RMSE_{CEF}$ values they line up along the diagonal, showing the AMC method to be a good approximation. The $RSME_{PPMF0}$ values are underestimated in some cases. Hexagons show location and density of the $22,105,741$ (301 queries x 73,441 geographies) points used for the plot.}
\includegraphics[width = \textwidth]{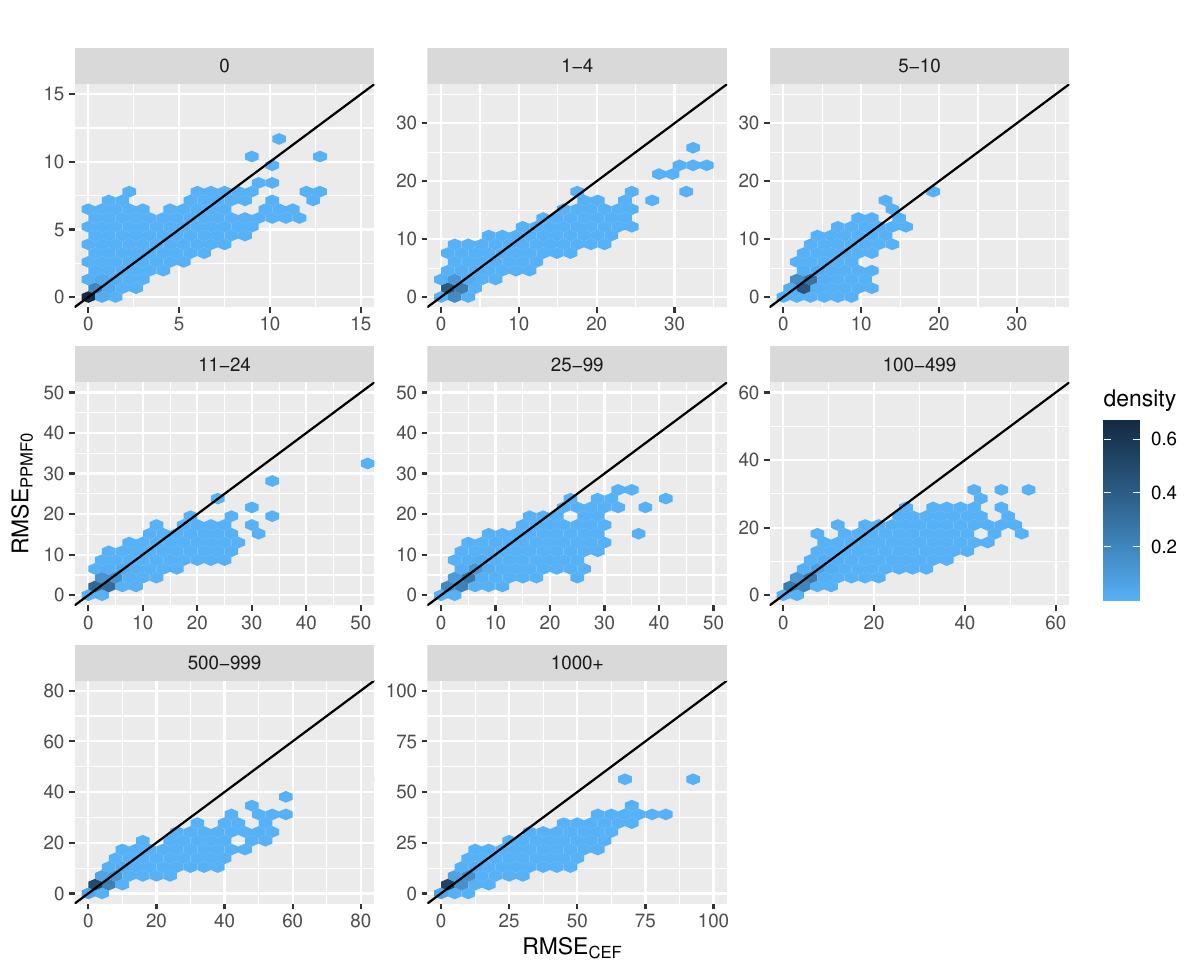}
\end{figure}

\begin{figure}[ht!]
\caption{Tract level estimated bias ($Bias_{PPMF0}$ vs. $Bias_{CEF}$), grouped by CEF-based query size. Each panel corresponds to certain sizes (i.e. counts) of the query (based on CEF). In the case when $Bias_{PPMF0}$ values are very similar to $Bias_{CEF}$ values they line up along the diagonal, showing the AMC method to be a good approximation. This is mostly the case with the data shown. For query size of 0, the estimated $Bias_{CEF}$ value can only be positive, due to the non-negativity constraint. For some of the larger query sizes (100-1000), sometimes the $Bias_{CEF}$ value is more strongly negative than the estimated $Bias_{PPMF0}$ value. Hexagons show location and density of the $22,105,741$ (301 queries x 73,441 geographies) points used for the plot.}
\includegraphics[width = \textwidth]{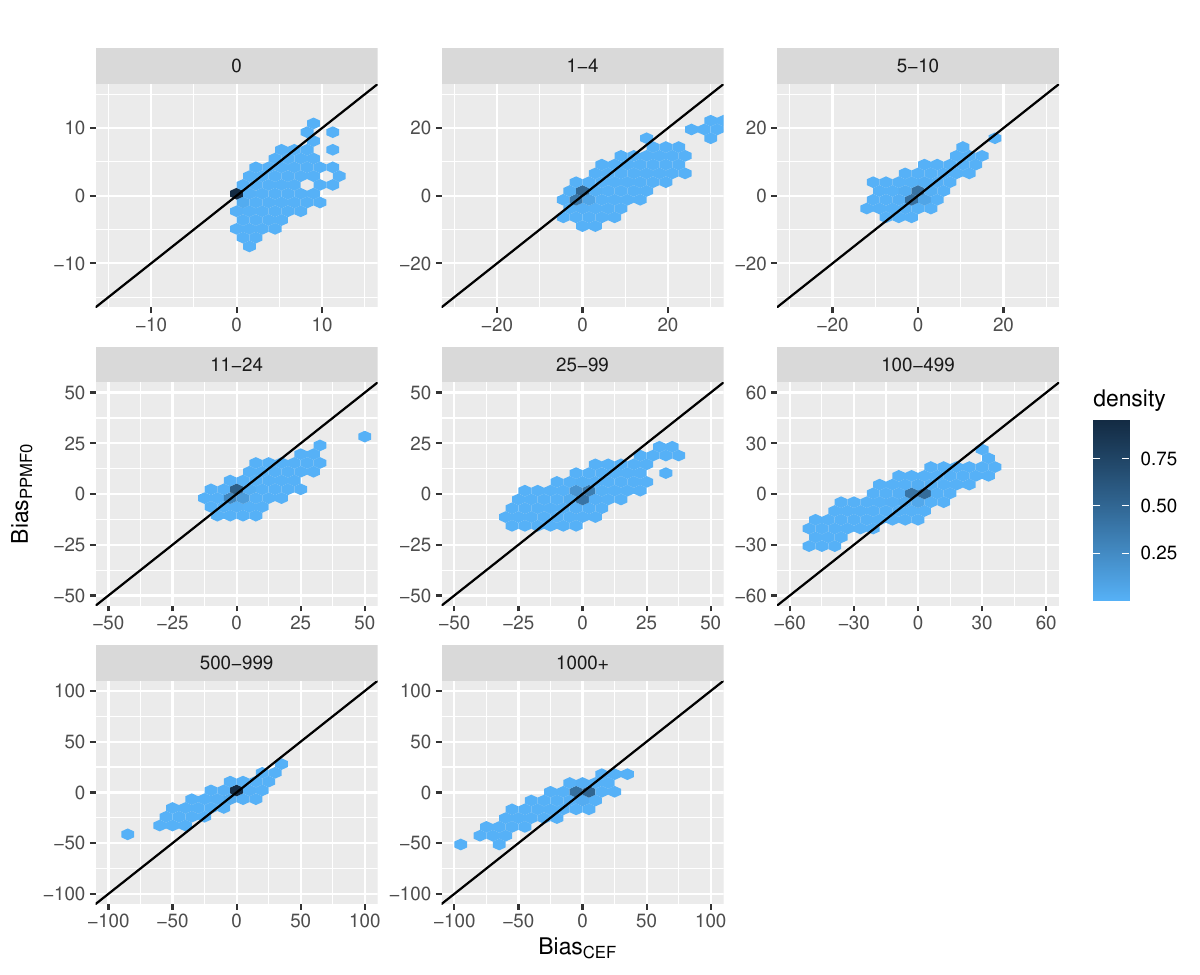}
\end{figure}

\begin{figure}[ht!]
\caption{Tract level estimated standard deviation ($SD_{PPMF0}$ vs. $SD_{CEF}$) by CEF query size grouping. Each panel corresponds to certain sizes (i.e. counts) of the Query (based on CEF). In the case when $SD_{PPMF0}$ values are very similar to $SD_{CEF}$ values they line up along the diagonal, showing the AMC method to be a good approximation. This is mostly the case with the data shown; however, for a few cases the $SD_{PPMF0}$ value appears very slightly underestimated for query sizes 5-10.  Hexagons show location and density of the $22,105,741$ (301 queries x 73,441 geographies) points used for the plot.}
\includegraphics[width = 0.75\textwidth]{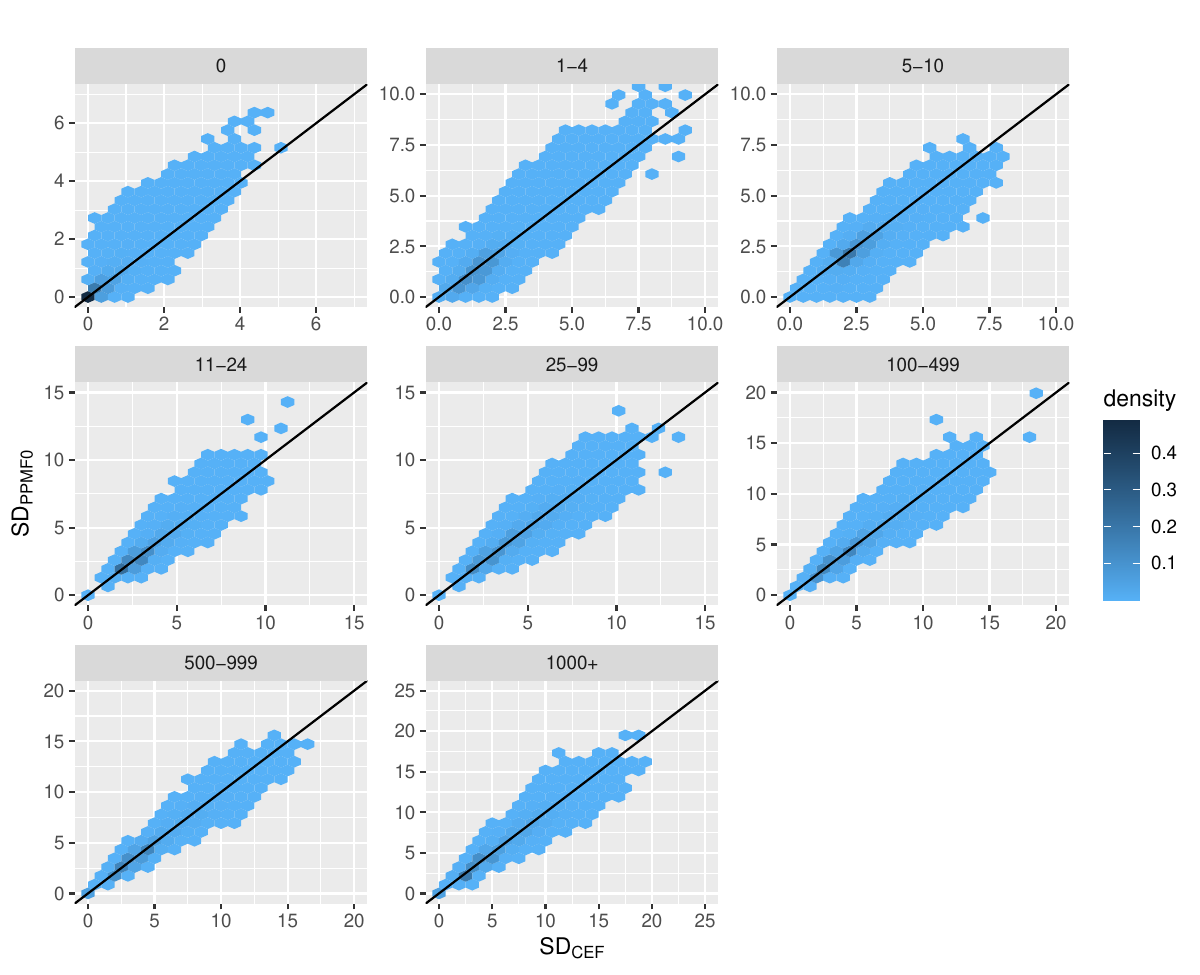}
\end{figure}

%Block

\begin{figure}[ht!]
\caption{Block level comparison of estimated RMSE ($RSME_{PPMF0}$ vs. $RMSE_{CEF}$). Each panel corresponds to certain sizes (i.e. counts) of the Query (based on CEF). When $RSME_{PPMF0}$ values are very similar to $RMSE_{CEF}$ values they line up along the diagonal, showing the AMC method to be a good approximation; however, the block level results show that the $RSME_{PPMF0}$ values are underestimated in many cases. Hexagons show location and density of the $780,493$ (301 queries x 2,593 sampled block geographies) points used for the plot.}
\includegraphics[width = \textwidth]{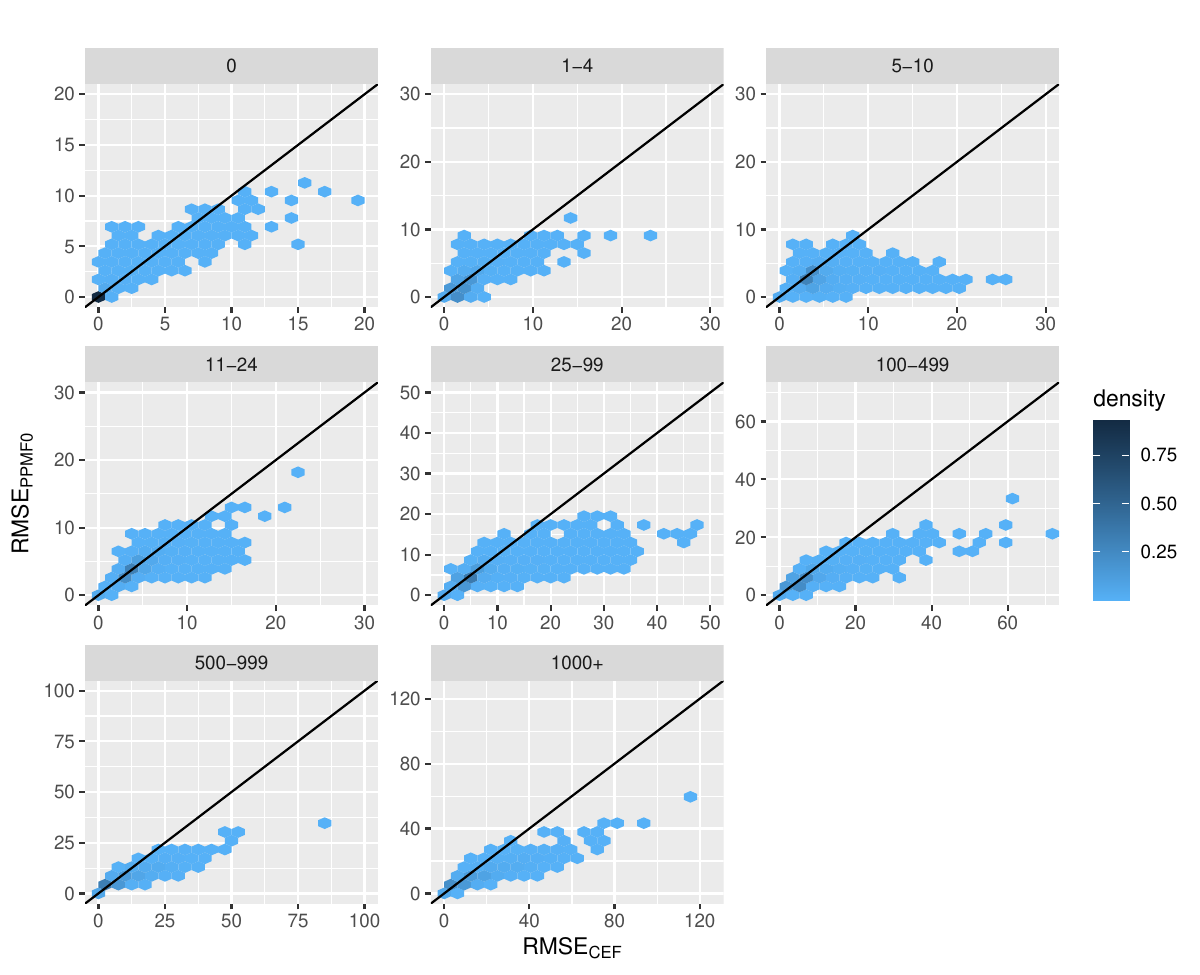}
\end{figure}

\begin{figure}[ht!]
\caption{Block level estimated bias ($Bias_{PPMF0}$ vs. $Bias_{CEF}$), grouped by CEF-based query size. Each panel corresponds to certain sizes (i.e. counts) of the Query (based on CEF). In the case when $Bias_{PPMF0}$ values are very similar to $Bias_{CEF}$ values they line up along the diagonal, showing the AMC method to be a good approximation. This is mostly the case with the data shown. For query size of 0, the estimated $Bias_{CEF}$ value can only be positive, due to the non-negativity constraint. The estimated $Bias_{CEF}$ value can tend to be more strongly negative than the estimated $Bias_{PPMF0}$ value. Hexagons show location and density of the $780,493$ (301 queries x 2,593 sampled block geographies) points used for the plot.}
\includegraphics[width = \textwidth]{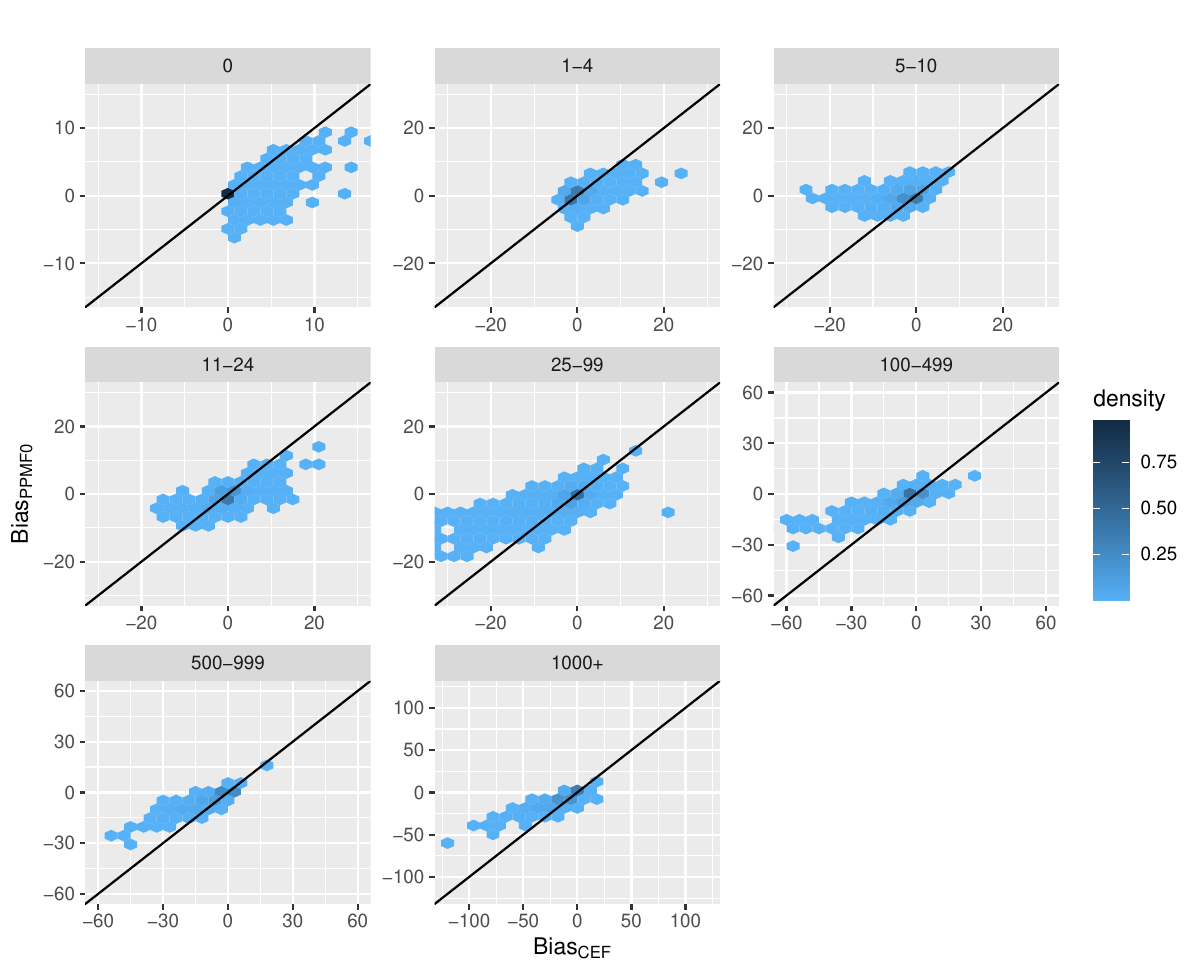}
\end{figure}

\begin{figure}[ht!]
\caption{Block level estimated standard deviation ($SD_{PPMF0}$ vs. $SD_{CEF}$) by CEF query size grouping. Each panel corresponds to certain sizes (i.e. counts) of the Query (based on CEF). In the case when $SD_{PPMF0}$ values are very similar to $SD_{CEF}$ values they line up along the diagonal, showing the AMC method to be a good approximation. This is mostly the case with the data shown; however, for a few cases the $SD_{PPMF0}$ value appears very slightly underestimated for query sizes 5-10. Hexagons show location and density of the $780,493$ (301 queries x 2,593 sampled block geographies) points used for the plot.}
\includegraphics[width = \textwidth]{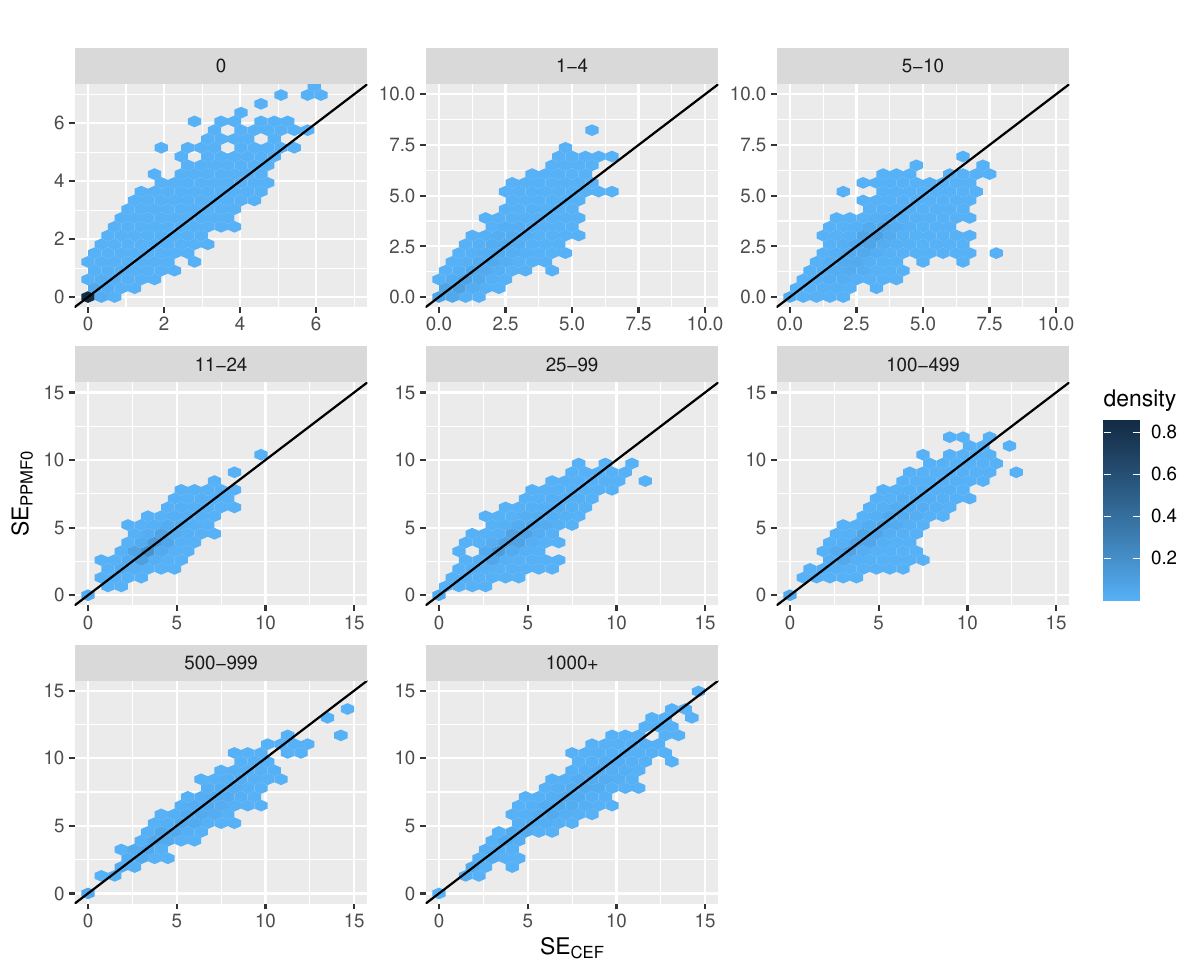}
\end{figure}

%AIAN Area

\begin{figure}[ht!]
\caption{AIAN area level comparison of estimated RMSE ($RSME_{PPMF0}$ vs. $RMSE_{CEF}$). Each panel corresponds to certain sizes (i.e. counts) of the Query (based on CEF). When $RSME_{PPMF0}$ values are very similar to $RMSE_{CEF}$ values they line up along the diagonal, showing the AMC method to be a good approximation. This is generally the case in the data shown. Hexagons show location and density of the $186,921$ (301 queries x 621 geographies) points used for the plot.}
\includegraphics[width = \textwidth]{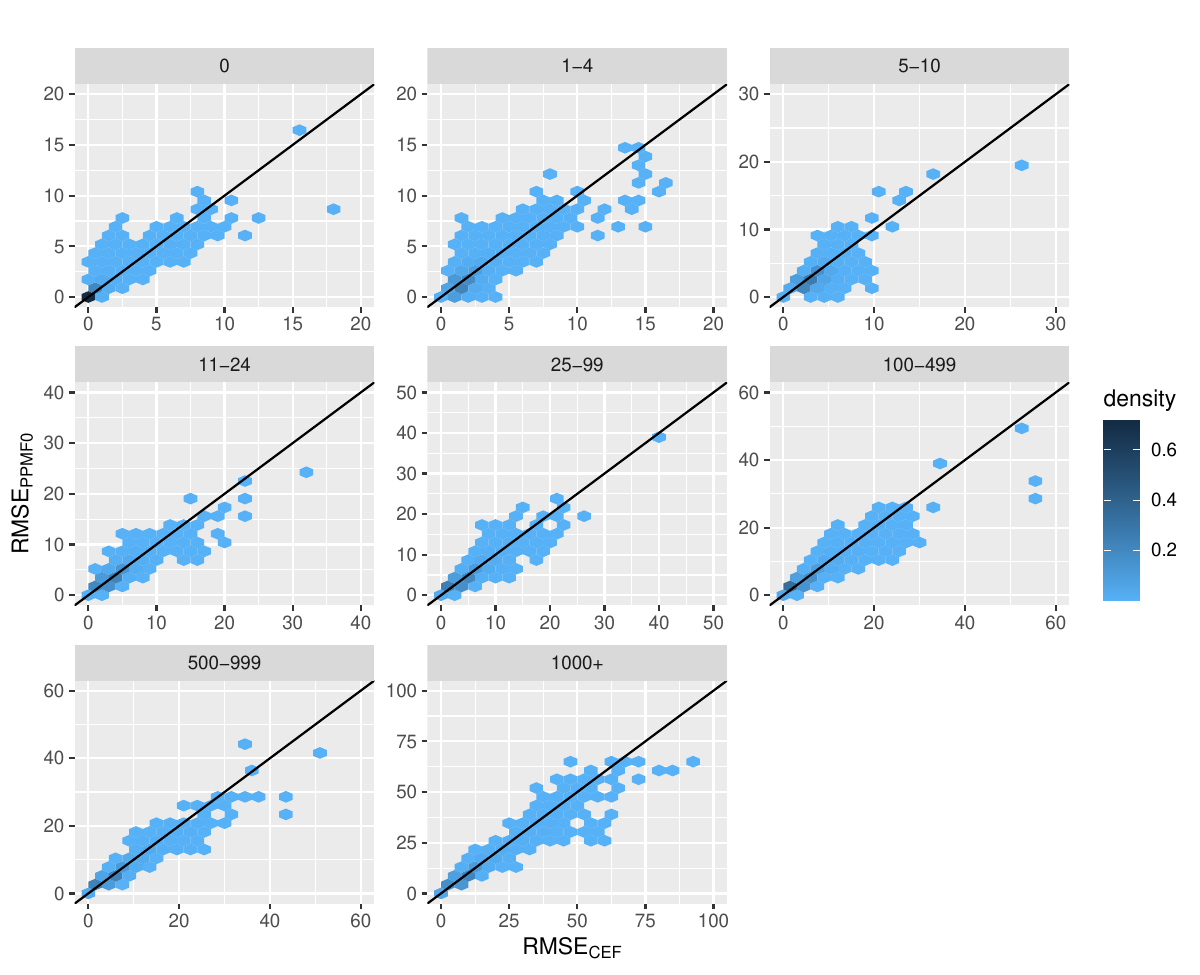}
\end{figure}

\begin{figure}[ht!]
\caption{AIAN area level estimated bias ($Bias_{PPMF0}$ vs. $Bias_{CEF}$), grouped by CEF-based query size. Each panel corresponds to certain sizes (i.e. counts) of the Query (based on CEF). In the case when $Bias_{PPMF0}$ values are very similar to $Bias_{CEF}$ values they line up along the diagonal, showing the AMC method to be a good approximation. This is mostly the case with the data shown. For query size of 0, the estimated $Bias_{CEF}$ value can only be positive, due to the non-negativity constraint. The estimated $Bias_{CEF}$ value can tend to be more strongly negative than the estimated $Bias_{PPMF0}$ value. Hexagons show location and density of the $186,921$ (301 queries x 621 geographies) points used for the plot.}
\includegraphics[width = \textwidth]{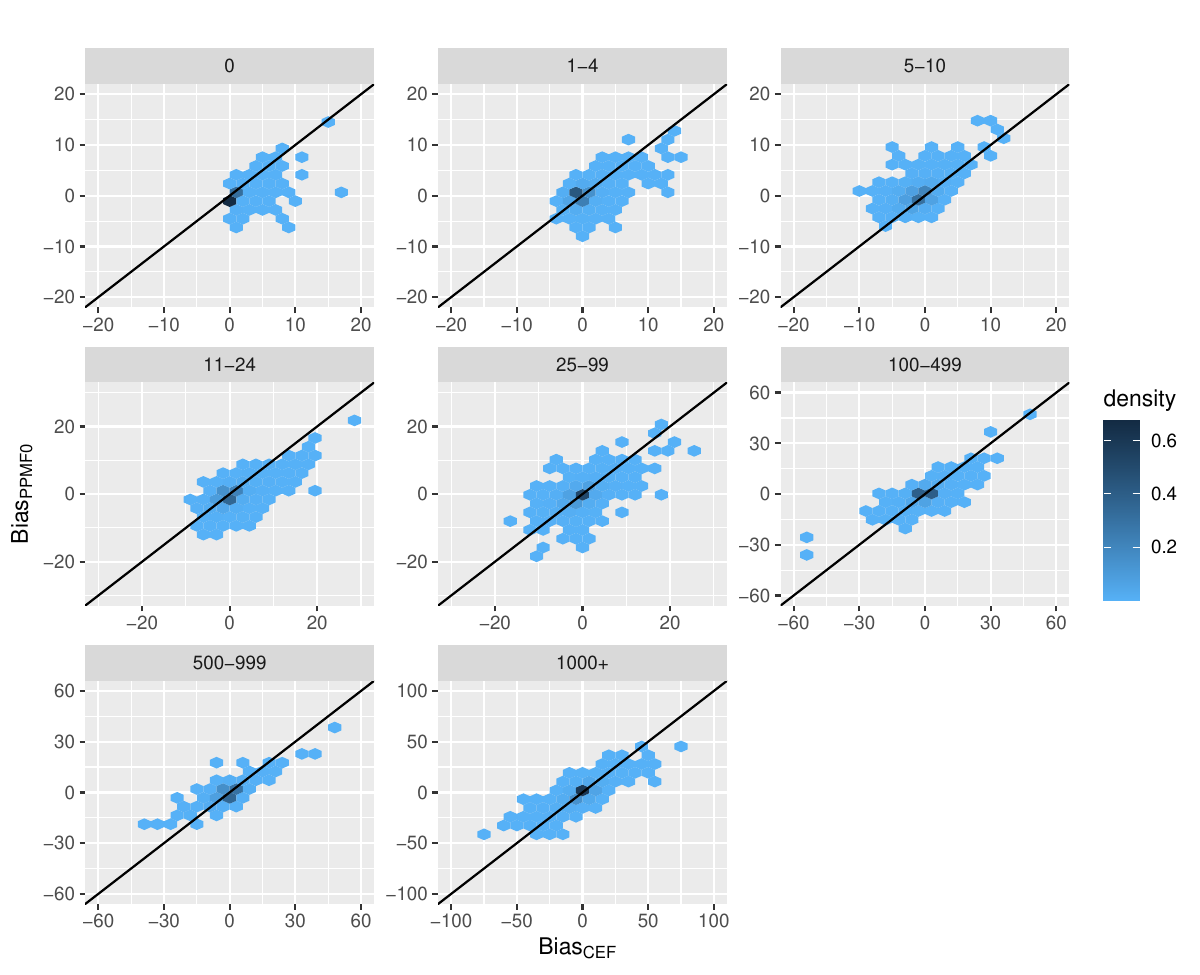}
\end{figure}

\begin{figure}[ht!]
\caption{AIAN area level estimated standard deviation ($SD_{PPMF0}$ vs. $SD_{CEF}$) by CEF query size grouping. Each panel corresponds to certain sizes (i.e. counts) of the Query (based on CEF). In the case when $SD_{PPMF0}$ values are very similar to $SD_{CEF}$ values they line up along the diagonal, showing the AMC method to be a good approximation. This is generally the case with the data shown. Hexagons show location and density of the $186,921$ (301 queries x 621 geographies) points used for the plot.}
\includegraphics[width = \textwidth]{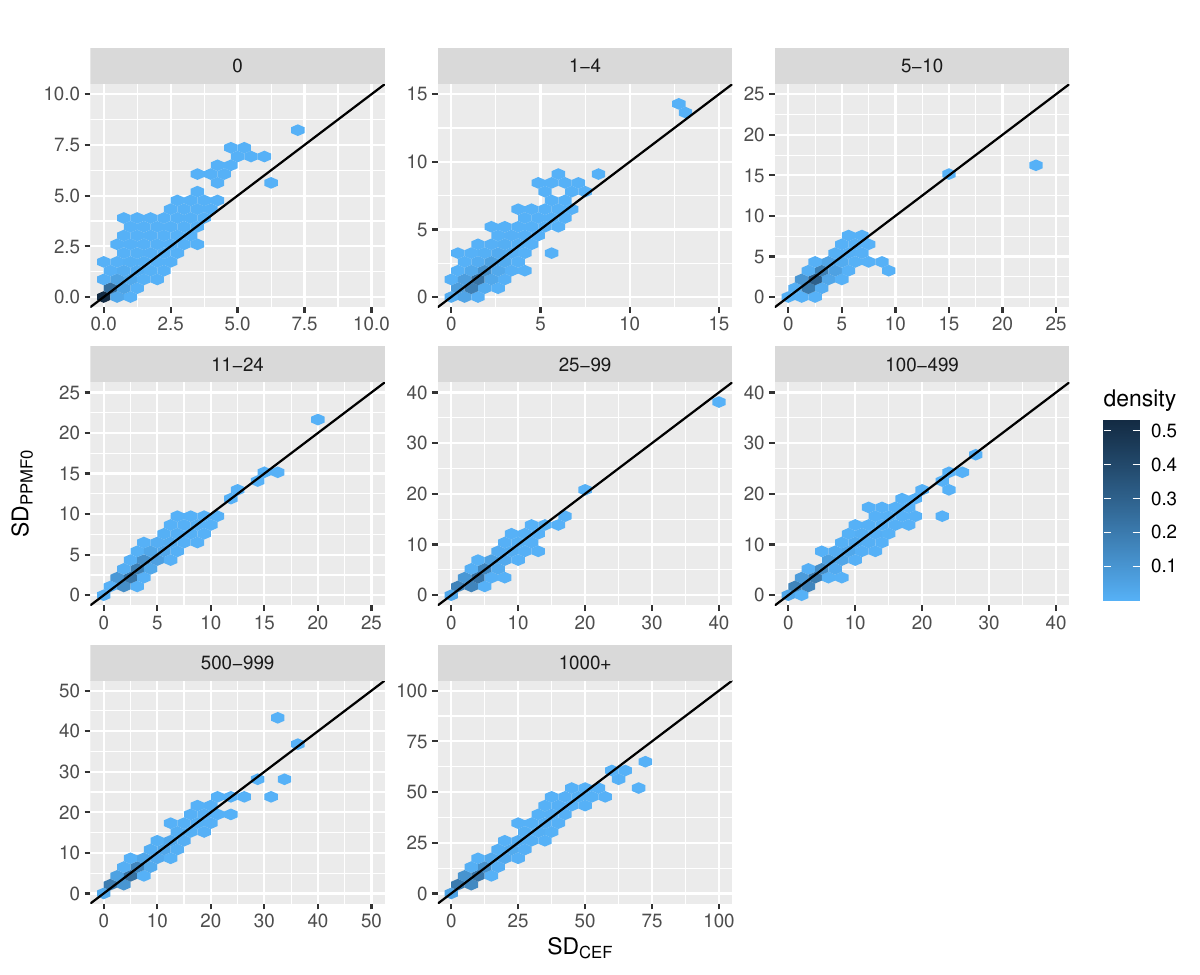}
\end{figure}

%School District

\begin{figure}[ht!]
\caption{Elementary school district level comparison of estimated RMSE ($RSME_{PPMF0}$ vs. $RMSE_{CEF}$). Each panel corresponds to certain sizes (i.e. counts) of the Query (based on CEF). When $RSME_{PPMF0}$ values are very similar to $RMSE_{CEF}$ values they line up along the diagonal, showing the AMC method to be a good approximation; however,  the $RSME_{PPMF0}$ value appears to be underestimated in many cases. Hexagons show location and density of the $693,504$ (301 queries x 2,304 geographies) points used for the plot.}
\includegraphics[width = \textwidth]{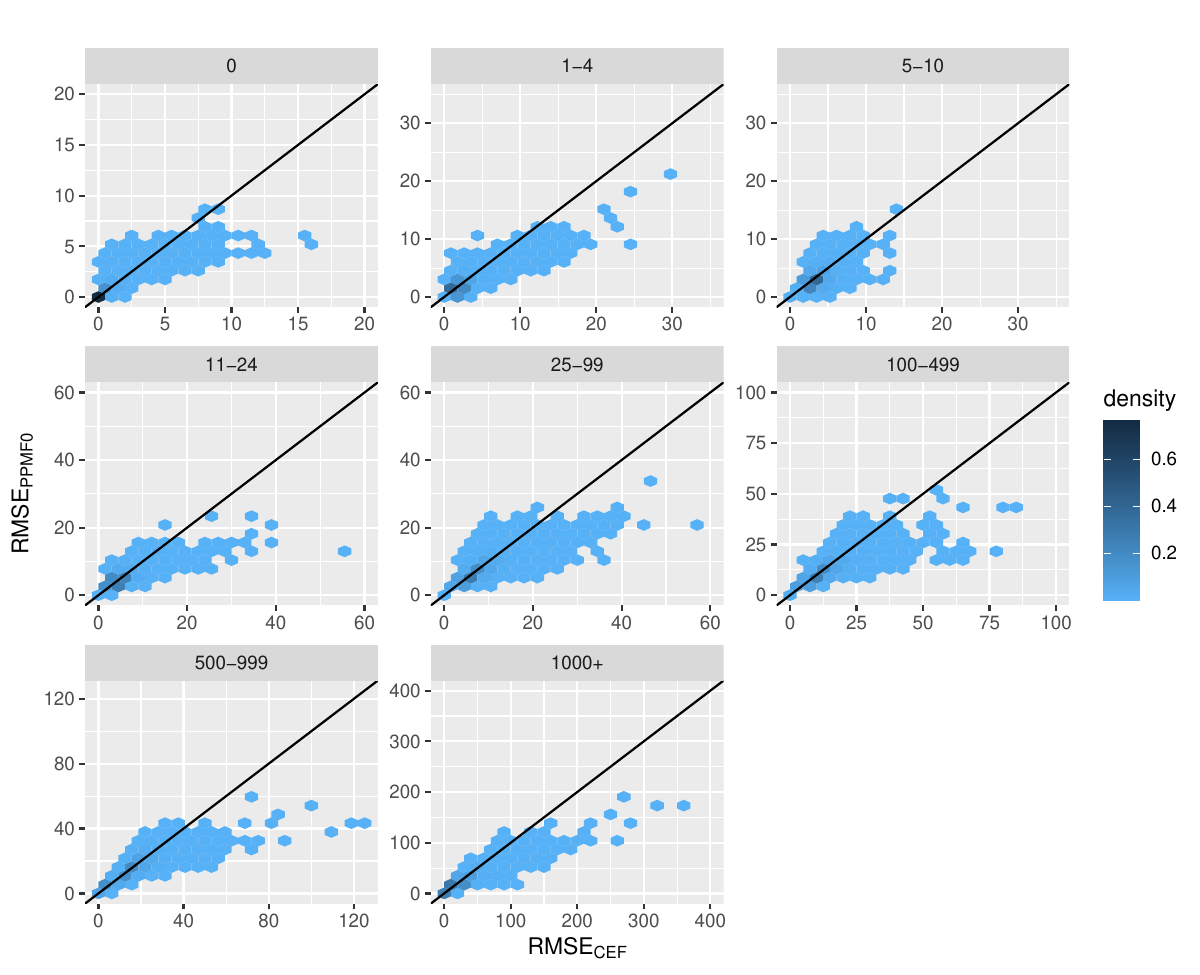}
\end{figure}

%%%%%%%%%%%%%%%%%%%%%%%%%%%%%
%CIs
%%%%%%%%%%%%%%%%%%%%%%%%%%%%%

\begin{figure}[ht!]
\caption{The proportion of $15,652$ (301 queries x 52 geographies) 90\% confidence intervals that contained the CEF value for state level PL94 queries aggregated by query size (the bracketed number in the lower right corner indicates the proportion of the queries in that size category, e.g. in this figure the plurality, 0.31, have value of 1000 or larger). Non-parametric CIs perform the most poorly with respect to containing the CEF value. For the Wald-based CIs, the ones using Student's $t$ distribution performed slightly better than the ones using the normal distribution. Conditional bias correction (ct) performs similarly to not using bias correction (t)  with both meeting the 0.90 target in all query size groupings.}
\includegraphics[width =\textwidth]{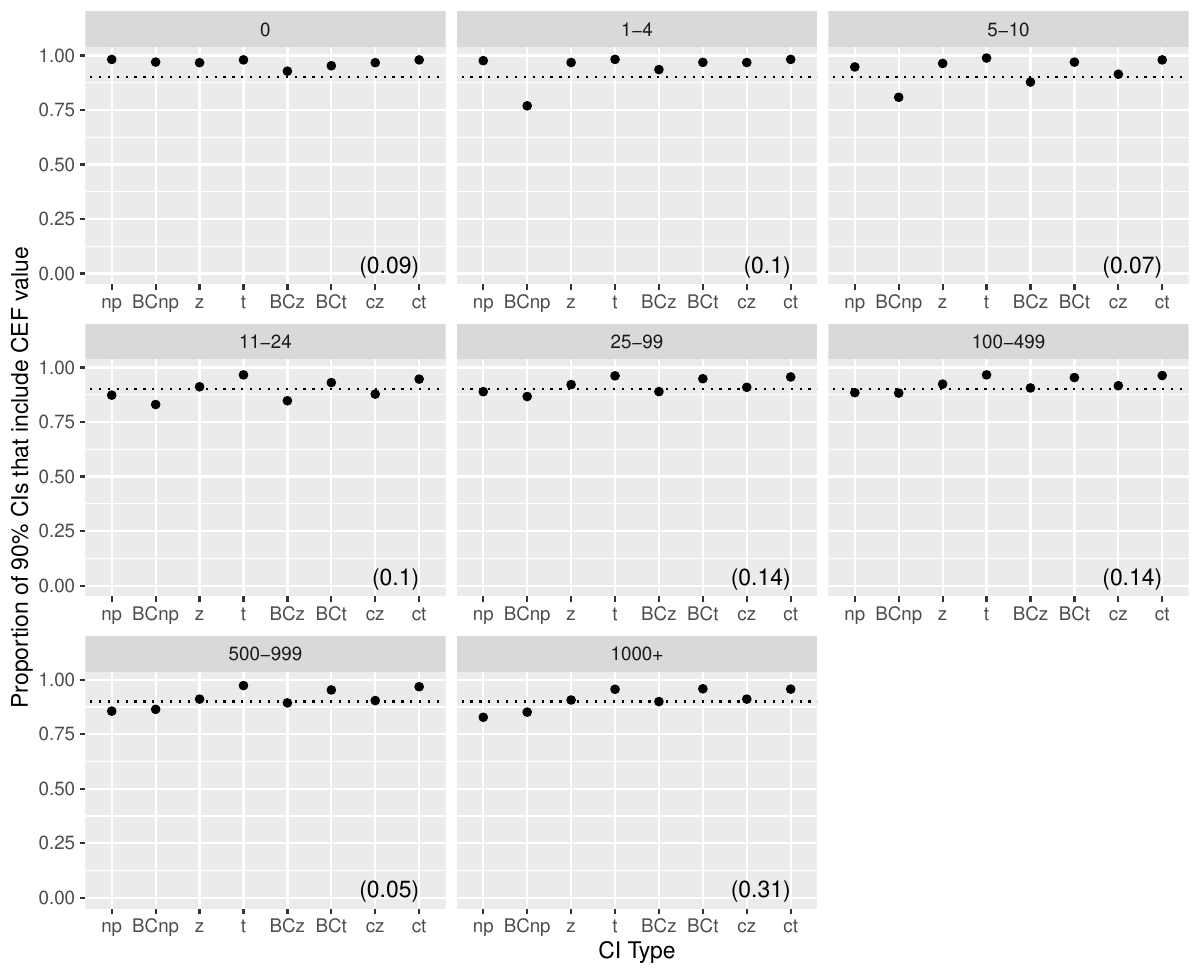}
\end{figure}

\begin{figure}[ht!]
\caption{The proportion of $969,521$ (301 queries x 3,221 geographies) 90\% confidence intervals that contained the CEF value for county level PL94 queries aggregated by query size (the bracketed number in the lower right corner indicates the proportion of the queries in that size category, e.g. in this figure the majority, 0.58, have value of 0). Non-parametric CIs perform the most poorly with respect to containing the CEF value. For the Wald-based CIs, the ones using Student's $t$ distribution performed slightly better than the ones using the normal distribution. Conditional bias correction (ct) performs similarly to not using bias correction (t)  with both meeting the 0.90 target in all query size groupings.}
\includegraphics[width =\textwidth]{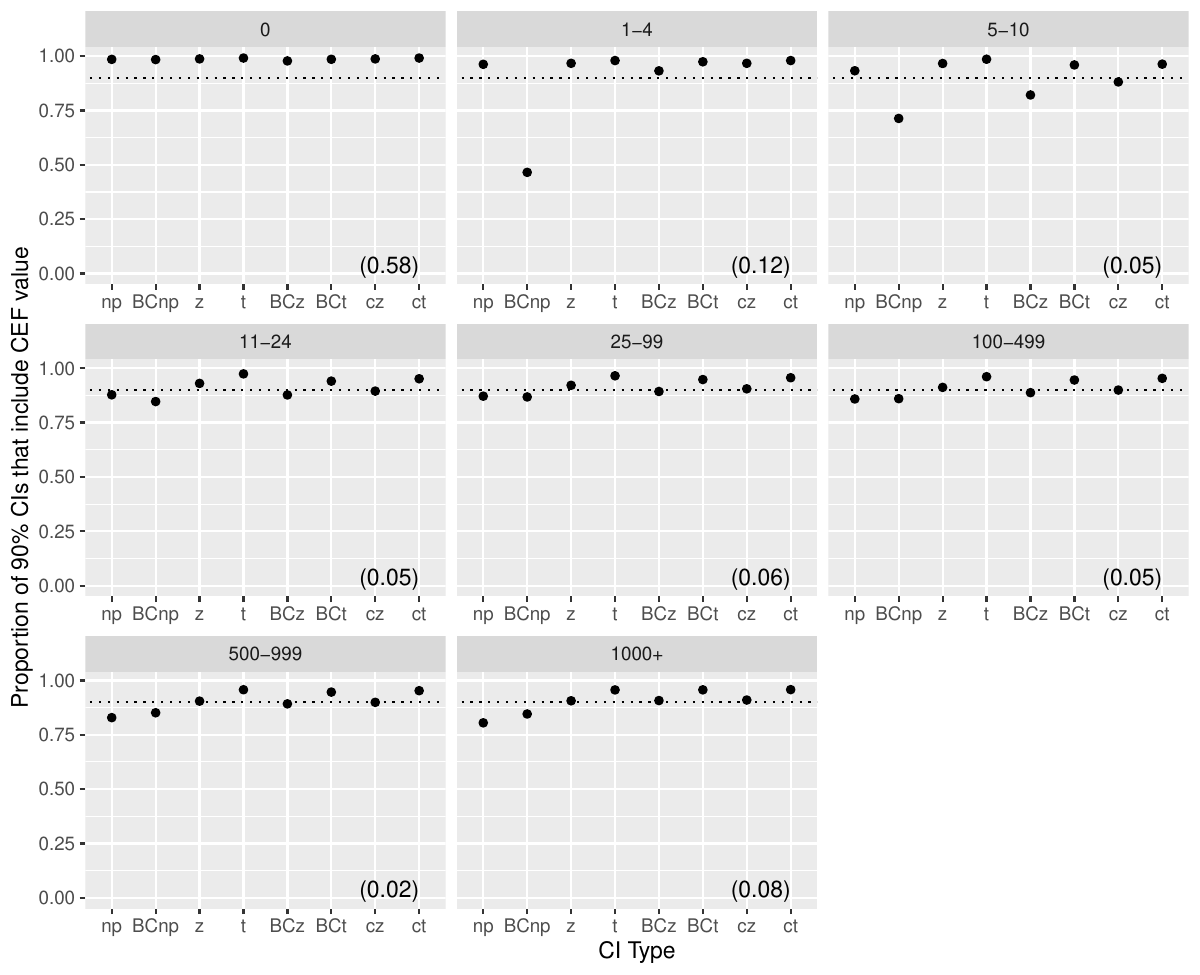}
\end{figure}

\begin{figure}[ht!]
\caption{The proportion of $22,105,741$ (301 queries x 73,441 geographies) 90\% confidence intervals that contained the CEF value for tract level PL94 queries aggregated by query size (the bracketed number in the lower right corner indicates the proportion of the queries in that size category, e.g. in this figure the majority, 0.71, have value of 0). Non-parametric CIs perform the most poorly with respect to containing the CEF value. For the Wald-based CIs, the ones using Student's $t$ distribution performed slightly better than the ones using the normal distribution. Conditional bias correction (ct) performs similarly to not using bias correction (t)  with both meeting the 0.90 target in all query size groupings.}
\includegraphics[width =\textwidth]{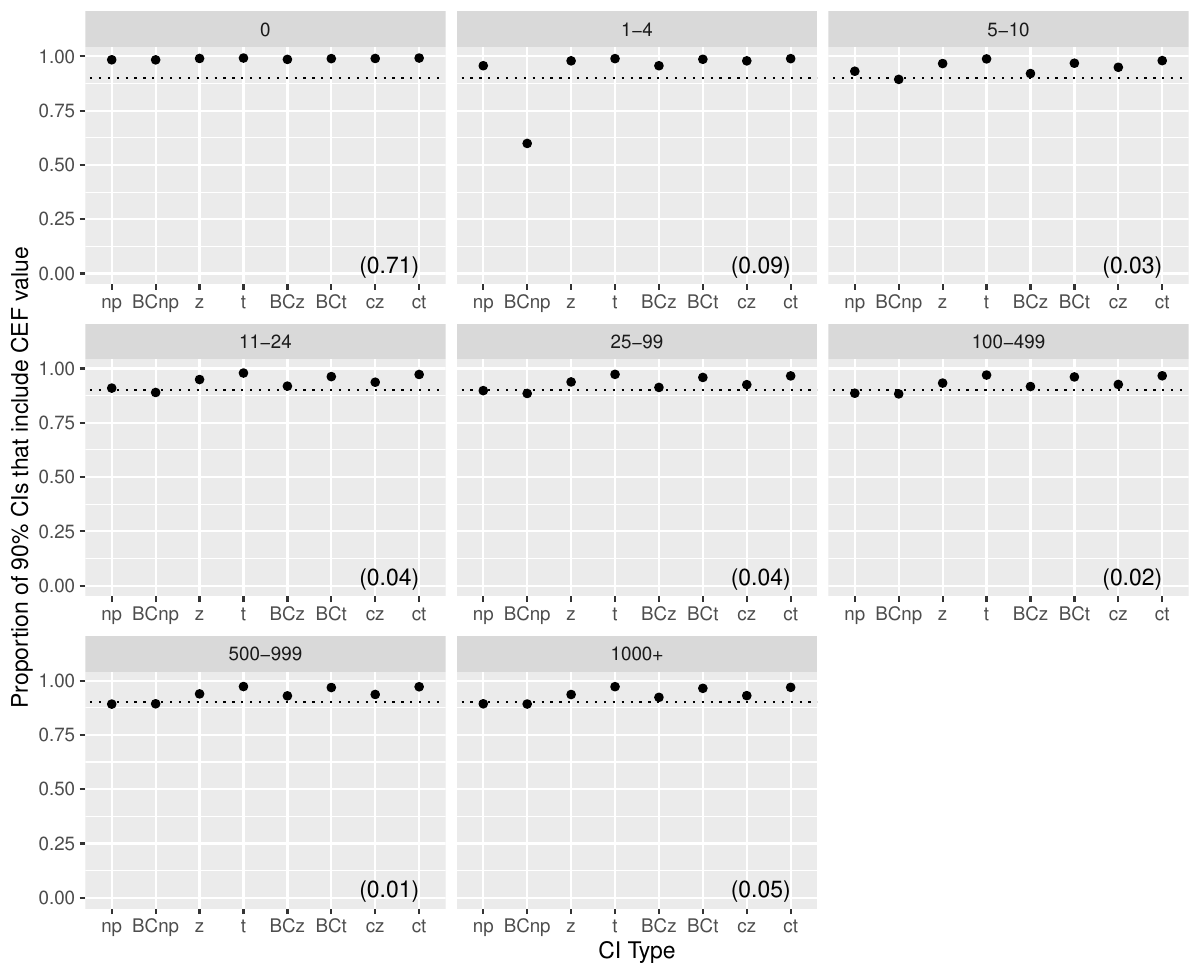}
\end{figure}

%This One in the main text

% \begin{figure}[ht!]
% \caption{The proportion of 90\% CIs that contained the CEF value for 2010 Block level PL94 queries aggregated by query count (size)}
% \includegraphics[width =0.75\textwidth]{appendix_graphics/PL94cis/block_pl94_cis_coverage.pdf}
% \end{figure}

\begin{figure}[ht!]
\caption{The proportion of $186,921$ (301 queries x 621 geographies) 90\% confidence intervals that contained the CEF value for AIAN area level PL94 queries aggregated by query size (the bracketed number in the lower right corner indicates the proportion of the queries in that size category, e.g. in this figure the majority, 0.79, have value of 0). Non-parametric CIs perform the most poorly with respect to containing the CEF value. For the Wald-based CIs, the ones using Student's $t$ distribution performed slightly better than the ones using the normal distribution. Conditional bias correction (ct) performs similarly to not using bias correction (t)  with both meeting the 0.90 target in all query size groupings.}
\includegraphics[width =\textwidth]{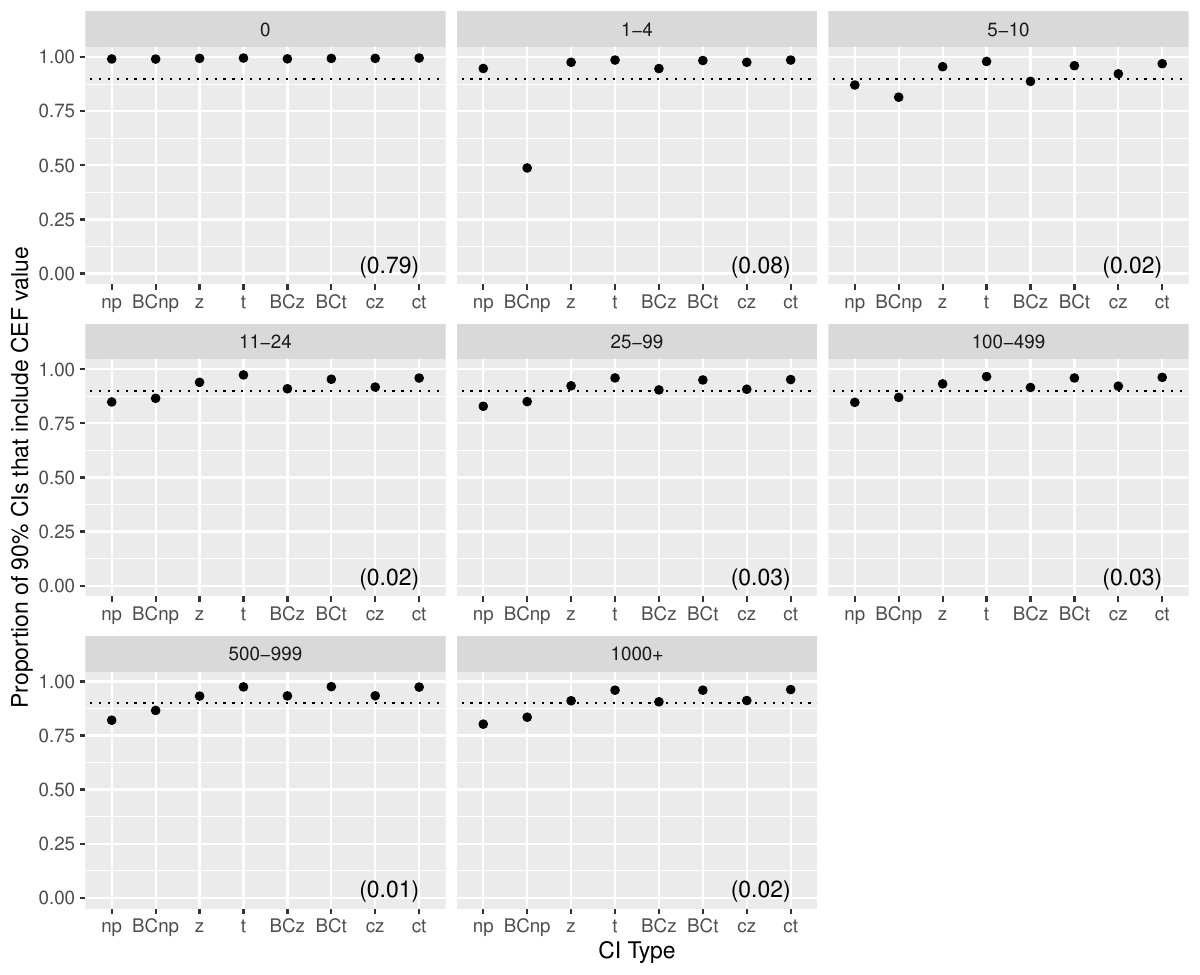}
\end{figure}

%%%%%%%%%%%%%%%%%%%%%%%%%%%%
%CI Widths
%%%%%%%%%%%%%%%%%%%%%%%%%%%%

\begin{figure}[ht!]
\caption{The distribution of $15,652$ (301 queries x 52 geographies) 90\% confidence interval widths for state PL94 queries aggregated by query size. As is to be expected, $t$ and BC$t$ intervals are slightly wider than the $z$ and BC $z$ intervals; also the confidence interval width generally increases as the size of the query increases.}
\includegraphics[width =0.75\textwidth]{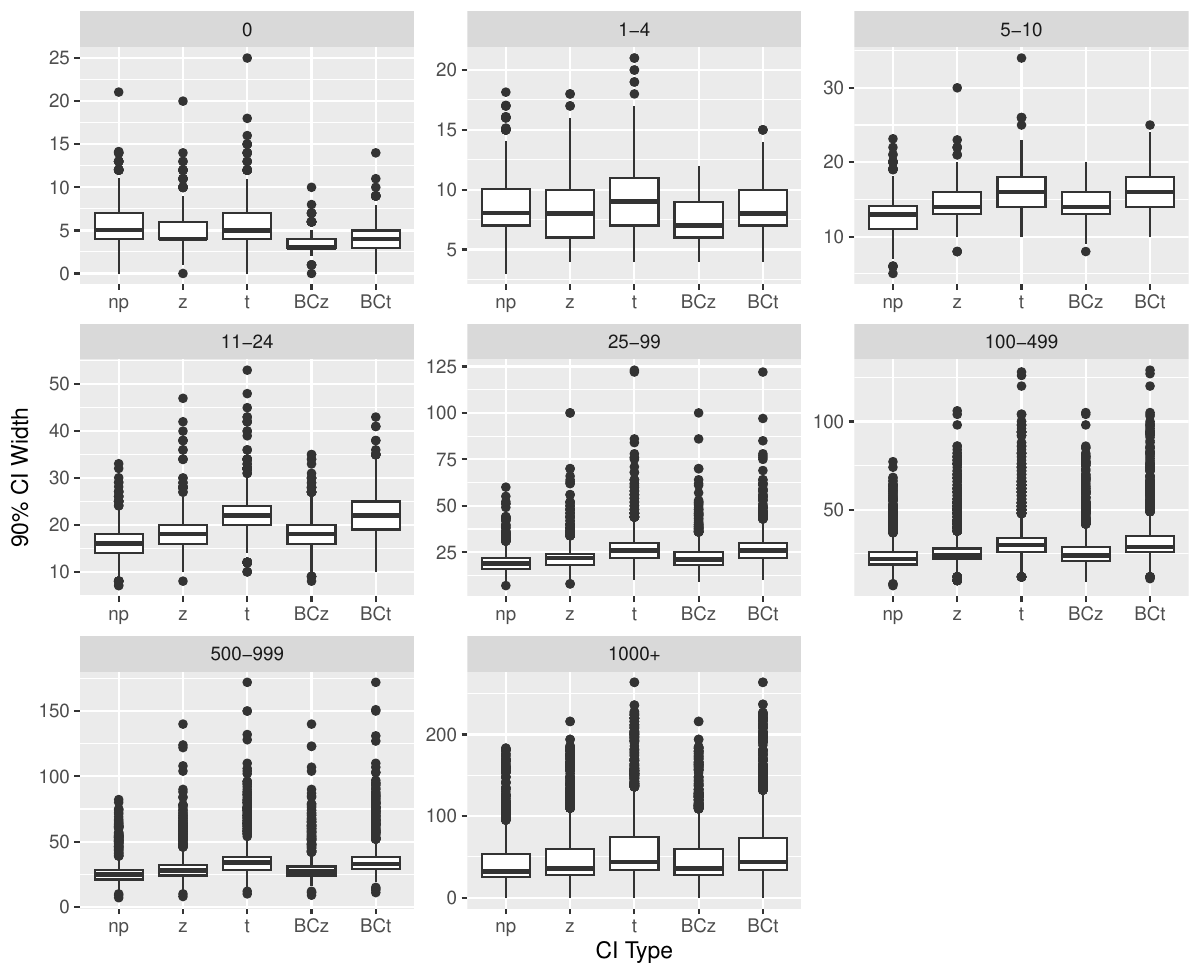}
\end{figure}

%This one in the main text
% \begin{figure}[ht!]
% \caption{The distribution of 90\% CI widths for 2010 PL94 County level queries aggregated by query size}
% \includegraphics[width =0.75\textwidth]{appendix_graphics/PL94cis/county_pl94_cis_width.pdf}
% \end{figure}

\begin{figure}[ht!]
\caption{The distribution of $22,105,741$ (301 queries x 73,441 geographies) 90\% confidence interval widths for tract PL94 queries aggregated by query size. As is to be expected, $t$ and BC$t$ intervals are slightly wider than the $z$ and BC $z$ intervals; also the confidence interval width generally increases as the size of the query increases.}
\includegraphics[width =0.75\textwidth]{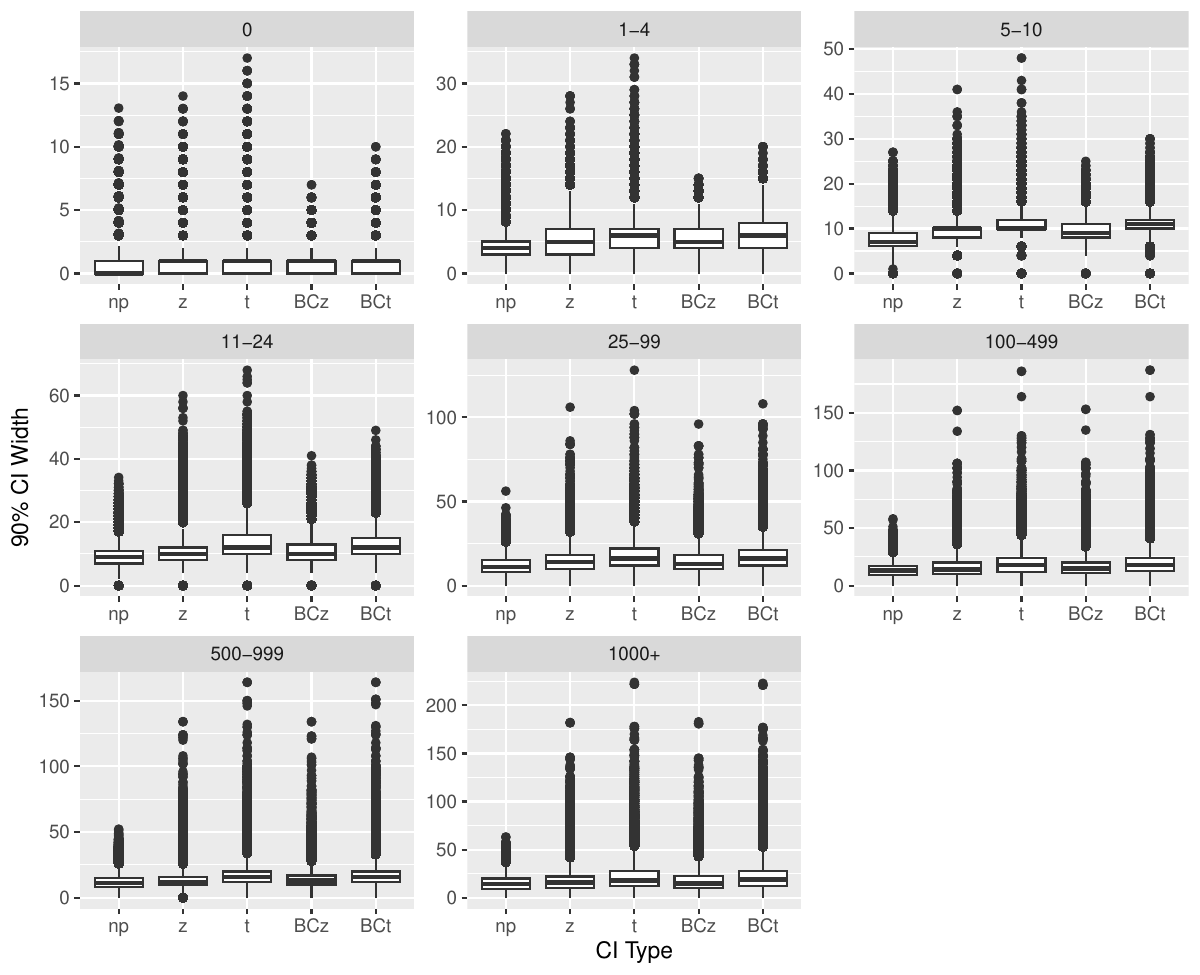}
\end{figure}

\begin{figure}[ht!]
\caption{The distribution of $780,493$ (301 queries x 2,593 geographies) 90\% confidence interval widths for block PL94 queries aggregated by query size. As is to be expected, $t$ and BC$t$ intervals are slightly wider than the $z$ and BC $z$ intervals; also the confidence interval width generally increases as the size of the query increases.}
\includegraphics[width =0.75\textwidth]{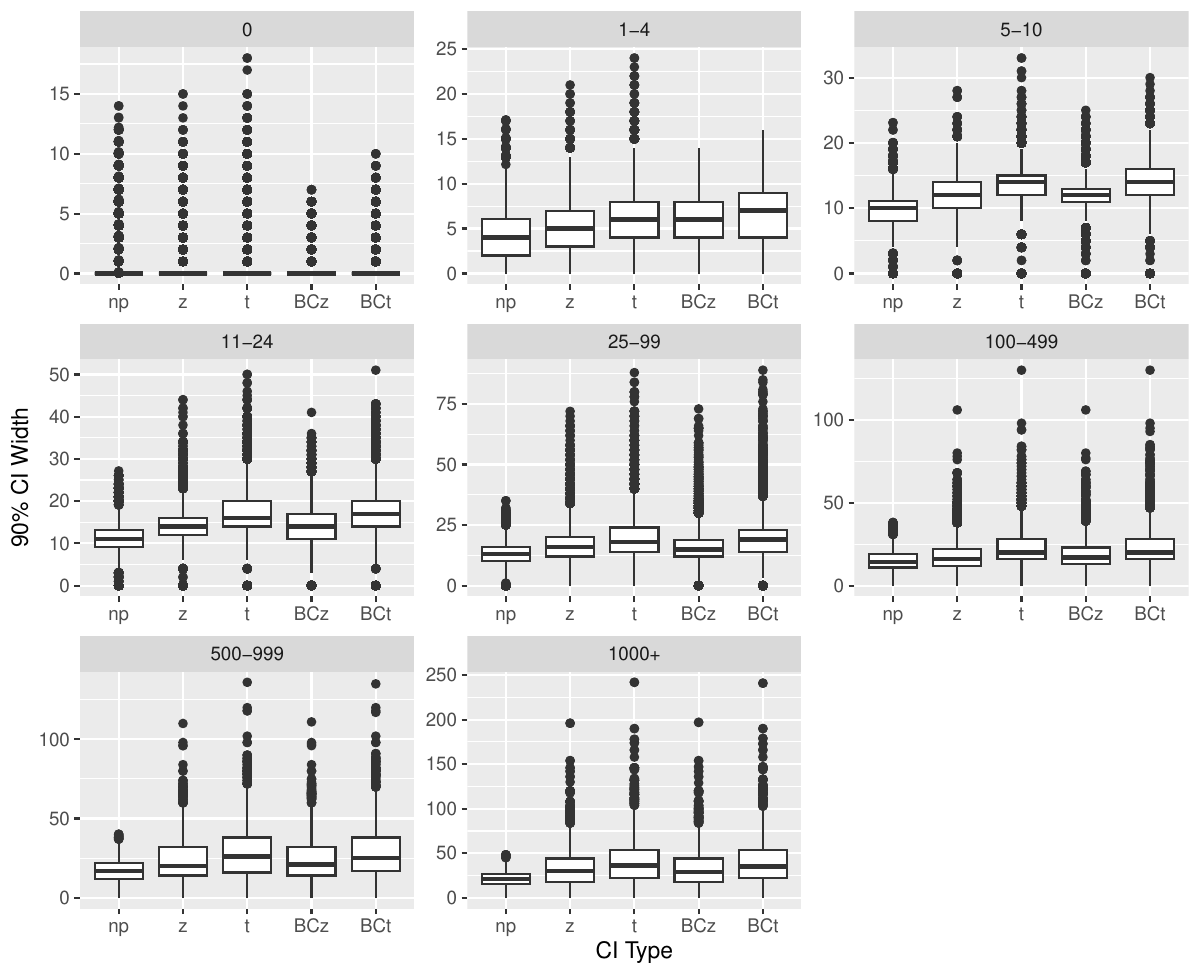}
\end{figure}

\begin{figure}[ht!]
\caption{The distribution of $186,921$ (301 queries x 621 geographies) 90\% confidence interval widths for AIAN area PL94 queries aggregated by query size. As is to be expected, $t$ and BC$t$ intervals are slightly wider than the $z$ and BC $z$ intervals; also the confidence interval width generally increases as the size of the query increases.}
\includegraphics[width =0.75\textwidth]{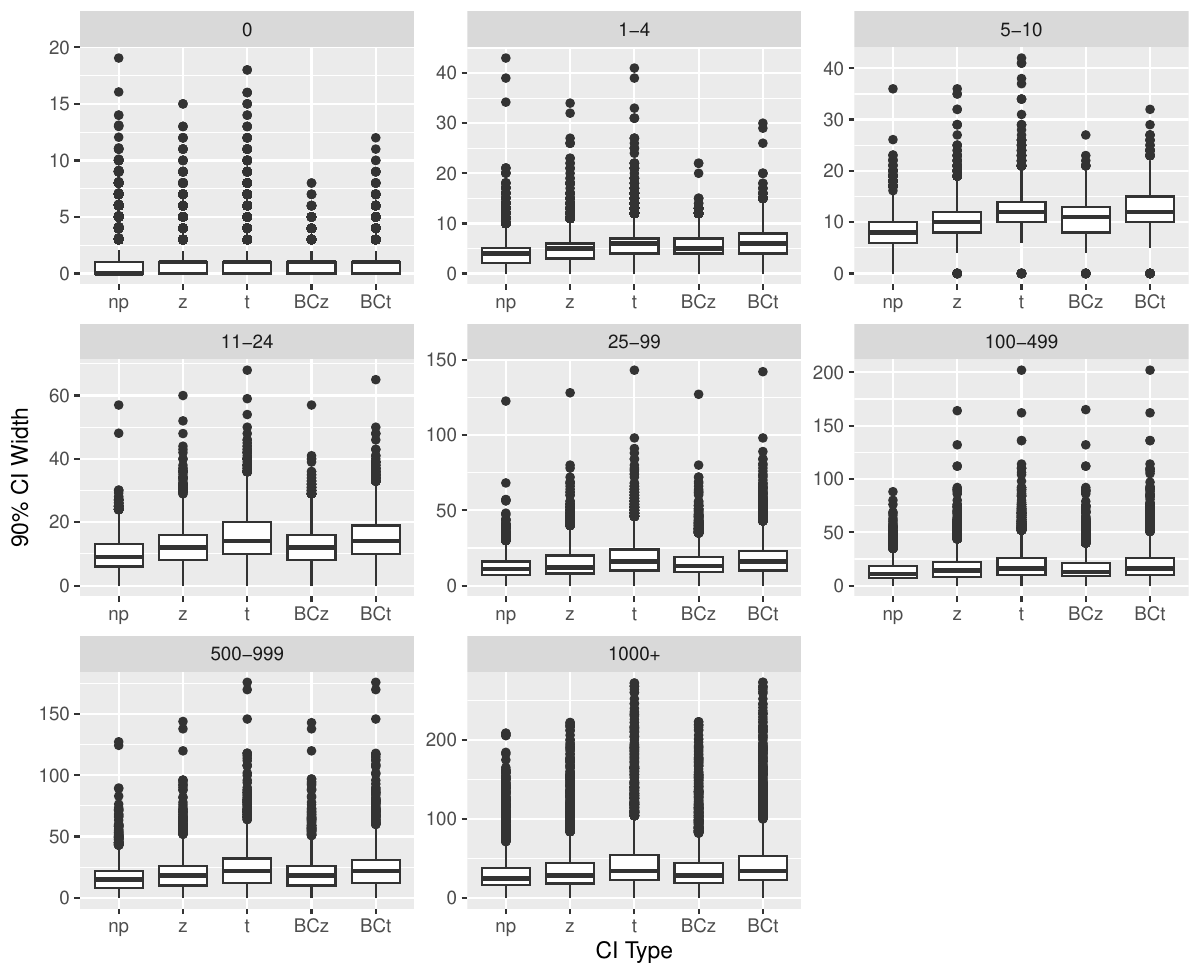}
\end{figure}

%%%%%%%%%%%%%%%%%%
%Coverage DHCP
%%%%%%%%%%%%%%%%%

%%%
\begin{figure}[ht!]
\caption{The proportion of $21,060$ (405 queries x 52 geographies) 90\% confidence intervals that contained the CEF value for state level DHCP queries aggregated by query size (the bracketed number in the lower right corner indicates the proportion of the queries in that size category, e.g. in this figure the majority, 0.73, have value of 1,000 or larger). The $t$ distribution based confidence intervals performed better than the ones using the normal distribution. Conditional bias correction (ct) performed similarly to not using bias correction (t) with both meeting the 0.90 benchmark in all query size groupings.}
\includegraphics[width =\textwidth]{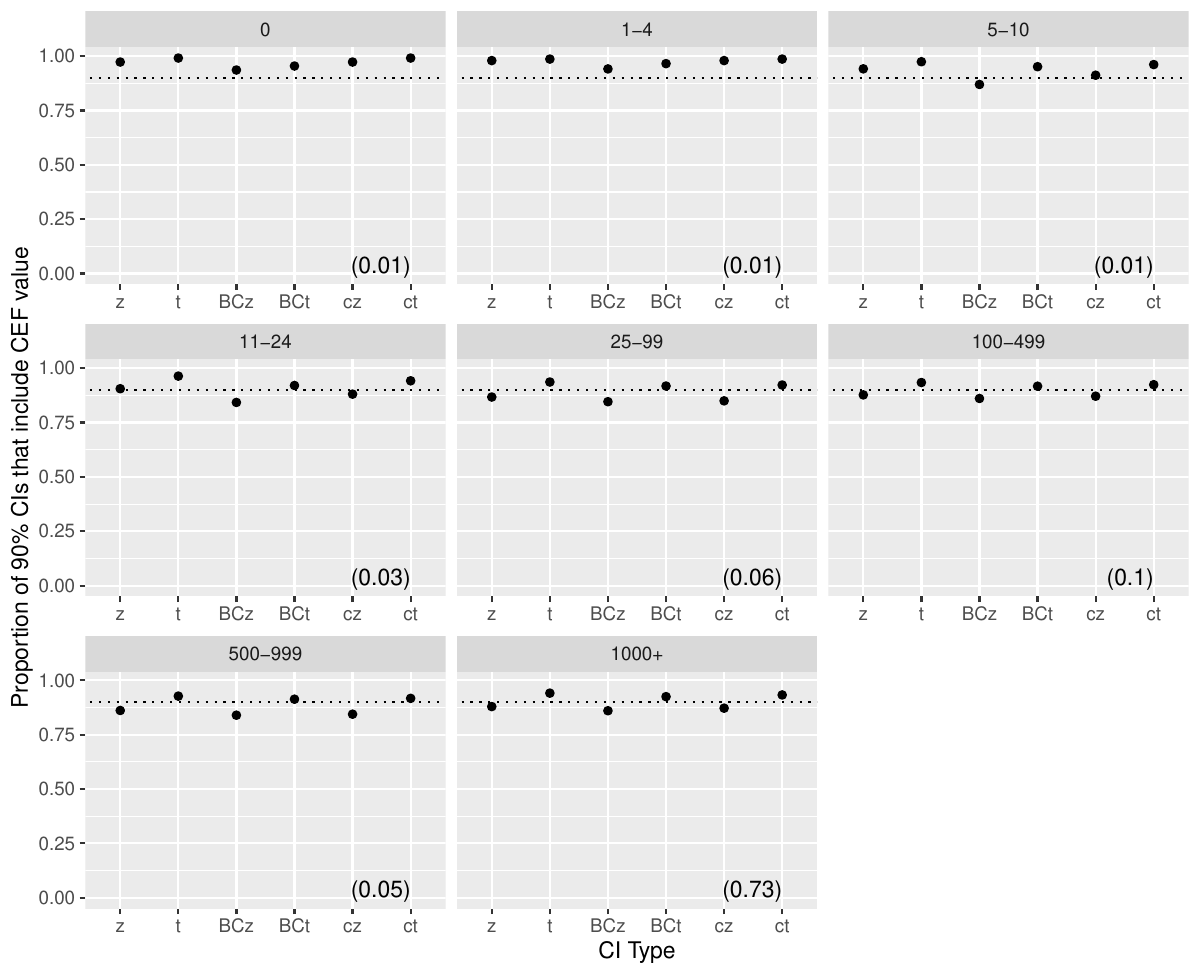}
\end{figure}

%%% This one is in the main text
% \begin{figure}[ht!]
% \caption{The proportion of 90\% CIs that contained the CEF value for 2010 County level DHCP queries aggregated by query size.}
% \includegraphics[width =0.75\textwidth]{appendix_graphics/DHCcis/county_dhcp_coverage.pdf}
% \end{figure}

%%%
\begin{figure}[ht!]
\caption{The proportion of $1,052,595$ (405 queries x 2,599 sampled geographies) 90\% confidence intervals that contained the CEF value for tract level DHCP queries aggregated by query size (the bracketed number in the lower right corner indicates the proportion of the queries in that size category, e.g. in this figure the plurality, 0.3, have value of 0). The $t$ distribution based confidence intervals performed better than the ones using the normal distribution. Conditional bias correction (ct) performed similarly to not using bias correction (t) with both meeting the 0.90 benchmark in all query size groupings.}
\includegraphics[width =\textwidth]{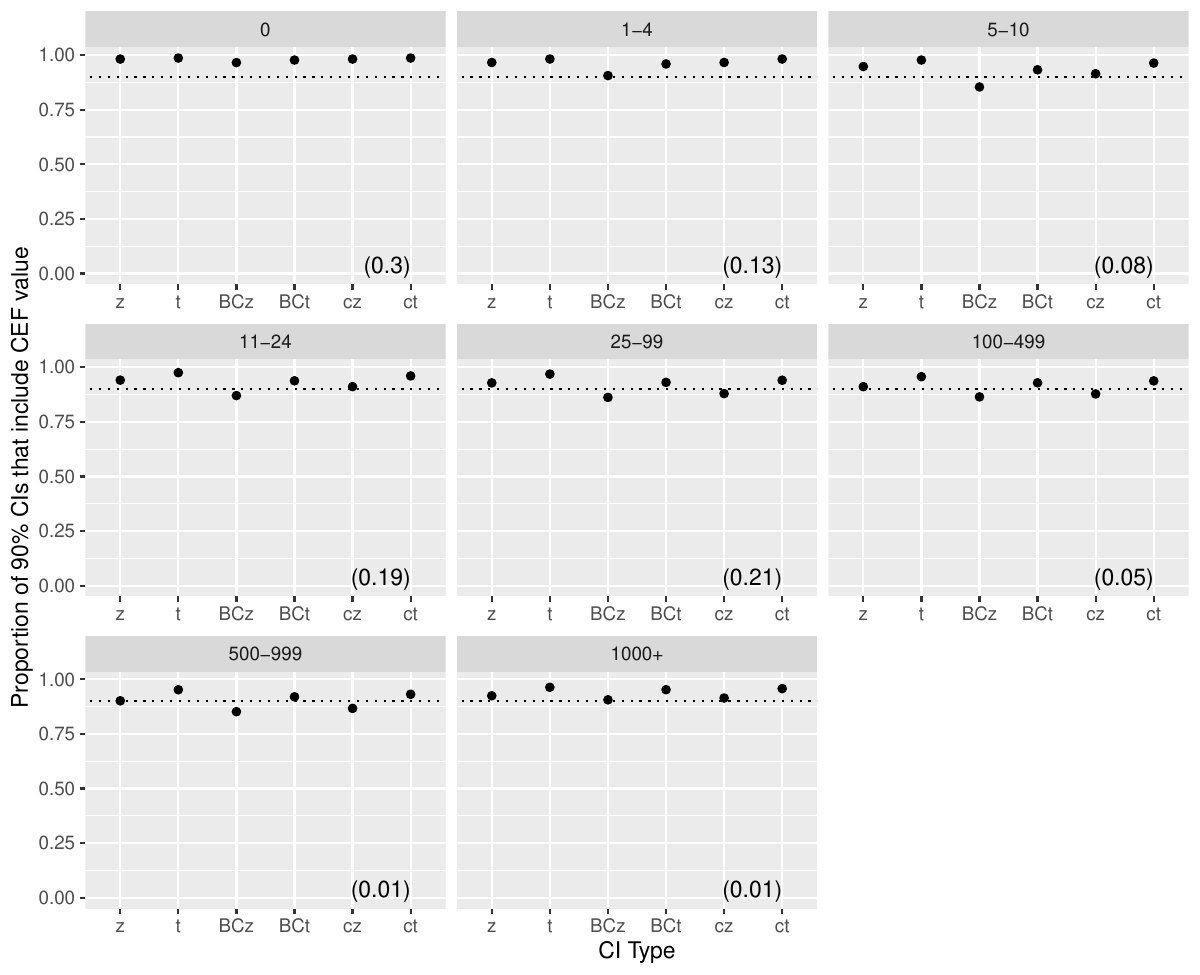}
\end{figure}

%%%
\begin{figure}[ht!]
\caption{The proportion of $1,050,165$ (405 queries x 2,593 sampled geographies) 90\% confidence intervals that contained the CEF value for block level DHCP queries aggregated by query size (the bracketed number in the lower right corner indicates the proportion of the queries in that size category, e.g. in this figure the majority, 0.85, have value of 0). The $t$ distribution based confidence intervals performed better than the ones using the normal distribution. Conditional bias correction (ct) performed substantially better than the (t) interval for 500-99 and 1000+ panels with the (z) and (t) intervals meaningfully below the $0.90$ target}
\includegraphics[width =\textwidth]{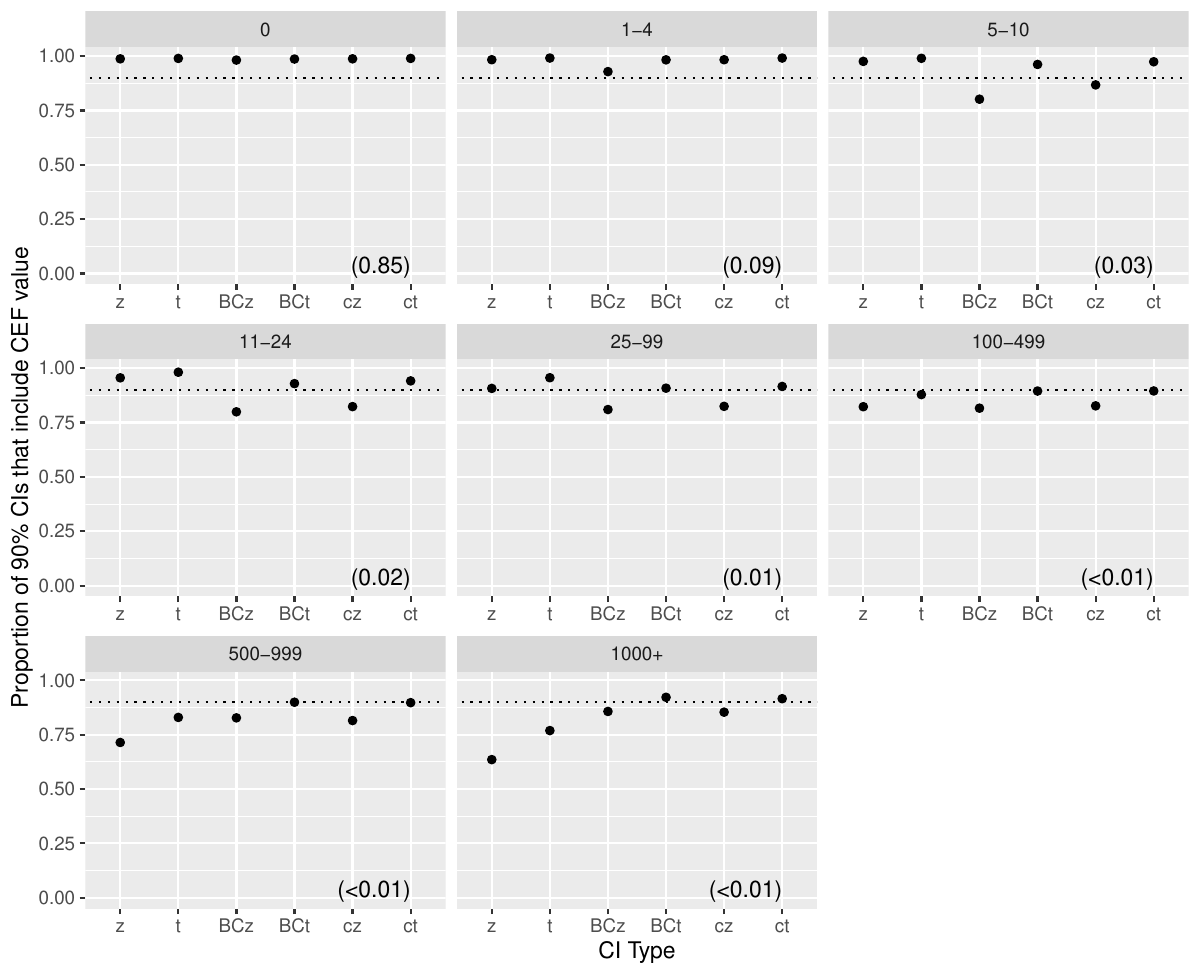}
\end{figure}

%%%
\begin{figure}[ht!]
\caption{The proportion of $251,505$ (405 queries x 621 geographies) 90\% confidence intervals that contained the CEF value for AIAN area level DHCP queries aggregated by query size (the bracketed number in the lower right corner indicates the proportion of the queries in that size category, e.g. in this figure the majority, 0.57, have value of 0). The $t$ distribution based confidence intervals performed better than the ones using the normal distribution. Conditional bias correction (ct) performed similarly to not using bias correction (t) with both meeting the 0.90 target in all query size groupings.}
\includegraphics[width =\textwidth]{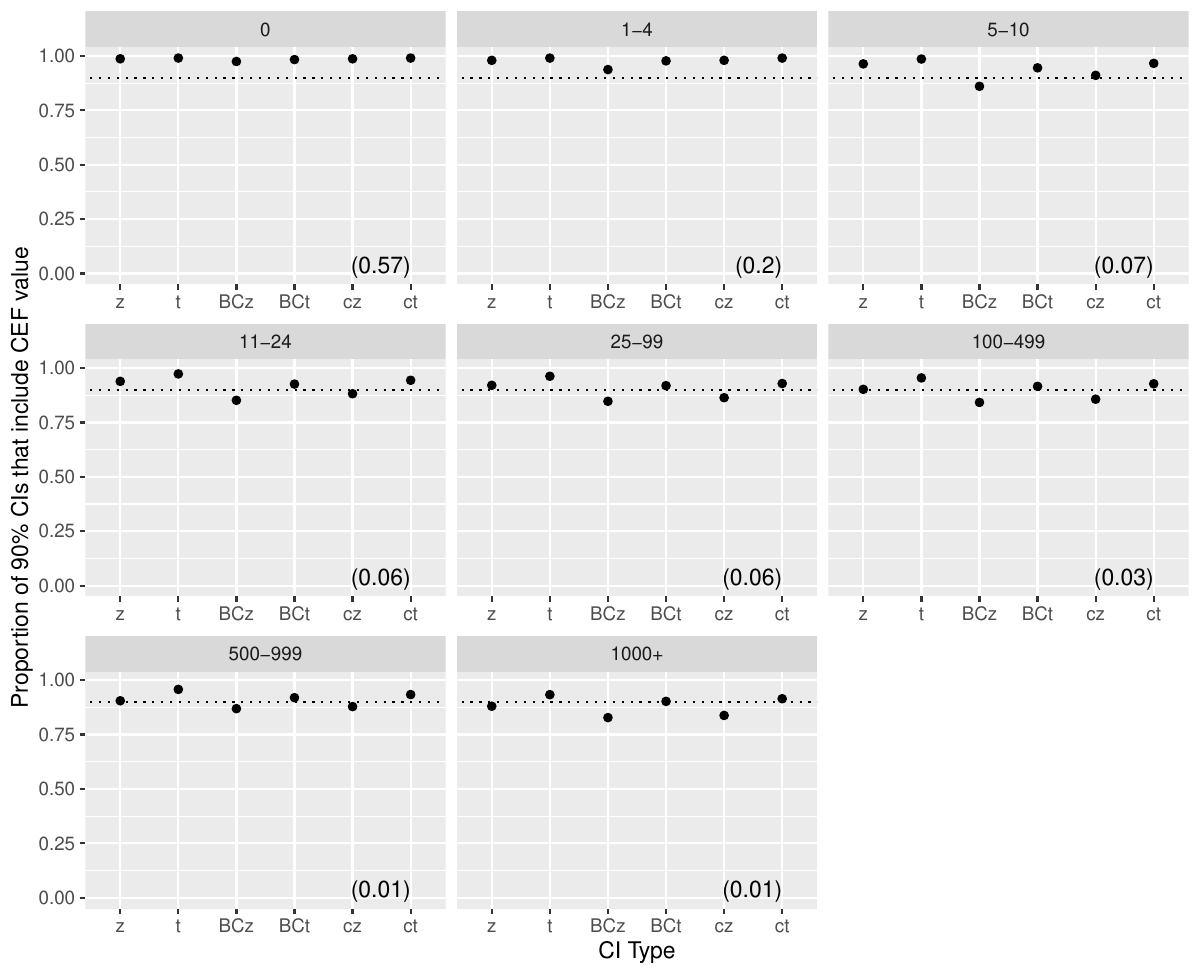}
\end{figure}

%%%

%%%%%%%%%%%%%%%%%%
%Coverage DHCH
%%%%%%%%%%%%%%%%%

%This one in main tex
% \begin{figure}[ht!]
% \caption{The proportion of 90\% CIs that contained the CEF value for 2010 County level DHCH queries aggregated by query size.}
% \includegraphics[width =0.75\textwidth]{appendix_graphics/DHCcis/county_dhch_coverage.pdf}
% \end{figure}

\begin{figure}[ht!]
\caption{The proportion of $1,050,165$ (405 queries x 2,593 sampled geographies) 90\% confidence intervals that contained the CEF value for block level DHCH queries aggregated by query size (the bracketed number in the lower right corner indicates the proportion of the queries in that (size category, e.g. in this figure the majority, 0.82, have value of 0). The $t$ distribution based confidence intervals performed substantially better than the ones using the normal distribution for many of the panels. The t, BCt, and ct were all t or near the $0.90$ target for all panels.}
\includegraphics[width =\textwidth]{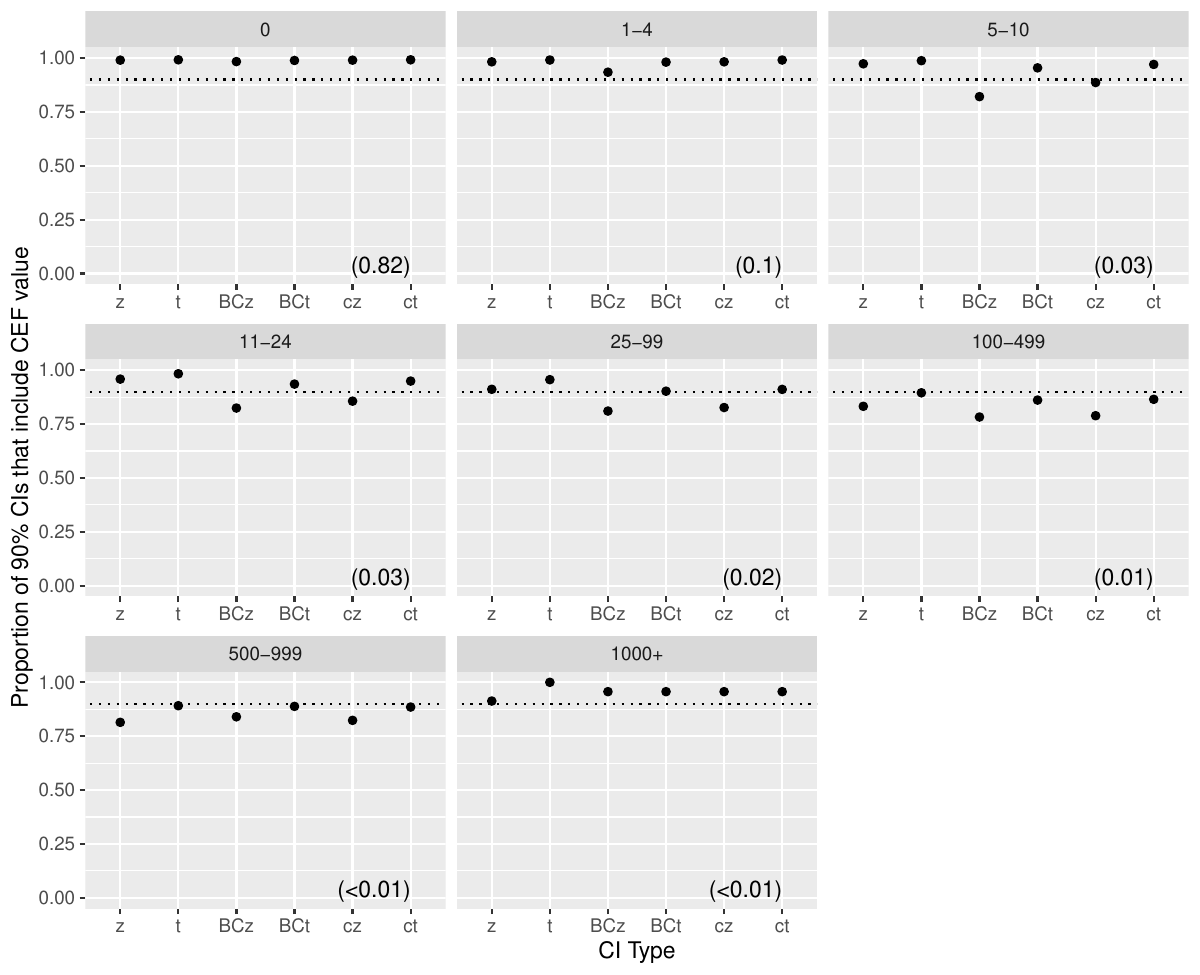}
\end{figure}

\begin{figure}[ht!]
\caption{The proportion of $251,505$ (405 queries x 621 geographies) 90\% confidence intervals that contained the CEF value for AIAN area level DHCH queries aggregated by query size (the bracketed number in the lower right corner indicates the proportion of the queries in that (size category, e.g. in this figure the majority, 0.55, have value of 0). The $t$ distribution based confidence intervals performed substantially better than the ones using the normal distribution for many of the panels. The t, BCt, and ct were all t or near the $0.90$ target for all panels.}
\includegraphics[width =\textwidth]{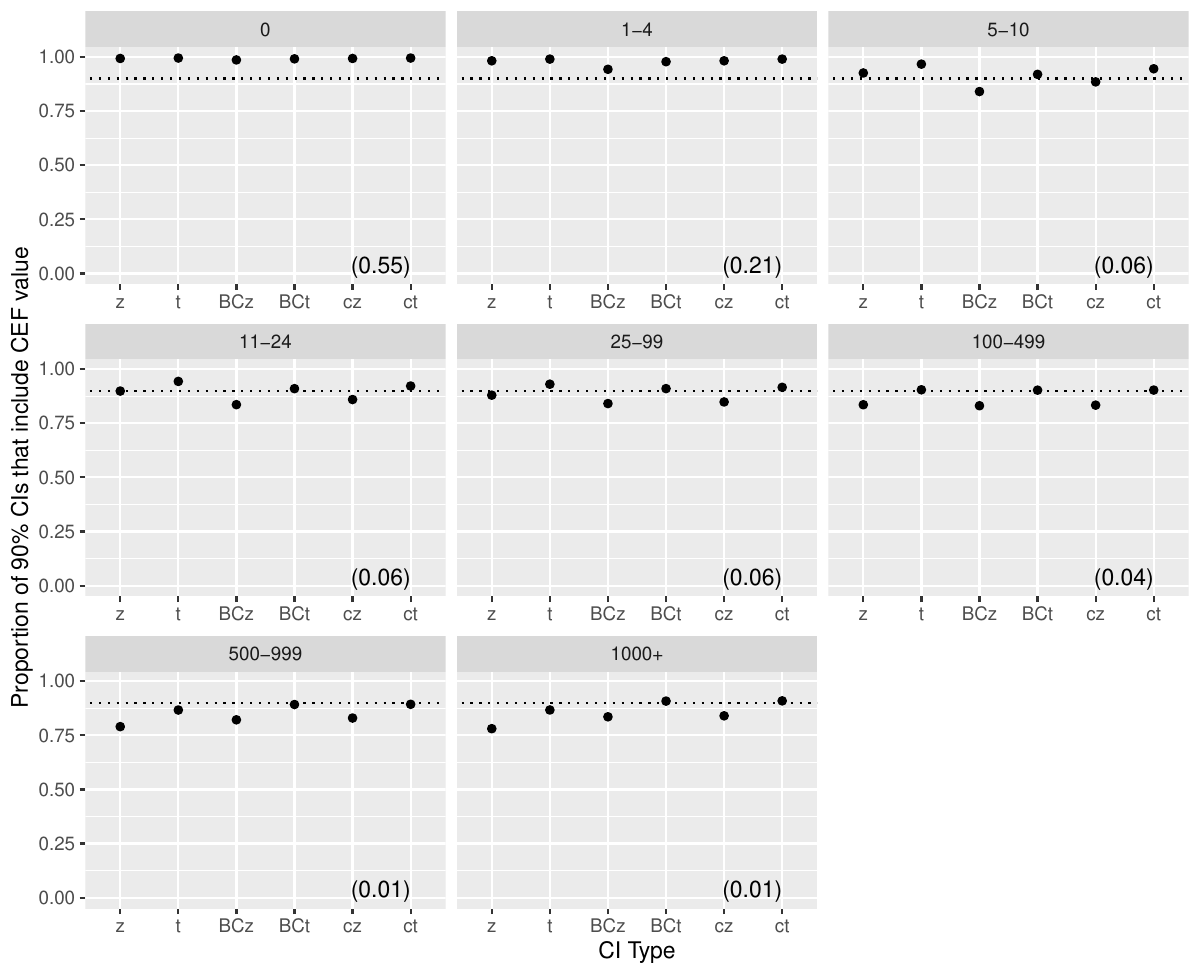}
\end{figure}

\FloatBarrier

\begin{table}[ht!]
\caption{The 1st, 50th and 99th percentile estimated bias of DHCP queries by query size for U.S. \& Puerto Rico, state, county, tract, block, AIAN area, and elementary school district geographies. There are moderate estimated biases at the extremes for  the U.S. \& Puerto Rico level for the largest query size groups. The estimated bias distributions are more concentrated around 0 for the other geographies.}
\centering
\begin{tabular}{l|rrr|rrr|rrr|rrr}
  \hline
 & \multicolumn{3}{c|}{U.S. \& Puerto Rico} & \multicolumn{3}{c|}{State} & \multicolumn{3}{c|}{County} & \multicolumn{3}{c}{Tract} \\
Query Size & 1st & 50th & 99th &  1st & 50th & 99th &  1st & 50th & 99th &  1st & 50th & 99th \\ 
0 & 0.4 & 1.1 & 1.8 & 0.1 & 1.1 & 5.4 & 0.0 & 0.2 & 3.2 & 0.0 & 0.0 & 2.3 \\ 
  1-4 & 0.4 & 1.0 & 3.5 & -1.9 & 1.0 & 6.8 & -2.6 & 0.0 & 5.0 & -2.4 & -0.1 & 4.0 \\ 
  5-10 & -0.7 & -0.2 & 4.7 & -3.5 & 0.7 & 7.5 & -4.5 & -0.2 & 6.2 & -3.8 & 0.0 & 5.1 \\ 
  11-24 & -7.5 & -1.1 & 5.7 & -6.5 & 0.8 & 9.8 & -6.5 & -0.2 & 7.2 & -5.2 & 0.0 & 5.6 \\ 
  25-99 & -9.2 & -2.9 & 14.7 & -16.5 & 0.0 & 16.3 & -8.7 & -0.1 & 8.4 & -7.4 & -0.3 & 6.0 \\ 
  100-499 & -19.4 & 0.3 & 42.8 & -59.7 & -0.4 & 34.0 & -12.2 & -0.1 & 11.3 & -8.4 & -0.0 & 7.8 \\ 
  500-999 & -29.0 & 3.5 & 47.8 & -71.8 & 0.4 & 84.3 & -17.0 & -0.1 & 16.2 & -9.6 & 0.0 & 8.9 \\ 
  1000+ & -1241.1 & -2.2 & 2310.4 & -55.5 & -1.0 & 61.2 & -27.5 & -0.1 & 26.9 & -8.0 & -0.0 & 7.7 \\ 
   \hline
    & \multicolumn{3}{c|}{Block} & \multicolumn{3}{c|}{AIAN Area} & \multicolumn{3}{c|}{School District} \\
Query Size & 1st & 50th & 99th &  1st & 50th & 99th &  1st & 50th & 99th \\ 
  \hline
0 & 0.0 & 0.0 & 2.0 & 0.0 & 0.0 & 1.9 & 0.0 & 0.0 & 2.4 \\ 
  1-4 & -3.3 & -0.6 & 5.4 & -2.3 & -0.3 & 3.3 & -2.4 & -0.0 & 4.2 \\ 
  5-10 & -6.4 & -1.6 & 7.0 & -4.1 & -0.5 & 4.6 & -4.0 & -0.2 & 5.3 \\ 
  11-24 & -9.8 & -2.0 & 7.6 & -5.8 & -0.4 & 6.4 & -6.0 & -0.2 & 6.5 \\ 
  25-99 & -14.4 & -1.7 & 9.4 & -9.0 & -0.3 & 8.5 & -9.4 & -0.2 & 9.9 \\ 
  100-499 & -19.2 & -2.9 & 9.0 & -14.5 & -0.2 & 13.4 & -17.5 & -0.2 & 17.6 \\ 
  500-999 & -26.3 & -7.1 & 4.2 & -22.8 & -0.2 & 18.1 & -26.4 & -0.2 & 25.5 \\ 
  1000+ & -42.0 & -9.4 & 2.7 & -38.6 & -0.1 & 36.4 & -43.2 & -0.3 & 38.1 \\ 
   \hline
\end{tabular}
\label{tab:dhcp_bias_tab}
\end{table}

\end{document}